\documentclass[aip,jcp,reprint,superscriptaddress,footinbib,floatfix]{revtex4-2}

\usepackage{graphicx}
\usepackage{bm}
\usepackage{amssymb}
\usepackage{amsbsy}
\usepackage{amsmath}
\usepackage{mathtools}
\usepackage{xcolor}
\usepackage{stmaryrd}
\usepackage{mathrsfs}
\usepackage{tikz}
\usetikzlibrary{shapes}
\usepackage{cancel}
\usepackage[normalem]{ulem}
\usepackage{hyperref}

\makeatletter
\let\cat@comma@active\@empty
\makeatother

\makeatletter 
\gdef\@ptsize{2}
\let\@currsize\normalsize 
\makeatother 

\newcommand{\markerfour}{\raisebox{-0.2pt}{\tikz{\node[draw=black,scale=0.6,diamond,fill=none](){};}}}

\begin{document}
\title{Shear viscosity for finitely extensible chains with fluctuating internal friction and hydrodynamic interactions}
\author{R. Kailasham}
\email{rkailash@andrew.cmu.edu}
\affiliation{Department of Chemical Engineering, Carnegie Mellon University, Pittsburgh, Pennsylvania -  15213, USA}
\author{Rajarshi Chakrabarti}
\email{rajarshi@chem.iitb.ac.in}
\affiliation{Department of Chemistry, Indian Institute of Technology Bombay, Mumbai, Maharashtra -  400076, India}
\author{J. Ravi Prakash}
\email{ravi.jagadeeshan@monash.edu}
\affiliation{Department of Chemical Engineering, Monash University,
Melbourne, VIC 3800, Australia}

\begin{abstract}
An exact solution of coarse-grained polymer models with fluctuating internal friction and hydrodynamic interactions has not been proposed so far due to a one-to-all coupling between the connector vector velocities that precludes the formulation of the governing stochastic differential equations. A methodology for the removal of this coupling is presented, and the governing stochastic differential equations, obtained by attaching a kinetic interpretation to the Fokker-Planck equation for the system, are integrated numerically using Brownian dynamics simulations. The proposed computational route eliminates the calculation of the divergence of the diffusion tensor which appears in models with internal friction, and is about an order of magnitude faster than the recursion-based algorithm for the decoupling of connector-vector velocities previously developed~[J. Rheol. \textbf{65}, 903 (2021)] for the solution of freely draining models with internal friction. The effects of the interplay of various combinations of finite extensibility, internal friction and hydrodynamic interactions on the steady-shear-viscosity is examined. While finite extensibility leads solely to shear-thinning, both internal friction and hydrodynamic interactions result in shear-thinning followed by shear-thickening. The shear-thickening induced by internal friction effects are more pronounced than that due to hydrodynamic interactions.
\end{abstract}

\maketitle

\section{\label{sec:intro} Introduction}

The time rate of spatial reorganization in polymer molecules is modulated both by solvent drag, and solvent-viscosity-independent intramolecular interactions, collectively termed as ``internal friction'' or ``internal viscosity" (IV). For example, the timescale of protein folding and reconfiguration~\cite{Ansari1992,Hagen2010385,Samanta2014,Samanta2016,Das2022}, the mechanical response of polysaccharides~\cite{Khatri20071825} and collapsed DNA globules~\cite{Murayama2007} to force spectroscopy, and the coil-stretch transition of polymer chains in turbulent flow~\cite{Vincenzi2020}, have all been shown to be affected by the presence of IV. Parallelly, the importance of accounting for solvent-mediated momentum transfer between polymer chain segments, also known as hydrodynamic interactions (HI), on the dynamics of macromolecules is also well-documented~\cite{Jendrejack2002,Larson2005,Schroeder2004,Prakash2019}. An exact solution to coarse-grained polymer models with arbitrary degrees of freedom that incorporate both IV and HI effects has so far remained elusive. In this paper, we present the derivation of the governing stochastic differential equations for such a model, outline an algorithm for its exact solution, derive a thermodynamically consistent stress tensor expression for this model, and use it to predict the steady-shear viscosity of polymer chain models with fluctuating internal friction and hydrodynamic interactions using Brownian dynamics (BD) simulations.

Polymer chains are capable of being stretched and reoriented, undergoing configurational changes both at equilibrium (representing the chain suspended in a quiescent fluid) and in the presence of a flow field. These macromolecules are routinely modeled~\cite{Bird1987b} as a linear sequence of massless beads connected by springs, where the former represent centres of friction, and the latter model the entropic elasticity of the polymer chain. The Rouse model for polymers, which consists of beads connected by Hookean springs, is solvable analytically, and predicts a non-zero value for the first normal stress difference observed in non-Newtonian polymer solutions. This rudimentary model, however, fails to predict the shear-thinning of viscosity in dilute polymer solutions. Additional constraints and mechanisms, such as the finite extensibility (FE) of the chain, the solvent-mediated transfer of momentum between chain segments, also referred to as hydrodynamic interactions (HI), and the solvent quality, have been cited and invoked in attempts to build more accurate polymer models~\cite{Bird1987b,Petera1999,Prakash2002}. The inclusion of such non-linear effects into the standard bead-spring-chain framework, however, renders an analytical solution intractable, and necessitates the use of numerical methods for its solution. Using BD simulations, it has been possible by incorporating these molecular scale phenomena, to obtain a quantitative, parameter-free agreement between computational predictions and experimental observations for the extension of DNA solutions in elongational flow~\cite{Sunthar2005,Sunthar2005b,Saadat2015,Sasmal2016}. None of the nonlinear effects mentioned above, however, can explain the discontinuous jump in stress of polymer solutions at the inception of flow (termed ``stress jump")~\cite{Mackay1992,liang1993stress}, which is hypothesized to originate from the frictional resistance offered by the polymer chain to short-time-scale variations in its conformation~\cite{kuhn1945bedeutung,Peterlin1967,Manke1988,Fixman1988}.

The energy difference between the $trans$ and $gauche$ conformations determines the static flexibility, or persistence length of polymers~\cite{Rubinstein2003}. The activation barrier separating these two states, represents the resistance to dihedral angle rotations, and determines the timescale below which the molecule appears rigid and resists changes to its configuration~\cite{degennes,Manke1985}. This resistance to dihedral angle rotations had been suspected, since the inception of polymer kinetic theory, to be a source of internal friction~\cite{kuhn1945bedeutung}. Molecular dynamics simulations~\cite{DeSancho2014,Echeverria2014} and recent experiments on intrinsically disordered proteins~\cite{Das2022} appear to confirm this notion. This rate-dependent force that resists relative motion between chain segments is incorporated into the bead-spring-chain model by the addition of viscous dashpots~\cite{kuhn1945bedeutung,Booij1970,Hua1995,Manke1988,ravibook} or dampers in parallel with the springs. The inclusion of dashpots, however, results in a coupling of the velocities of the connector vectors that join adjacent beads, rendering an exact solution infeasible for all but the simplest case of a dumbbell model~\cite{Hua1995,Hua1996,Kailasham2018} (two beads connected by a spring). Such a coupling precludes both the formulation of a Fokker-Planck equation for the configurational distribution function of the chain, and by extension, the derivation of the stochastic differential equations governing the motion of beads. 

~\citet{Manke1988} derived a semi-analytical approximation for the stress jump of free-draining bead-spring-dashpot chains, using a recursive-algorithm for the decoupling of bead velocities, leveraging the fact that in the absence of HI only the velocities of nearest neighbors in the chain are coupled, and restricted their analysis to the linear viscoelastic regime. In a prior work~\cite{Kailasham2021}, we used the decoupling methodology developed by~\citet{Manke1988} to derive and solve the exact set of stochastic differential equations governing the motion of free-draining bead-spring-dashpot chains that is valid both at equilibrium and in the presence of flow. By comparison against BD simulation results, it was established that the~\citet{Manke1988} prediction for the stress jump improves with the number of beads in the chain. The framework developed by Manke and Williams~\cite{Manke1988,Manke1992,Dasbach1992} is applicable, however, only for linear viscoelastic predictions, and numerical simulations are required, as explained in ref.~\citenum{Kailasham2021}, for the calculation of viscometric functions in the presence of a flow field.

The exact formulation and solution of free-draining bead-spring-dashpot chains permitted an investigation~\cite{Kailasham2021sm} of the Rouse model with internal friction~\cite{Khatri2007rif,Khatri20071825} (RIF), a widely used theoretical framework for the interpretation of internal friction effects in biophysical contexts which relies on a preaveraged treatment of internal friction. While the RIF model predicts that the relaxation time of the end-to-end vector in a chain diverges in the asymptotic limit of infinite IV, the incorporation of fluctuations in the IV force results, more realistically, in a finite value for the relaxation time under the same limit. Furthermore, the viscosity of the preaveraged model is found to be independent of the shear rate, while the exact model predicts a shear-rate dependent viscosity that undergoes both thinning and thickening. Additionally, the importance of accounting for fluctuations in HI for the estimation of viscometric functions is now well-understood~\cite{Zylka1989,Zylka1991,Prabhakar2006}.

The simultaneous inclusion of fluctuating internal friction and hydrodynamic interaction effects, however, results in a one-to-all coupling of the connector vector velocities, which does not permit the use of the decoupling approach developed previously for free-draining models with internal friction. Fixman remarked~\cite{Fixman1986} that the solution of such a model would require ``complicated matrix operations that may well be impractical unless preaveraging of the matrices is introduced." Using a preaveraged version of the hydrodynamic interaction tensor, ~\citet{Manke1992} applied the decoupling algorithm to solve for the stress jump of such chains in planar and elongational flow, and also derived semi-analytical results for material functions in small amplitude oscillatory shear flow~\cite{Dasbach1992}. The accuracy of these approximate results have remained untested from a theoretical standpoint, due to the unavailability of exact solutions to the bead-spring-dashpot chain model with fluctuating hydrodynamic interactions. In this paper, we prescribe an exact solution methodology for this model, by developing a numerically efficient algorithm for computing the inverse of the effective friction tensor, and present preliminary rheological results. The present algorithm may also be applied to the free-draining IV model and is found to be an order of magnitude faster than the decoupling-methodology-based implementation developed recently~\cite{Kailasham2021}.

A distinguishing feature of models with internal friction is the appearance of the divergence of the diffusion tensor in the governing stochastic differential equation obtained from an It\^{o} interpretation of the underlying Fokker-Planck equation. When only first order interactions are retained in the expression for the hydrodynamic mobility tensor, this divergence vanishes, but such a simplification does not hold when IV is included. ~\citet{Hutter1998} prescribe a numerical algorithm for the solution of such stochastic differential equations that relies on the kinetic interpretation~\cite{Klimontovich1990,KLIMONTOVICH1992,Schieber1992} of the governing Fokker-Planck equation. This route replaces the calculation of the divergence of the diffusion tensor by that of its inverse, and has been recommended as being numerically more efficient except for cases in which a closed-form expression for the divergence is available. This approach has been used predominantly in simulations of colloidal particles and suspensions~\cite{Chau2008,DeCorato2015,DeCorato2016}, and polymer chain models with constraints~\cite{Lang2014}, such as the bead-rod-model. We show that the coupling of connector vector velocities in our model may be removed by the use of simple linear algebra, and employ the technique suggested by~\citet{Hutter1998} for the numerical integration of the resulting stochastic differential equations. 

While the time-evolution of the configuration of a polymer chain model with fluctuating IV and HI may be obtained from simulations without explicit calculation of the divergence of the diffusion tensor, the thermodynamically consistent stress tensor expression for this model, however, contains the divergence term. We employ the random finite difference (RFD) algorithm, developed by Donev and coworkers~\cite{Sprinkle2017,Sprinkle2019} for the calculation of this divergence term. We discuss later in Sec.~\ref{sec:eqns}, why nevertheless, the kinetic interpretation is still preferable to evaluating the divergence term, while solving the governing stochastic differential equation.

The most iconic rheological characteristic of dilute polymer solutions, namely, the shear-thinning of viscosity, has been attributed to several nonlinear intramolecular effects, including but not limited to: finite extensibility of the polymer chain, excluded volume effects~\cite{Petera1999,Prakash2002} (EV), hydrodynamic interactions~\cite{Prabhakar2004,Pincus2020}, and internal friction. Preliminary investigations of the steady-shear rheology of bead-spring-dashpot dumbbell models relied on a Gaussian approximation of IV~\cite{Wedgewood1993,Schieber1993,Sureshkumar1995}, predicting a shear-thinning of viscosity following a constant Newtonian plateau. Subsequent (exact) BD simulations~\cite{Hua1995,Hua1996,Kailasham2018,Kailasham2021} established that internal friction causes shear-thinning at low-to-moderate shear rates, followed by a thickening of the viscosity at higher shear rates. Finite extensibility is known to exclusively induce shear-thinning~\cite{Bird1987b}, while the inclusion of hydrodynamic interactions in coarse-grained models with greater than six beads has been shown to result in shear-thinning followed by thickening~\cite{Kishbaugh1990,Zylka1991,Prabhakar2006}. In this paper, we use an exact model that accounts for fluctuations in internal viscosity and hydrodynamic interactions, to systematically disentangle the impact of FE, IV, and HI on the steady-shear viscosity profile of bead-spring-dashpot chains.

 \begin{figure}[t]
\centering
\includegraphics[width=3.3in,height=!]{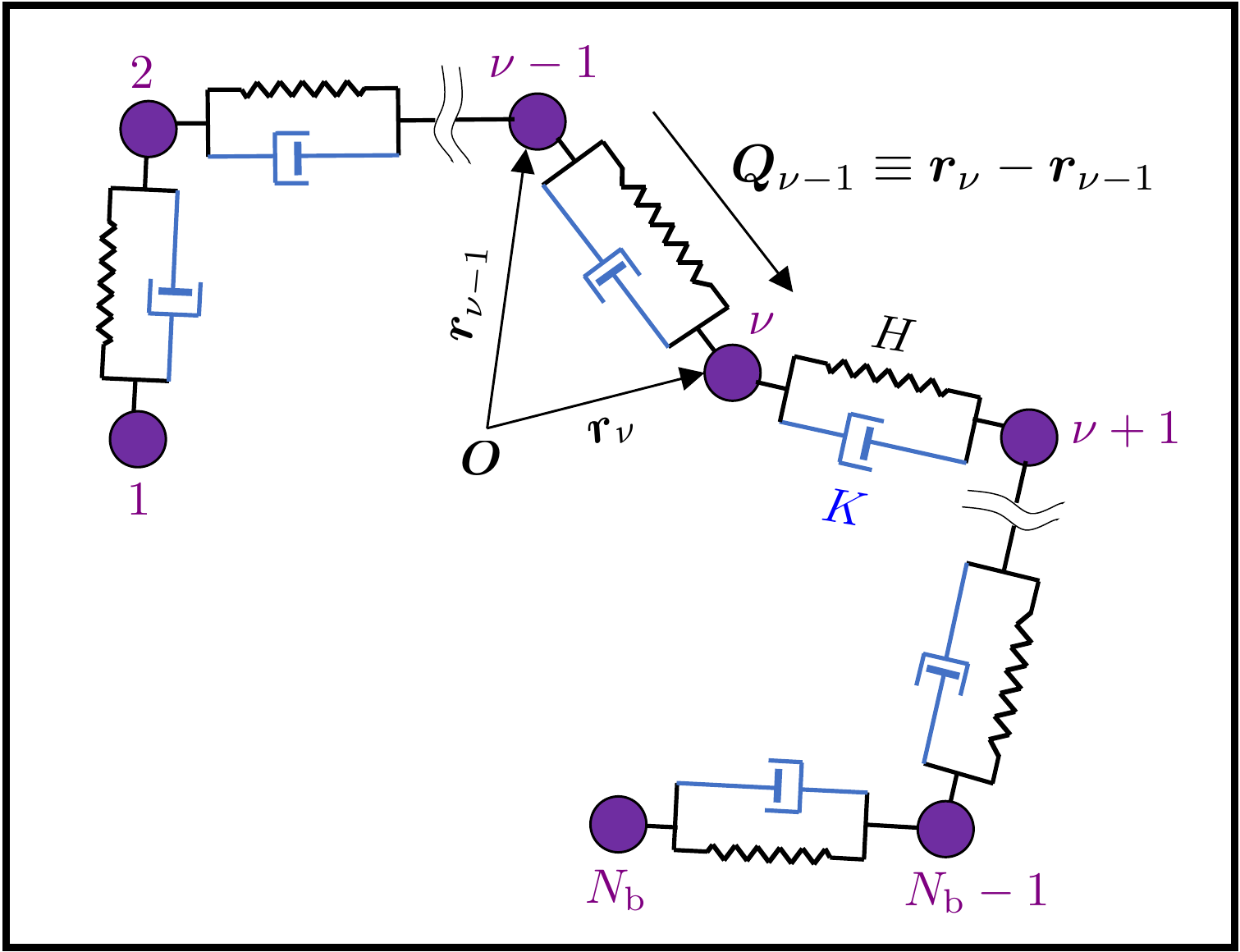}  
\caption{Micromechanical model for a polymer chain, consisting of a sequence of beads connected by spring-dashpots. Each spring is associated with a Hookean spring constant $H$, and the damping coefficient of each dashpot is $K$.}
\label{fig:model}
\end{figure}

The rest of the paper is structured as follows. Sec.~\ref{sec:eqns} describes the bead-spring-dashpot chain model for a polymer, presents the governing stochastic differential equations and the stress tensor expression, outlines simulation details pertaining to the numerical integration of the governing equations, and discusses the rationale behind the choice of model parameters. Sec.~\ref{sec:results}, which is a compilation of our results and the relevant discussion, is divided into four sections; Sec.~\ref{sec:code_valid} deals with code validation, Sec.~\ref{sec:sjump} presents results for the stress jump at the inception of shear flow, Sec.~\ref{sec:steady_sr} contains results for steady shear viscometric functions, followed by a comparison to experimental data in Sec.~\ref{sec:exp_compare}. We conclude in Sec.~\ref{sec:conclusions}. 

\section{\label{sec:eqns} Numerical Algorithm}

\subsection{\label{sec:gov_eq} Governing equations}

We consider a micromechanical model for a linear polymer chain that consists of $N_{\text{b}}$ massless beads, each of radius $a$, joined by $N\equiv\left(N_{\text{b}}-1\right)$ springs. The Hookean spring constant associated with each spring is $H$, and the dashpot in parallel with each spring has a damping coefficient of $K$, as shown in Fig.~\ref{fig:model}. The position of the $i^{\text{th}}$ bead is denoted as $\bm{r}_{i}$, and the connector vector joining adjacent beads is represented as $\bm{Q}_{i-1}\equiv\bm{r}_{i}-\bm{r}_{i-1}$. The chain, as shown in Fig.~\ref{fig:model}, is suspended in a Newtonian solvent of viscosity $\eta_{\text{s}}$ where the velocity $\bm{v}_{\text{f}}$ at any location $\bm{r}_{\text{f}}$ in the fluid is given by
$\bm{v}_{\text{f}}(\bm{r}_{\text{f}},t)\equiv\,\bm{v}_{0}+\boldsymbol{\kappa}(t)\cdot\bm{r}_{\text{f}}$, where $\bm{v}_{0}$ is a constant vector, and the transpose of the velocity gradient tensor is denoted as $\boldsymbol{\kappa}\equiv\left(\nabla\bm{v}_{\text{f}}\right)^{T}$. The chain is assumed to have equilibrated in momentum space, and its normalized configurational distribution function at any time $t$ is given by $\Psi\equiv\Psi\left(\bm{r}_1,\bm{r}_2,...,\bm{r}_{{N_{\text{b}}}},t\right)=\left(1/\mathcal{Z}\right)\exp\left[-\phi/k_BT\right]$, where $\phi$ represents the intramolecular potential energy stored in the springs joining the beads, $k_B$ is Boltzmann's constant, $T$ the absolute temperature, and the partition function $\mathcal{Z}=\int\exp\left[-\phi/k_BT\right]d\bm{Q}_1d\bm{Q}_2\dots\,d\bm{Q}_{N}$. The force on a bead $i$ due to the spring potential is given by $\bm{F}^{(\phi)}_{i}=-\partial \phi/\partial \bm{r}_i$, while that in the $k^{\text{th}}$ connector vector, joining the $k^{\text{th}}$ and $(k+1)^{\text{th}}$ bead, is denoted as $\bm{F}^{(\text{c})}_{k}=\partial \phi/\partial \bm{Q}_k$~\cite{Bird1987b}. Both Hookean and finitely extensible nonlinear elastic (FENE) springs are considered in this work, and the functional form of these spring force laws are provided below. The expression for the internal viscosity force, $\bm{F}^{\text{IV}}_{k}$, in the $k^{\text{th}}$ connector vector may be written as 
$\bm{F}^{\text{IV}}_{k}=K\left({\bm{Q}_k\bm{Q}_k}/{Q_k^2}\right)\cdot\llbracket\dot{\bm{Q}}_k\rrbracket$,
where $\llbracket\dots\rrbracket$ denotes an average over momentum-space. Within the framework of polymer kinetic theory~\cite{Bird1987b}, the Fokker-Planck equation for the configurational distribution function is obtained by combining a force-balance on the beads (or connector vectors) with a continuity equation in probability space. The force-balance dictates that the sum of the restoring force from the spring, the internal friction force due to the dashpot, the random Brownian force arising from collisions with solvent molecules, and the hydrodynamic force which represents the solvent's resistance to the motion of the bead, equals zero. The friction coefficient associated with each bead is given by $\zeta=6\pi\eta_{\text{s}}\,a$, and is used to define the timescale for the model, $\lambda_{H}=\zeta/4H$. The length-scale is taken to be $l_{H}=\sqrt{k_BT/H}$. Dimensionless quantities are denoted with an asterisk as superscript, for example, $a^{*}=a/l_{H}$.

In the following derivation, summations are indicated explicitly, and the Einstein convention is not adopted. For a chain with $N$ springs, the following equations of motion for the velocity of the connector vector and the centre of mass has been derived in Ref.~\citenum{ravibook}, 
\begin{widetext}
 \begin{align}\label{eq:qdot}
 \llbracket\dot{\bm{Q}}_{j}\rrbracket&=\boldsymbol{\kappa}\cdot\bm{Q}_j-\dfrac{1}{\zeta}\sum^{N}_{k=1}\tilde{\bm{A}}_{jk}\cdot\left(k_BT\dfrac{\partial \ln \Psi}{\partial \bm{Q}_k}+\dfrac{\partial \phi}{\partial \bm{Q}_k}+K\dfrac{\bm{Q}_k\bm{Q}_k}{\bm{Q}^2_k}\cdot\llbracket\dot{\bm{Q}}_{k}\rrbracket\right)
 \end{align}
  \begin{align}\label{eq:rc_dot}
 \llbracket\dot{\bm{r}}_{\text{c}}\rrbracket&=\boldsymbol{v}_{o}+\boldsymbol{\kappa}\cdot{\bm{r}}_{\text{c}}-\dfrac{1}{N_{\text{b}}\zeta}\sum^{N_{\text{b}}}_{\mu,\nu=1}\sum^{N}_{k=1}\bar{B}_{k\mu}\left(\delta_{\mu\nu}\boldsymbol{\delta}+\zeta\boldsymbol{\Omega}_{\mu\nu}\right)\cdot\left(k_BT\dfrac{\partial \ln \Psi}{\partial \bm{Q}_k}+\dfrac{\partial \phi}{\partial \bm{Q}_k}+K\dfrac{\bm{Q}_k\bm{Q}_k}{\bm{Q}^2_k}\cdot\llbracket\dot{\bm{Q}}_{k}\rrbracket\right)
 \end{align}
 where 
 \begin{align}\label{eq:a_tilde_def}
 \tilde{\bm{A}}_{jk}&={A}_{jk}\bm{\delta}+\zeta\left(\bm{\Omega}_{j,k}+\bm{\Omega}_{j+1,k+1}-\bm{\Omega}_{j,k+1}-\bm{\Omega}_{j+1,k}\right)\\[5pt]
 \bar{B}_{k\mu}&=\delta_{k+1,\mu}-\delta_{k\mu}
 \end{align}
 and ${A}_{jk}$ are the elements of the Rouse matrix, given as
 \begin{align}\label{eq:a_matrix_def}
A_{jk}= \left\{
\begin{array}{ll}
       2; &  j=k \\[15pt]
      -1; & |j-k|=1 \\[15pt]
       0 ; & \text{otherwise}
\end{array} 
\right. 
\end{align}
The general form of the symmetric hydrodynamic interaction tensor, $\boldsymbol{\Omega}_{\mu\nu}$, is given by
\begin{align}\label{eq:hi_tensor_gen}
\boldsymbol{\Omega}_{\mu\nu}= \dfrac{3a}{4\zeta\,r_{\mu\nu}}\Biggl\{\mathscr{A}\boldsymbol{\delta}+\mathscr{B}\dfrac{\boldsymbol{r}_{\mu\nu}\boldsymbol{r}_{\mu\nu}}{\boldsymbol{r}^2_{\mu\nu}}\Biggr\}
 \end{align}
 with $\boldsymbol{r}_{\mu\nu}=\boldsymbol{r}_{\nu}-\boldsymbol{r}_{\mu}$ denoting the interbead separation
and the coefficients $\mathscr{A}$ and $\mathscr{B}$, according to the Rotne-Prager-Yamakawa (RPY) definition~\citep{Rotne1969,Yamakawa}, are given by 
 \begin{equation}\label{eq:compdef_rpy}
 \begin{split}
 \mathscr{A}= \left(1+\dfrac{2a^2}{3r^{2}_{\mu\nu}}\right),\quad\mathscr{B}=\left(1-\dfrac{2a^2}{r^2_{\mu\nu}}\right)\quad \text{for}\quad r_{\mu\nu} \geq 2a\\[5pt]
 \mathscr{A}=\dfrac{r_{\mu\nu}}{2a}\left(\dfrac{8}{3}-\dfrac{3r_{\mu\nu}}{4a}\right),\quad\mathscr{B}=\dfrac{1}{8}\left(\dfrac{r_{\mu\nu}}{a}\right)^2\quad\text{for}\quad r_{\mu\nu} < 2a
 \end{split}
 \end{equation}
 Self-interactions are suppressed by requiring that $\boldsymbol{\Omega}_{\mu\nu}=\bm{0}$ for $\mu=\nu$. We define the collective coordinates
\begin{align}\label{eq:coll_coord}
\bm{\mathcal{Q}}&\equiv\left[\bm{Q}_1,\,\bm{Q}_2,\,...,\,\bm{Q}_N\right]\nonumber\\[5pt]
&\equiv\left[Q^{1}_1,Q^{
2}_1,Q^{3}_1,Q^{1}_2,Q^{2}_2,\,...,\,Q^{3}_{N}\right]
\end{align}
and write $\mathcal{Q}_{i}=Q^{\omega}_{j}$, where $j=1,2,...,N$ and $\omega=1,2,3$ represent Cartesian components in the $x,y,z$ directions, respectively, with $i$ related to $j$ and $\omega$ as $i=3\left(j-1\right)+\omega$. Similarly, $\bm{\mathcal{F}}^{(\text{c})}\equiv\left[\bm{F}^{(\text{c})}_{1},\,\bm{F}^{(\text{c})}_{2},\,...,\,\bm{F}^{(\text{c})}_{N}\right]$, with $\bm{F}^{(\text{c})}_{j}=\left(\partial \phi/\partial \bm{Q}_{j}\right)$. 
In terms of the collective coordinates, eq.~(\ref{eq:qdot}) may be recast as,
\begin{equation}\label{eq:rev_1_1}
\llbracket\dot{\bm{\mathcal{Q}}}\rrbracket=\bm{\mathcal{K}}\cdot\bm{\mathcal{Q}} - \bm{\mathcal{A}}\cdot\left(\dfrac{k_BT}{\zeta}\dfrac{\partial \ln\Psi}{\partial\bm{\mathcal{Q}}}+\dfrac{1}{\zeta}\bm{\mathcal{F}}^{(\text{c})}+\bm{\mathcal{V}}\cdot\llbracket\dot{\bm{\mathcal{Q}}}\rrbracket\right),
\end{equation}
where $\bm{\mathcal{K}}, \bm{\mathcal{A}}, $ and $\bm{\mathcal{V}}$ are block matrices of size $N\times N$. The entries of $\bm{\mathcal{A}}$ are given by eq.~(\ref{eq:a_tilde_def}), while $\bm{\mathcal{K}}$ and $\bm{\mathcal{V}}$ are block-diagonal matrices with the off-diagonal components set to $\bm{0}$, and diagonal entries given by $\boldsymbol{\kappa}$ and $\left[\varphi\left(\bm{Q}_{j}\bm{Q}_{j}/Q^2_{j}\right)\right]$, respectively, where we have defined $\varphi=\left(K/\zeta\right)$.
Simplifying eq.~(\ref{eq:rev_1_1}), we get 
\begin{equation}\label{eq:rev_1_3}
\left(\bm{\mathcal{A}}^{-1}+\bm{\mathcal{V}}\right)\cdot\llbracket\dot{\bm{\mathcal{Q}}}\rrbracket=\bm{\mathcal{A}}^{-1}\cdot\left(\bm{\mathcal{K}}\cdot\bm{\mathcal{Q}}\right) - \dfrac{k_BT}{\zeta}\dfrac{\partial \ln\Psi}{\partial\bm{\mathcal{Q}}}-\dfrac{1}{\zeta}\bm{\mathcal{F}}^{(\text{c})},
\end{equation}
and finally
\begin{equation}\label{eq:rev_1_4}
\llbracket\dot{\bm{\mathcal{Q}}}\rrbracket=\left(2\bm{\mathcal{D}}\cdot\bm{\mathcal{A}}^{-1}\right)\cdot\left(\bm{\mathcal{K}}\cdot\bm{\mathcal{Q}}\right) - \dfrac{2k_BT}{\zeta}\bm{\mathcal{D}}\cdot\dfrac{\partial \ln\Psi}{\partial\bm{\mathcal{Q}}}-\dfrac{2}{\zeta}\bm{\mathcal{D}}\cdot\bm{\mathcal{F}}^{(\text{c})},
\end{equation}
where 
\begin{equation}\label{eq:inv_d_calc}
\bm{\mathcal{D}}=\dfrac{1}{2}\left(\bm{\mathcal{A}}^{-1}+\bm{\mathcal{V}}\right)^{-1}.
\end{equation}
Eq.~(\ref{eq:rev_1_4}) represents the fully decoupled form of the equation for the connector vector velocity, and upon substitution into the equation of continuity, it may be shown that the appropriate diffusion tensor is given by $\bm{\mathcal{D}}$.  Since $\bm{\mathcal{A}}$ is symmetric and positive-definite, while $\bm{\mathcal{V}}$ is symmetric and positive semi-definite, it follows readily that $\bm{\mathcal{D}}$ is also symmetric and positive-definite. The elements of the diffusion tensor do not vanish as $\varphi\to\infty$. This is readily illustrated for the case of a free-draining dumbbell with internal friction, for which the diffusion tensor may be evaluated analytically~\cite{Hua1995,Kailasham2018,Kailasham2021}. We have, for a dumbbell, $\bm{\mathcal{D}}=\boldsymbol{\delta}-\left[2\varphi/\left(2\varphi+1\right)\right]\bm{Q}\bm{Q}/Q^2$, and it is clear that the elements of $\bm{\mathcal{D}}$ remain finite-valued even as $\varphi\to\infty$. This property holds true even for bead-spring-dashpot chains with $N>1$, but is less apparent from the inverse-based approach for the calculation of $\bm{\mathcal{D}}$, and may be demonstrated more clearly using the recursive algorithm developed in ref.~\citenum{Kailasham2021} that does not rely on numerical matrix inversion.

While the method for the construction of the diffusion tensor given by eq.~(\ref{eq:inv_d_calc}) clearly identifies the contribution to the effective friction from hydrodynamic interactions and internal friction and also analytically establishes its symmetricity and positive-definiteness, it is less efficient computationally since it requires the evaluation of two matrix inverses at each time-step. We therefore present the following alternative derivation for the diffusion tensor which requires just one matrix inversion per step, even though it is less transparent about the physical properties of the diffusion tensor. Returning to Eq.~(\ref{eq:qdot}), we obtain
\begin{align}\label{eq:qdot2}
 \llbracket\dot{\bm{Q}}_{j}\rrbracket&=\boldsymbol{\kappa}\cdot\bm{Q}_j-\dfrac{1}{\zeta}\sum^{N}_{k=1}\tilde{\bm{A}}_{jk}\cdot\left(k_BT\dfrac{\partial \ln \Psi}{\partial \bm{Q}_k}+\dfrac{\partial \phi}{\partial \bm{Q}_k}\right)-\varphi\sum^{N}_{k=1}\tilde{\bm{Z}}_{jk}\cdot\llbracket\dot{\bm{Q}}_{k}\rrbracket
 \end{align}
 where
 \begin{equation}\label{eq:z_tilde_def}
 \tilde{\bm{Z}}_{jk}=\tilde{\bm{A}}_{jk}\cdot\left(\dfrac{\bm{Q}_k\bm{Q}_k}{\bm{Q}^2_k}\right)
 \end{equation}
 Further analysis is simplified if the $k=j$ and the $k\neq\,j$ cases in the last term on the RHS of Eq.~(\ref{eq:qdot2}) are treated separately
\begin{align}\label{eq:qdot3}
 \llbracket\dot{\bm{Q}}_{j}\rrbracket&=\boldsymbol{\kappa}\cdot\bm{Q}_j-\dfrac{1}{\zeta}\sum^{N}_{k=1}\tilde{\bm{A}}_{jk}\cdot\left(k_BT\dfrac{\partial \ln \Psi}{\partial \bm{Q}_k}+\dfrac{\partial \phi}{\partial \bm{Q}_k}\right)-\varphi\tilde{\bm{Z}}_{jj}\cdot\llbracket\dot{\bm{Q}}_{j}\rrbracket-\varphi\sum^{N}_{k\neq\,j}\tilde{\bm{Z}}_{jk}\cdot\llbracket\dot{\bm{Q}}_{k}\rrbracket
 \end{align}
 Grouping the terms containing $ \llbracket\dot{\bm{Q}}_{j}\rrbracket$ and simplifying,
 \begin{align}\label{eq:qdot5}
\llbracket\dot{\bm{Q}}_{j}\rrbracket&=\bm{Y}_{jj}\cdot\left(\boldsymbol{\kappa}\cdot\bm{Q}_j\right)-\dfrac{1}{\zeta}\sum^{N}_{k=1}\left(\bm{Y}_{jj}\cdot\tilde{\bm{A}}_{jk}\right)\cdot\left(k_BT\dfrac{\partial \ln \Psi}{\partial \bm{Q}_k}+\dfrac{\partial \phi}{\partial \bm{Q}_k}\right)-\varphi\sum^{N}_{k\neq\,j}\left(\bm{Y}_{jj}\cdot\tilde{\bm{Z}}_{jk}\right)\cdot\llbracket\dot{\bm{Q}}_{k}\rrbracket
 \end{align}
 where
 \begin{align}\label{eq:def_y}
 \bm{Y}_{jj}= \left(\boldsymbol{\delta}+\varphi\tilde{\bm{Z}}_{jj}\right)^{-1}
 \end{align}
 \begin{equation}\label{eq:process_z}
 \begin{split}
 \tilde{\bm{Z}}_{jj}&=2\left(\boldsymbol{\delta}-\zeta\boldsymbol{\Omega}_{j,j+1}\right)\cdot\left(\dfrac{\bm{Q}_j\bm{Q}_j}{\bm{Q}^2_j}\right)=2\beta_{j}\left(\dfrac{\bm{Q}_j\bm{Q}_j}{\bm{Q}^2_j}\right)
 \end{split}
 \end{equation}
 with
 \begin{equation}\label{eq:beta_def}
 \beta_{j}=\left[1-\dfrac{3a}{4Q_j}\left(\mathscr{A}+\mathscr{B}\right)\right]
 \end{equation}
 Using Eqs.~(\ref{eq:process_z}) and (~\ref{eq:beta_def}), Eq.~(\ref{eq:def_y}) is simplified to give
 \begin{equation}\label{eq:inv_y}
  \bm{Y}_{jj}=\left[\boldsymbol{\delta}+\epsilon\beta_{j}\left(\dfrac{\bm{Q}_j\bm{Q}_j}{\bm{Q}^2_j}\right)\right]^{-1}=\left[\boldsymbol{\delta}-\dfrac{\epsilon\beta_{j}}{\epsilon\beta_{j}+1}\dfrac{\bm{Q}_j\bm{Q}_j}{\bm{Q}^2_j}\right]
 \end{equation}
 where $\epsilon=2\varphi$, and the inverse has been obtained analytically using the Sherman-Morrison theorem~\cite{press2007numerical}, as described in~Ref.\citenum{Kailasham2018}. 
We may further rewrite Eq.~(\ref{eq:qdot5}) as
 \begin{align}\label{eq:intro_e_mat}
\sum_{k=1}^{N}\bm{J}_{jk}\cdot\llbracket\dot{\bm{Q}}_{k}\rrbracket&=\bm{Y}_{jj}\cdot\left(\boldsymbol{\kappa}\cdot\bm{Q}_j\right)-\dfrac{1}{\zeta}\sum^{N}_{k=1}\bm{X}_{jk}\cdot\left(k_BT\dfrac{\partial \ln \Psi}{\partial \bm{Q}_k}+\dfrac{\partial \phi}{\partial \bm{Q}_k}\right) 
\end{align}
with 
\begin{equation}\label{eq:x_def}
\bm{X}_{jk}=\bm{Y}_{jj}\cdot\tilde{\bm{A}}_{jk}
\end{equation}
and
 \begin{align}\label{eq:e_def}
\boldsymbol{J}_{jk}= \left\{
\begin{array}{ll}
       \boldsymbol{\delta}; &  j=k \\[15pt]
      \varphi\bm{Y}_{jj}\cdot\tilde{\bm{Z}}_{jk}; & j\neq\,k 
\end{array} 
\right. 
\end{align}
We next define the dimensionless block matrices $\bm{\mathcal{Y}},\,\bm{\mathcal{X}},\,\text{and}\,\bm{\mathcal{J}}$, each of size $N\times\,N$, and whose elements are the $3\times3$ matrices given by Eqs.~(\ref{eq:inv_y}), (\ref{eq:x_def}), and~(\ref{eq:e_def}), respectively. Note that the diagonal elements of $\bm{\mathcal{Y}}$ are given by Eq.~(\ref{eq:inv_y}), with the off-diagonal elements set to $\bm{0}$.


In terms of collective coordinates, we define
\begin{equation}\label{eq:m_def}
\bm{\mathcal{M}}=\bm{\mathcal{J}}^{-1}\cdot\bm{\mathcal{Y}}
\end{equation}
\begin{equation}\label{eq:diff_mat_def}
\bm{\mathcal{D}}=\dfrac{1}{2}\left(\bm{\mathcal{J}}^{-1}\cdot\bm{\mathcal{X}}\right)
\end{equation}
and rewrite Eq.~(\ref{eq:intro_e_mat}) as
\begin{equation}\label{eq:qdot_coll_cord2}
\llbracket\dot{\bm{\mathcal{Q}}}\rrbracket=\bm{\mathcal{M}}\cdot\left(\bm{\mathcal{K}}\cdot\bm{\mathcal{Q}}\right)-\dfrac{2k_BT}{\zeta}\dashuline{\bm{\mathcal{D}}\cdot\left(\dfrac{\partial \ln \Psi}{\partial \bm{\mathcal{Q}}}\right)}-\dfrac{2}{\zeta}\bm{\mathcal{D}}\cdot\bm{\mathcal{F}}^{(\text{c})}
\end{equation}
which represents a fully decoupled expression for $\llbracket\dot{\bm{\mathcal{Q}}}\rrbracket$ that may be substituted into the equation of continuity. Eq.~(\ref{eq:qdot_coll_cord2}) is rewritten in terms of the individual connector vectors as follows
\begin{equation}\label{eq:qdot_indiv_decoupled}
\llbracket\dot{\bm{Q}}_j\rrbracket=\sum_{k=1}^{N}\bm{M}_{jk}\cdot\left(\boldsymbol{\kappa}\cdot\bm{Q}_{k}\right)-\dfrac{2k_BT}{\zeta}\sum_{k=1}^{N}\bm{D}_{jk}\cdot\left(\dfrac{\partial \ln \Psi}{\partial \bm{Q}_k}\right)-\dfrac{2}{\zeta}\sum_{k=1}^{N}\bm{D}_{jk}\cdot\left(\dfrac{\partial \phi}{\partial \bm{Q}_{k}}\right)
\end{equation}
where $\bm{M}_{jk}$ and $\bm{D}_{jk}$ are $3\times3$ matrices which are elements of $\bm{\mathcal{M}}$ and $\bm{\mathcal{D}}$, respectively. The $j^{\text{th}}$ element of the underlined term in Eq.~(\ref{eq:qdot_coll_cord2}) is to be interpreted as
\begin{equation}
\left[\bm{\mathcal{D}}\cdot\left(\dfrac{\partial \ln \Psi}{\partial \bm{\mathcal{Q}}}\right)\right]_{j}=\sum_{k=1}^{N}\bm{D}_{jk}\cdot\left(\dfrac{\partial \ln \Psi}{\partial \bm{Q}_k}\right)
\end{equation}
Substituting Eq.~(\ref{eq:qdot_indiv_decoupled}) into the equation of continuity,
\begin{equation}\label{eq:cont_eq}
\dfrac{\partial \Psi}{\partial t}=-\sum_{j=1}^{N}\dfrac{\partial}{\partial \bm{Q}_{j}}\cdot\left\{\llbracket\dot{\bm{Q}}_j\rrbracket\Psi\right\},
\end{equation}
 the Fokker-Planck equation is obtained as
 \begin{equation}\label{eq:fp_sdot}
 \begin{split}
 \dfrac{\partial \Psi}{\partial t}&=-\sum_{j=1}^{N}\dfrac{\partial}{\partial \bm{Q}_{j}}\cdot\Biggl\{\Biggl[\sum_{k=1}^{N}\bm{M}_{jk}\cdot\left(\boldsymbol{\kappa}\cdot\bm{Q}_{k}\right)-\dfrac{2}{\zeta}\sum_{k=1}^{N}\bm{D}_{jk}\cdot\left(\dfrac{\partial \phi}{\partial \bm{Q}_{k}}\right)\Biggr]\Psi\Biggr\}+\left(\dfrac{2k_BT}{\zeta}\right)\sum_{j,k=1}^{N}\dfrac{\partial}{\partial \bm{Q}_{j}}\cdot\bm{D}_{jk}\cdot\dfrac{\partial \Psi}{\partial \bm{Q}_{k}}
 \end{split}
 \end{equation}
The second term on the RHS of Eq.~(\ref{eq:fp_sdot}) must be processed further in order to cast the Fokker-Planck equation in a form amenable to the It\^{o} interpretation. Following the procedure outlined in Ref.~\citenum{Kailasham2021}, the dimensionless governing Fokker-Planck equation is obtained as
\begin{equation}\label{eq:fp_dimless_in_d}
 \begin{split}
 \dfrac{\partial \Psi^{*}}{\partial t^{*}}&=-\sum_{j=1}^{N}\dfrac{\partial}{\partial \bm{Q}^{*}_{j}}\cdot\Biggl\{\Biggl[\sum_{k=1}^{N}\bm{M}_{jk}\cdot\left(\boldsymbol{\kappa}^{*}\cdot\bm{Q}^{*}_{k}\right)-\dfrac{1}{2}\sum_{k=1}^{N}\bm{D}_{jk}\cdot\left(\dfrac{\partial \phi^{*}}{\partial \bm{Q}^{*}_{k}}\right)\\[5pt]
 &+\dfrac{1}{2}\sum_{k=1}^{N}\dfrac{\partial}{\partial \bm{Q}^{*}_{k}}\cdot\bm{D}^{T}_{jk}\Biggr]\Psi^{*}\Biggr\}+\dfrac{1}{2}\sum_{j,k=1}^{N}\dfrac{\partial}{\partial \bm{Q}^{*}_{j}}\dfrac{\partial}{\partial \bm{Q}^{*}_{k}}:\left[\bm{D}_{jk}^{T}\Psi^{*}\right]
 \end{split}
 \end{equation}
 where
 \begin{equation}\label{eq:dimless_exp}
 \Psi^{*}=\Psi l^3_{H},\,t^{*}=t/\lambda_{H},\,\phi^{*}=\phi/k_BT,\quad\text{and}\quad \bm{Q}^{*}_{j}=\bm{Q}_{j}/l_{H}.
 \end{equation}
 Setting $N=1$ in eq.~(\ref{eq:fp_dimless_in_d}), followed by simplification, yields the previously derived~\cite{Kailasham2018} governing Fokker-Planck equation for a dumbbell with fluctuating IV and HI, as shown in Sec.~{SII} of the Supplementary Material. 
 
The dimensionless form of the force in a connector vector $k$ corresponding to the two types of springs considered in this work is given as follows:
  \begin{align}\label{eq:force_con_def}
\bm{F}^{*(\text{c})}_k= \left\{
\begin{array}{ll}
       \bm{Q}^{*}_{k}; & \text{Hookean}\\[15pt]
      \dfrac{\bm{Q}^{*}_{k}}{1-{Q}^{*2}_{k}/b}; & \text{FENE}
\end{array} 
\right. 
\end{align}
where the parameter $b$ denotes the square of the maximum permissible stretch of the spring in dimensionless units.
 
The stochastic differential equation corresponding to Eq.~(\ref{eq:fp_dimless_in_d}) under the It\^{o} interpretation may be written in terms of collective coordinates to be
 \begin{equation}\label{eq:sde_collective_dimless}
 \begin{split}
 d\bm{\mathcal{Q}}^{*}&=\left[\bm{\mathcal{C}}^{*}(\bm{\mathcal{Q}}^{*})+\dfrac{1}{2}\boldsymbol{\nabla}^{*}\cdot\bm{\mathcal{D}}\right]dt^{*}+\bm{\mathcal{B}}\cdot\,d\bm{\mathcal{W}}^{*}
 \end{split}
 \end{equation}
where
\begin{align}
&\bm{\mathcal{C}}^{*}(\bm{\mathcal{Q}}^{*})=\bm{\mathcal{M}}\cdot\left(\bm{\mathcal{K}}^{*}\cdot\bm{\mathcal{Q}}^{*}\right)-\dfrac{1}{2}\bm{\mathcal{D}}\cdot\bm{\mathcal{F}}^{*(\text{c})}
\end{align}
\begin{align}\label{eq:fdt_b_mat}
&\bm{\mathcal{B}}\cdot\bm{\mathcal{B}}^{T}=\bm{\mathcal{D}}
\end{align}
and $\bm{\mathcal{W}}^{*}$ is a $3N$-dimensional Wiener process. The $j^{\text{th}}$ element of $\left(\boldsymbol{\nabla}^{*}\cdot\bm{\mathcal{D}}\right)$ is the three-component vector denoted by
\begin{equation}
\left(\boldsymbol{\nabla}^{*}\cdot\bm{\mathcal{D}}\right)_{j}=\sum_{k=1}^{N}\dfrac{\partial}{\partial \bm{Q}^{*}_{k}}\cdot\bm{D}_{kj}
\end{equation}
While there exist infinitely many choices for the block matrix $\bm{\mathcal{B}}$ that satisfies the fluctuation-dissipation theorem given in eq.~(\ref{eq:fdt_b_mat}), we set it to be the square root of the block diffusion tensor, $\bm{\mathcal{B}}=\bm{\mathcal{D}}^{1/2}$, and evaluate it using Cholesky decomposition whose computational cost scales as $n^3$ where $n=3N$ is the size of the matrix. BD simulations with fluctuating hydrodynamic interactions routinely employ~\cite{Fixman1986hi,Laso2000,Jendrejack2000,Prabhakar2004} a Chebyshev polynomial based method for the computation of the matrix square root, which scales as $n^2\ell$, where $\ell$ is related to the square root of the ratio of the maximum and minimum eigenvalues of the diffusion tensor. It was shown, to a good approximation~\cite{Laso2000}, that $\ell\sim N^{1/4}$, resulting in an overall scaling of $n^{2.25}$ that is less expensive than the Cholesky method. In the absence of a detailed analysis of the spectral properties of the diffusion tensor for a system with fluctuating IV and HI, it is difficult to estimate $\textit{a priori}$ the computational savings resulting from opting for the Chebyshev method over Cholesky's. Cognizant of the relatively short chain lengths studied in the present work ($N_{\text{b}}\leq 10$), we have therefore chosen to use the latter for the evaluation of $\bm{\mathcal{D}}^{1/2}$, and note that it would be worthwhile to examine the improvement of the numerical performance of the current approach by using alternative methods to evaluate the matrix square root.

As described in detail by~\citet{Hutter1998}, it is computationally more efficient to solve the SDE obtained from the kinetic interpretation of Eq.~(\ref{eq:fp_dimless_in_d}), given by
\begin{equation}\label{eq:sde_kinetic}
\begin{split}
d\bm{\mathcal{Q}}^{*}&=\bm{\mathcal{C}}^{*}(\bm{\mathcal{Q}}^{*})dt^{*}+\dfrac{1}{2}\left[\bm{\mathcal{D}}\left(\bm{\mathcal{Q}}^{*}+d\bm{\mathcal{Q}}^{*}\right)\cdot\bm{\mathcal{D}}^{-1}\left(\bm{\mathcal{Q}}^{*}\right)+\bm{\mathcal{I}}\right]\cdot\bm{\mathcal{D}}^{1/2}\cdot\,d\bm{\mathcal{W}}^{*}\\[5pt]
&=\bm{\mathcal{C}}^{*}(\bm{\mathcal{Q}}^{*})dt^{*}+\bm{\mathcal{D}}^{1/2}\,\protect\markerfour\,d\bm{\mathcal{W}}^{*},
\end{split}
\end{equation}
as compared to the numerical integration of the equivalent It\^{o} interpretation given by Eq.~(\ref{eq:sde_collective_dimless}). This manoeuvre replaces the calculation of the divergence of the diffusion tensor by that of its inverse. The quantity $\bm{\mathcal{I}}$ that appears in eq.~(\ref{eq:sde_kinetic}) is a block matrix of size $N\times N$, whose each entry is the $3\times3$ identity matrix $\boldsymbol{\delta}$.


A predictor-corrector algorithm for the numerical integration of Eq.~(\ref{eq:sde_kinetic}) is constructed next, following the steps detailed in Ref.~\citenum{Hutter1998}. Equations~(\ref{eq:pred_equation})-(\ref{eq:corr_equation}) are in their dimensionless form, with the asterisk superscript omitted for notational simplicity.

\noindent \textbf{Predictor step}
\begin{equation}\label{eq:pred_equation}
\begin{split}
\bm{\mathcal{Q}}^{\text{(p)}}(t_{i+1})&=\bm{\mathcal{Q}}(t_{i})+\bm{\mathcal{C}}\left(\bm{\mathcal{Q}}(t_{i})\right)dt+\bm{\mathcal{D}}^{1/2}\left(\bm{\mathcal{Q}}(t_{i})\right)\cdot\Delta \bm{\mathcal{W}}
\end{split}
\end{equation}

\noindent \textbf{Corrector step}
\begin{equation}\label{eq:corr_equation}
\begin{split}
\bm{\mathcal{Q}}(t_{i+1})&=\bm{\mathcal{Q}}(t_{i})+\dfrac{1}{2}\left\{\bm{\mathcal{C}}\left(\bm{\mathcal{Q}}^{\text{(p)}}(t_{i+1})\right)+\bm{\mathcal{C}}\left(\bm{\mathcal{Q}}(t_{i})\right)\right\}dt\\[5pt]
&+\dfrac{1}{2}\left\{\bm{\mathcal{D}}\left(\bm{\mathcal{Q}}^{\text{(p)}}(t_{i+1})\right)\cdot\bm{\mathcal{D}}^{-1}\left(\bm{\mathcal{Q}}(t_{i})\right)+\bm{\mathcal{I}}\right\}\cdot\bm{\mathcal{D}}^{1/2}\left(\bm{\mathcal{Q}}(t_{i})\right)\cdot\Delta \bm{\mathcal{W}}
\end{split}
\end{equation}
~\citet{Hutter1998} note that the two-step numerical integration scheme mentioned above is weakly convergent to first order in the time-step width $\Delta t$.

Eq.~\ref{eq:fdt_b_mat} ensures that the governing stochastic differential equation, given by eq.~(\ref{eq:sde_kinetic}), satisfies the fluctuation-dissipation theorem. We further illustrate this point in Sec.~{SIV} of the Supplementary Material, by verifying that the probability distribution of the lengths of the end-to-end vector of a ten-bead chain with IV and HI at equilibrium, obtained by numerically integrating eq.~(\ref{eq:sde_kinetic}) in the absence of flow, agrees with the analytical expression.


A thermodynamically consistent stress tensor expression for chains with fluctuating internal friction and hydrodynamic interactions may be derived using the Kramers-Kirkwood relation~\cite{Bird1987b}, as suggested by~\citet{Hua19961473}. We therefore have 
\begin{equation}\label{eq:kram_kirk_textbook}
\boldsymbol{\tau}_{\text{p}}=-{n_{\text{p}}}\sum_{\nu=1}^{N_{\text{b}}}\left<\bm{R}_{\nu}\bm{F}_{\nu}^{(h)}\right>
\end{equation}
where $\bm{R}_{\nu}=\bm{r}_{\nu}-\bm{r}_{\text{c}}$ is the position of the $\nu^{\text{th}}$ bead with respect to the centre of mass of the chain, and $\bm{F}_{\nu}^{(h)}$ is the hydrodynamic drag force on the $\nu^{\text{th}}$ bead. Equation~(\ref{eq:kram_kirk_textbook}) may be recast, after some algebra, in terms of the connector vectors,
\begin{equation}\label{eq:con_vec_kram_kirk}
\boldsymbol{\tau}_{\text{p}}=-{n_{\text{p}}}\sum_{k=1}^{N_{\text{b}}-1}\left<\bm{Q}_{k}\left[k_BT\,\dfrac{\partial \ln \Psi}{\partial\bm{Q}_{k}}+\bm{F}^{*(\text{c})}_k+K\left(\dfrac{\bm{Q}_k\bm{Q}_k}{Q^2_k}\right)\cdot\llbracket\dot{\bm{Q}}_k\rrbracket\right]\right>
\end{equation}
Substituting the expression for $\llbracket\dot{\bm{Q}}_k\rrbracket$ from Eq.~(\ref{eq:qdot_indiv_decoupled}) into the above equation and simplifying (as described in Sec.~{SVI} of the Supplementary Material), the dimensionless stress tensor expression is obtained as follows
\begin{align}\label{eq:stress_tensor_dimless}
&\dfrac{\boldsymbol{\tau}_{\text{p}}}{n_{\text{p}}k_BT}=\left(N_{\text{b}}-1\right)\boldsymbol{\delta}-\left<\sum_{k=1}^{N_{\text{b}}-1}\bm{Q}^{*}_{k}\bm{F}^{*(\text{c})}_k\right>-2\epsilon\left<\sum^{N_{\text{b}}-1}_{k,j=1}\left(\bm{M}_{kj}\cdot\boldsymbol{\kappa}^{*}\right)^{T}:\left[\dfrac{\bm{Q}^{*}_{k}\bm{Q}^{*}_{j}\bm{Q}^{*}_{k}\bm{Q}^{*}_{k}}{Q^{*2}_{k}}\right]\right>\nonumber\\[5pt]
&+{\epsilon}\left<\sum^{N_{\text{b}}-1}_{k,j=1}\left[\bm{D}^{T}_{kj}:\bm{Q}^{*}_{k}\bm{F}^{*(\text{c})}_{j}\right]\dfrac{\bm{Q}^{*}_{k}\bm{Q}^{*}_{k}}{Q^{*2}_{k}}\right>-\epsilon\left<\sum^{N_{\text{b}}-1}_{k,j=1}\left(\dfrac{\bm{Q}^{*}_{k}\bm{Q}^{*}_{k}}{Q^{*2}_{k}}\right)\bm{Q}^{*}_{k}\cdot\dashuline{\left[\dfrac{\partial}{\partial \bm{Q}^{*}_{j}}\cdot\bm{D}^{T}_{kj}\right]}\right>\\[5pt]
&-\epsilon\Biggl[\left<\sum_{k=1}^{N_{\text{b}}-1}\left[\text{tr}(\bm{D}_{kk})-2\bm{D}_{kk}:\dfrac{\bm{Q}^{*}_{k}\bm{Q}^{*}_{k}}{Q^{*2}_{k}}\right]\dfrac{\bm{Q}^{*}_{k}\bm{Q}^{*}_{k}}{Q^{*2}_{k}}\right>+\left<\sum_{k=1}^{N_{\text{b}}-1}\dfrac{\bm{Q}^{*}_{k}\bm{Q}^{*}_{k}}{Q^{*2}_{k}}\cdot\bm{D}_{kk}\right>+\left<\sum_{k=1}^{N_{\text{b}}-1}\bm{D}_{kk}\cdot\dfrac{\bm{Q}^{*}_{k}\bm{Q}^{*}_{k}}{Q^{*2}_{k}}\right>\Biggr]\nonumber
\end{align}

The underlined term in Eq.~(\ref{eq:stress_tensor_dimless}), when summed over the $j$ index, represents the divergence of the diffusion tensor, and is evaluated at each instance when the viscometric functions need to be computed. The method of random finite difference discussed in Refs.~\citenum{Sprinkle2017} \& ~\citenum{Sprinkle2019} is adopted for the calculation of this term. We provide a brief recap of this procedure below, wherein the divergence of a matrix may be evaluated as an ensemble average. 

Considering an arbitrary configuration-dependent tensor $\bm{\mathcal{S}}(\bm{\mathcal{Q}})$ of size $\mathcal{N}\times\mathcal{N}$, and two independent Gaussian random vectors $\boldsymbol{\rho}_A$ and $\boldsymbol{\rho}_B$ such that $\left<\boldsymbol{\rho}_A\boldsymbol{\rho}_B\right>_{\text{RFD}}=\boldsymbol{I}$ where $\boldsymbol{I}$ is an identity matrix of the same size as $\bm{\mathcal{S}}$, and $\left<\dots\right>_{\text{RFD}}$ denotes an ensemble-average, where the subscript is used to highlight that the ensemble-size for the RFD procedure may be chosen independently of the number of polymer chain trajectories used for the calculation of observables. We may thus write
\begin{equation}\label{eq:rfd_defn}
\begin{split}
&\lim_{\delta\to 0}\dfrac{1}{\delta}\left<\left\{\bm{\mathcal{S}}\left(\bm{\mathcal{Q}}+\dfrac{\delta}{2}\boldsymbol{\rho}_{A}\right)-\bm{\mathcal{S}}\left(\bm{\mathcal{Q}}-\dfrac{\delta}{2}\boldsymbol{\rho}_{A}\right)\right\}\boldsymbol{\rho}_{B}\right>_{\text{RFD}}=\left(\dfrac{\partial \bm{\mathcal{S}}}{\partial \bm{\mathcal{Q}}}\right):\left<\boldsymbol{\rho}_A\boldsymbol{\rho}_B\right>_{\text{RFD}}=\dfrac{\partial}{\partial \bm{\mathcal{Q}}}\cdot\bm{\mathcal{S}}
\end{split}
\end{equation}
A value of $\delta=10^{-5}$ is chosen after testing for convergence. Similarly, an ensemble size of $5\times10^3$ for the RFD calculation is found to suffice for obtaining convergent results, and is subsequently employed for all the simulation runs.

We have not considered excluded volume effects in the present work. While the decoupling procedure developed above would remain unaltered by the inclusion of an additional intramolecular potential, the expression for the stress tensor would need to be appropriately modified.
\end{widetext}

The dynamics of polymer chain models with fluctuating internal friction and hydrodynamic interactions may be computed by numerically integrating eq.~(\ref{eq:sde_kinetic}), where the use of the kinetic interpretation circumvents the calculation of the divergence of the diffusion tensor. In a prior work~\cite{Kailasham2021}, a recursion-based approach was used to decouple the equations of motion for a freely draining coarse-grained model with internal friction in order to obtain the governing Fokker-Planck equation for the system. The equivalent stochastic differential equation is obtained using an It\^{o} interpretation, and its numerical integration requires the calculation of the divergence, $\sum_{k=1}^{N}\left(\partial/\partial\bm{Q}_{k}^{*}\right)\cdot\bm{V}_{jk}^{T}$, at each timestep. The formal definition of $\bm{V}_{jk}^{T}$ is based on a recurrence relation that is quite involved, and is provided in ref.~\citenum{Kailasham2021}. A consequence of the decoupling methodology is that a closed form (but recursive) relationship for the dependence of $\bm{V}_{jk}$ on the connector vectors is obtained, which permitted the calculation of the divergence via a simple finite difference scheme. In the present work, however, there exists no closed form relationship for the diffusion tensor as a function of the chain configuration, and each evaluation of the block diffusion tensor would involve a matrix inversion (cf. eq.~(\ref{eq:diff_mat_def})), thus motivating the use of a divergence-free solution methodology. The algorithm used in the present work is an order of magnitude faster than that based on the decoupling method for the numerical solution of the governing stochastic differential equations, as illustrated in Sec.~{SIII} of the Supplementary Material.

The stress tensor expression for polymer models with internal friction (with or without hydrodynamic interactions), however, contains a divergence term whose calculation cannot be avoided by a different choice of interpretation of the governing Fokker-Planck equation. In the decoupling approach, the matrix whose divergence is computed for the solution of the governing stochastic differential equation ($\bm{V}_{jk}$) is different from that whose divergence is evaluated in the stress tensor calculation ($\boldsymbol{\mu}_{kl}$). The two quantities are related to each other by
\begin{equation}
\,\bm{V}_{jl}=\sum_{j=1}^{N}\bm{A}_{jk}\cdot\boldsymbol{\mu}_{kl}, 
\end{equation}
where $\bm{A}_{jk}$ is the Rouse matrix defined in eq.~(\ref{eq:a_matrix_def})). In the methodology used in the present work, however, the divergence of the same quantity appears in both the governing stochastic differential equation and the stress tensor expression. This divergence needs to be evaluated only at the sampling instances where the viscosity needs to be computed, and not at each timestep. Therefore, if the interest is to only simulate the dynamics of the polymer chain, recording the changes in its configuration as a function of time, then the kinetic interpretation offers a manifestly faster computational route over the conventional It\^{o} interpretation. For the calculation of the viscometric functions, the computational cost of using the methodology outlined in the present work would therefore scale with the number of sampling instances, unlike the recursion-based decoupling methodology. We reiterate that the one-to-all coupling between the connector vector velocities resulting from the inclusion of hydrodynamic interactions (cf. eq.~(\ref{eq:qdot_indiv_decoupled})) implies that the applicability of the decoupling methodology is restricted to the free-draining case.


The bead-spring-dashpot chain is subjected to steady simple shear flow. The flow tensor, $\boldsymbol{\kappa}$ has the following form
\begin{equation}\label{eq:kappa_shear_def}
\boldsymbol{\kappa} \equiv \boldsymbol{\kappa}^{*}\,\lambda^{-1}_{H}=   \dot{\gamma}
\begin{pmatrix}
0& 1 & {0}\\
0& 0& 0\\
{0} & 0 & 0 
\end{pmatrix}
\end{equation}
and the shear viscosity is defined as
\begin{align}\label{eq:visc_fun_def}
\eta_{\text{p}}=-\dfrac{\tau_{\text{p},xy}}{\dot{\gamma}}
\end{align}
where $\tau_{\text{p},xy}$ refers to the $xy$ element of the stress tensor.

\subsection{\label{sec:sim_details} Simulation details}

The timestep used in the numerical integration procedure is dependent upon the shear rate and the magnitude of the internal friction parameter, with higher values of these parameters necessitating the use of smaller timesteps. For the highest value of the internal friction parameter considered in this work, $\varphi=5.0$, values of $\Delta t^{*}=10^{-3}$ for simulations of $\lambda_{H}\dot{\gamma}\leq1.0$ and $\Delta t^{*}=10^{-4}$ for $1.0\leq\lambda_{H}\dot{\gamma}\leq100.0$ have been found to result in convergent results [see Sec.~{SV} of the Supplementary Material]. Variance reduction~\cite{Wagner1997} has been used in the evaluation of steady-shear viscometric functions at low shear rates ($\lambda_{H}\dot{\gamma}<0.1$). Since internal friction and hydrodynamic interactions do not affect the configurational distribution function of polymer chains at equilibrium, the simulations are initiated by drawing the bead positions from the appropriate equilibrium distribution corresponding to Hookean or FENE springs. Internal friction or hydrodynamic interaction effects are turned on at $t^{*}=0$, concomitantly as the shear flow is started. A rejection criterion~\cite{Ottinger1996} is employed that discards the instantaneous configuration if any spring is found to have exceeded its maximum permissible extension, $\sqrt{b}$. Each trajectory is run for a duration $t^{*}_{\text{max}}$, until the shear viscosity attains a constant value, and the length of the trajectory depends upon the shear rate. For example, runs at $\lambda_{H}\dot{\gamma}=10.0$ attain steady state by $t^{*}_{\text{max}}=60.0$, while the corresponding value at $\lambda_{H}\dot{\gamma}=100.0$ is $t^{*}_{\text{max}}=30.0$.  Simulations of Rouse chains, particularly with the inclusion of IV, are observed to require a longer time for the attainment of steady state, and hence we use $t^{*}_{\text{max}}=400.0$ for such chains at $\lambda_{H}\dot{\gamma}\geq1.0$ and $t^{*}_{\text{max}}=200.0$ at $\lambda_{H}\dot{\gamma}<1.0$. Averages are evaluated over an ensemble of $\mathcal{O}\left(10^{4}\right)-\mathcal{O}\left(10^{5}\right)$ individual trajectories.

Steady-shear viscosity profiles for the various models are scaled by their respective zero-shear rate value, $\eta^{*}_{\text{p},0}$, calculated as the error-weighted mean of the viscosity values computed at the four lowest dimensionless shear rates, namely, $\lambda_{\text{H}}\dot{\gamma}=0.001,0.002,0.005,0.01$, after ensuring that shear-thinning has not set in at these shear rates. Shear rates are scaled by the relaxation time defined by $\lambda_{\text{p}}=\eta_{\text{p},0}/n_{\text{p}}k_BT$.

\subsection{\label{sec:param_values}Choice of parameter values}

The value of the finite extensibility parameter $b$ in eq.~(\ref{eq:force_con_def}) for a given polymer may be found from its experimentally measured radius of gyration under $\theta$-solvent conditions, and its contour length~\citep{Sunthar2005}. We do not seek to model a specific polymer in this work, and use $b=100$ in all our simulations that employ FENE springs. The use of this value of $b$ appears to be prevalent in computational rheological studies~\cite{Ottinger1987,Wedgewood1988,Hua1995}.

The hydrodynamic interaction parameter is defined on the basis of the dimensionless bead radius, as $h^{*}=a^{*}/\sqrt{\pi}$. The mean-squared distance of a Hookean spring is given by $\left<Q^2\right>=3\left(k_BT/H\right)$, implying that the mean rest length of the spring is $\sqrt{\left<Q^2\right>}=\sqrt{3}l_{H}$. This sets the physically realistic condition that $2a^{*}<\sqrt{\left<Q^{*2}\right>}$, implying $0\leq h^{*}<0.5$. In Brownian dynamics simulations, values of the hydrodynamic interaction parameter in the range $0\leq h^{*}\leq0.3$ are typically used~\cite{Laso2000,Ottinger1987}.
The strength of hydrodynamic interactions for a polymer-solvent system close to equilibrium is determined by the draining parameter, defined as $h=h^{*}\sqrt{N}$, where $N$ is the number of springs in the model~\cite{Sunthar2005}. Several non-dimensional ratios of equilibrium and linear viscoelastic properties are known to attain universal values in the non-draining limit denoted by $h\to\infty$. For the calculation of such properties, it is clear that the actual value of $h^{*}$ is irrelevant provided that the simulation uses a chain with large enough number of springs~\cite{Ottinger1996}. It has been shown, using renormalization group theory~\cite{Ottinger1987jcp}, that several universal ratios follow the functional form,
\begin{equation}\label{eq:univ_rel}
U(h^{*},N)=\tilde{U}_{i}+\Lambda_{i}\left(\dfrac{1}{h^{*}_{i}}-\dfrac{1}{h^{*}}\right)\frac{1}{\sqrt{N}}+\mathcal{O}\left(\frac{1}{N}\right)
\end{equation}
where $\tilde{U}_{i}$ is the universal value of the ratio, the subscript $i$ runs over the various possible universal ratios, and $\Lambda_{i}$ is a constant~\cite{Laso2000,Sunthar2005}. Simulation-based estimates of universal ratios proceed by their calculation for chains of various values of $N$, followed by extrapolation to the limit of $N\to\infty$, or $N^{-1/2}\to0$. It is apparent from eq.~(\ref{eq:univ_rel}), that the limiting, non-draining value is quickly attained if the $h^{*}$ is chosen closed to the fixed point given by $h^{*}_{i}$, which has been shown to be close to $0.25$ for a host of universal ratios~\cite{Ottinger1987jcp,Laso2000}. 
For these reasons, we choose $h^{*}=0$ to model free-draining chains, and $h^{*}=0.3$ as a reasonable value for this parameter to model chains with hydrodynamic interactions.

The rationale for the selection of values for the internal friction parameter $\varphi$ is explained in detail next. The rheological consequences of internal friction are the appearance of a stress jump at the inception and cessation of flow, and a high frequency asymptote in the dynamic viscosity, $\eta'$. While these quantities have been experimentally measured, the extraction of an internal friction coefficient from these rheological signatures has remained a scarcely-attempted exercise. ~\citet{Massa1971} fit Peterlin's theory~\cite{Peterlin1967} to $G'$ and $G''$ data on polystyrene solutions in Aroclor, and conclude that the internal friction parameter is in the range of $1.5$ to $2.5$. These numbers must be considered with caution, since Peterlin's theory uses the linearized rotational velocity (LRV) approximation for the internal friction force, which has been shown by Manke and Williams~\cite{Manke1988,Dasbach1992} to result in physically unrealistic predictions in the $\varphi\to\infty$ limit. The numerical algorithm developed in the present work offers an opportunity to revisit their experimental data on linear viscoelastic material functions and extract the internal friction parameter.
While internal viscosity has been known to induce stress jumps at the startup of shear flow, the difficulties associated with the estimation of this quantity from experiments has been documented in ref.~\citenum{Mackay1992}. In ref.~\citenum{liang1993stress}, Liang and Mackay subjected xanthan-gum solutions to cessation of shear flow experiments and measured the ratio of the shear-stress in the solution at the instance of cessation, to the polymer contribution to the total shear stress in the fluid before cessation. This quantity is denoted by $R^{-}\left(t;\gamma_{\text{s}},\dot{\gamma}_{\text{s}}\right)$ in their paper, where $t$ represents the time of cessation of flow, $\gamma_{\text{s}}$ the amount of shear strain in the solution prior to cessation of flow, and $\dot{\gamma}_{\text{s}}$ the steady shear-rate to which the solution is subjected before flow is turned off. The quantity $R^{-}\left(0;\infty,\dot{\gamma}_{\text{s}}\right)$ is reported in fig. 10 of ref.~\citenum{liang1993stress}, which is obtained from measurements in which the flow is turned off at $t=0$, after allowing the solution to reach steady-state ($\gamma_{\text{s}}\to\infty$) at various values of the shear rate $\dot{\gamma}_{\text{s}}$. Liang and Mackay use the semi-analytical theory of~\citet{Manke1992} to predict that a value of $\varphi=0.5$ would result in a value of about $0.608$ for $R^{-}\left(0;\infty,\dot{\gamma}_{\text{s}}\to0\right)$, with preaveraged hydrodynamic interactions. The experiments show that $R^{-}\left(0;\infty,\dot{\gamma}_{\text{s}}\right)$ has value of about $0.8$ at $1\,\text{s}^{-1}$, with data at smaller values of the shear-rate not reported due to the level of the noise in the system. The good qualitative agreement between the theoretical prediction and experimental observations suggests that the internal friction parameter in xanthan-gum solutions is $\mathcal{O}(1)$.

The distinction between ``dry" and ``wet" internal friction has been discussed in detail in ref.~\citenum{Kailasham2020}. The former refers to a mode of dissipation that exists independently of the solvent and couples additively to hydrodynamic friction. This type of internal friction is represented in coarse-grained models using dashpots. Wet internal friction has been attributed to the slowed diffusion of biomolecules on a rugged energy landscape, and the dissipation associated with the breakage of cohesive intramolecular interactions. This kind of internal friction couples multiplicatively with the solvent friction, and is incorporated in coarse-grained models as a non-bonded interaction potential. It is entirely plausible that a polymer molecule possesses both the types of internal friction simultaneously. It is possible to distinguish between these two types of internal friction in the extrapolated limit of zero solvent viscosity, since the wet component is expected to vanish in this limit. A methodology for the estimation of the dry internal friction coefficient from the average work dissipated in repeatedly stretching a polymer molecule ($\left<W_{\text{dis}}\right>$) calculated using the Jarzynski equality has also been illustrated in ref.~\citenum{Kailasham2020}. For a single-mode spring-dashpot, it may be shown that $\left<W_{\text{dis}}\right>_{\eta_{\text{s}}\to 0}=K\,vd$, where $\left<W_{\text{dis}}\right>_{\eta_{\text{s}}\to 0}$ is the average dissipation in the limit of zero solvent viscosity, with $v$ and $d$ denoting the constant velocity and distance of pulling. However, there currently exists no theoretical model to relate the average dissipation to an equivalent damping coefficient for each spring in a bead-spring-dashpot chain with greater than two beads.

Experimental attempts~\cite{Hagen2010385,Das2022} for the quantification of internal friction have traditionally relied on the calculation of reconfiguration or relaxation time of protein molecules suspended in solvents of varying viscosity using fluorescence spectroscopy, followed by a linear extrapolation to the zero solvent viscosity limit to isolate internal friction effects from those of the solvent. This timescale is taken to represent the resistances to protein reconfiguration that are solely intramolecular in origin and independent of the solvent viscosity, and is only a qualitative measure of internal friction~\citep{Qiu2004,Soranno2012,Soranno2017}. By mapping experimental results on the forced unfolding of DNA globules~\cite{Murayama2007} on to the rugged energy landscape model proposed by Zwanzig~\cite{Zwanzig1988}, ~\citet{Alexander-Katz2009} estimated that the friction due to the globules is about 2.6 times higher than that due to the solvent. This scenario, however, corresponds to the wet internal friction, and cannot be mapped on to an equivalent dashpot coefficient. All-atom molecular dynamics simulations of the force-induced unfolding of peptides~\cite{Schulz2015} conclude that the friction arising from intramolecular hydrogen bonds is about an order of magnitude or two higher than that due to the solvent. If one were to assume that the peptide may be modelled as a single-mode spring-dashpot, attributing the entire dissipation to a single dashpot, then its internal friction parameter would be in the range of $\varphi\approx\mathcal{O}(10)-\mathcal{O}(10^2)$. It may be anticipated, however, that apportioning the dissipation amongst a sequence of bead-spring-dashpots would result in an internal friction coefficient of $\varphi\approx\mathcal{O}(1)$, depending on the number of beads used in the coarse-grained representation.

Using the example of a single-mode spring-dashpot, we have discussed in detail~\cite{Kailasham2018,Kailasham2021sm} how in the limit of $\varphi\to\infty$, the governing stochastic differential equation of a dumbbell with IV approaches that of a rigid rod. In particular, we have shown in ref.~\citenum{Kailasham2018} that the stress jump for a FENE-spring-dashpot with $\varphi=5$ is within $10\%$ of the value for a dilute solution consisting of rigid rods whose lengths are drawn from a FENE distribution. With this rationale, we pick the values for $\varphi$ in the range $[0,5]$, as we believe that it sufficiently spans the range from low IV to high IV. We note that Schieber and coworkers~\cite{Hua1995,Hua1996} have also picked values for $\varphi$ in the same range, in their numerical investigations of internal friction.

\section{\label{sec:results} Results}
\subsection{\label{sec:code_valid} Code Validation}

\begin{figure}[t]
\begin{center}
\begin{tabular}{c}
\includegraphics[width=3.1in,height=!]{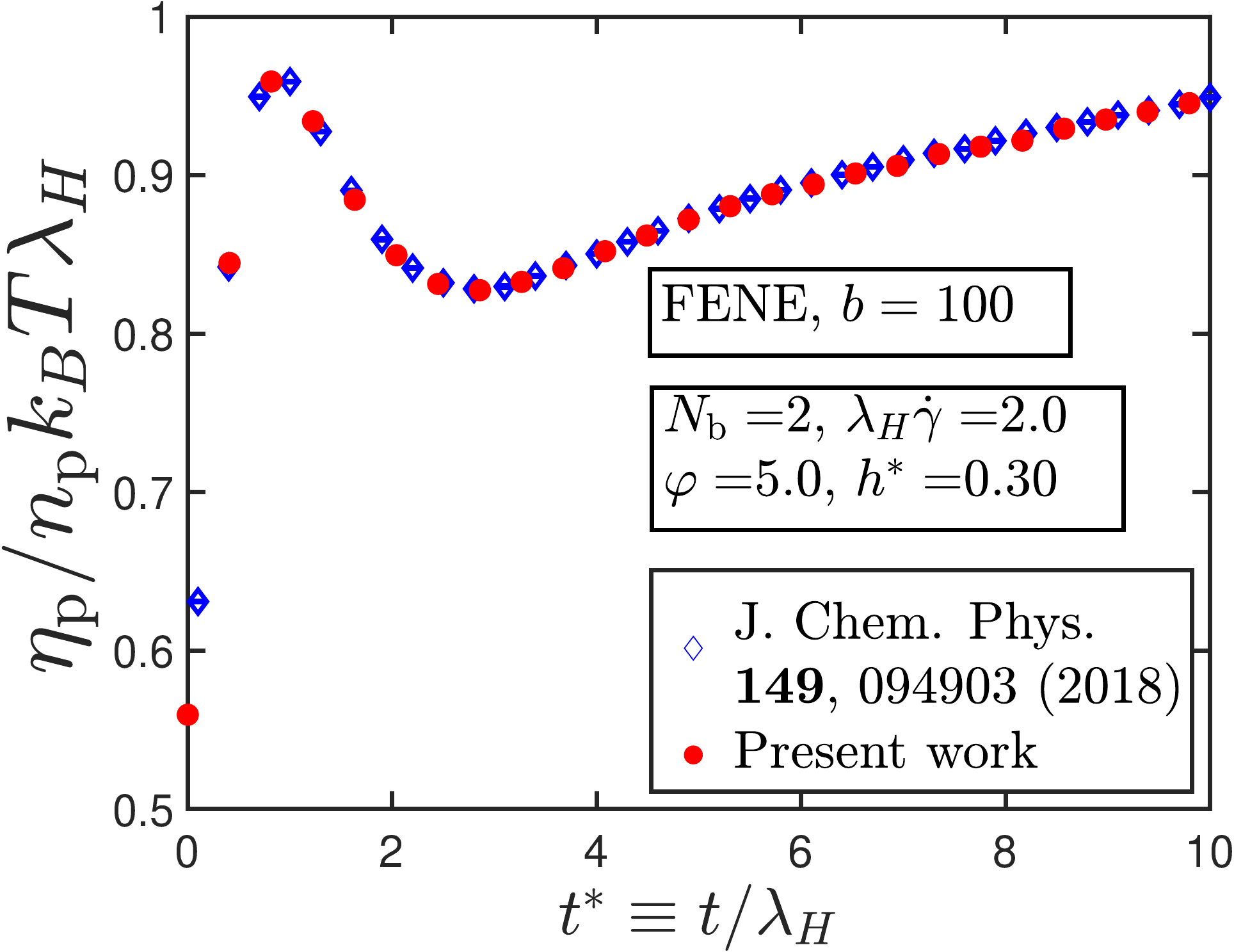}\\
(a)\\
\includegraphics[width=3.1in,height=!]{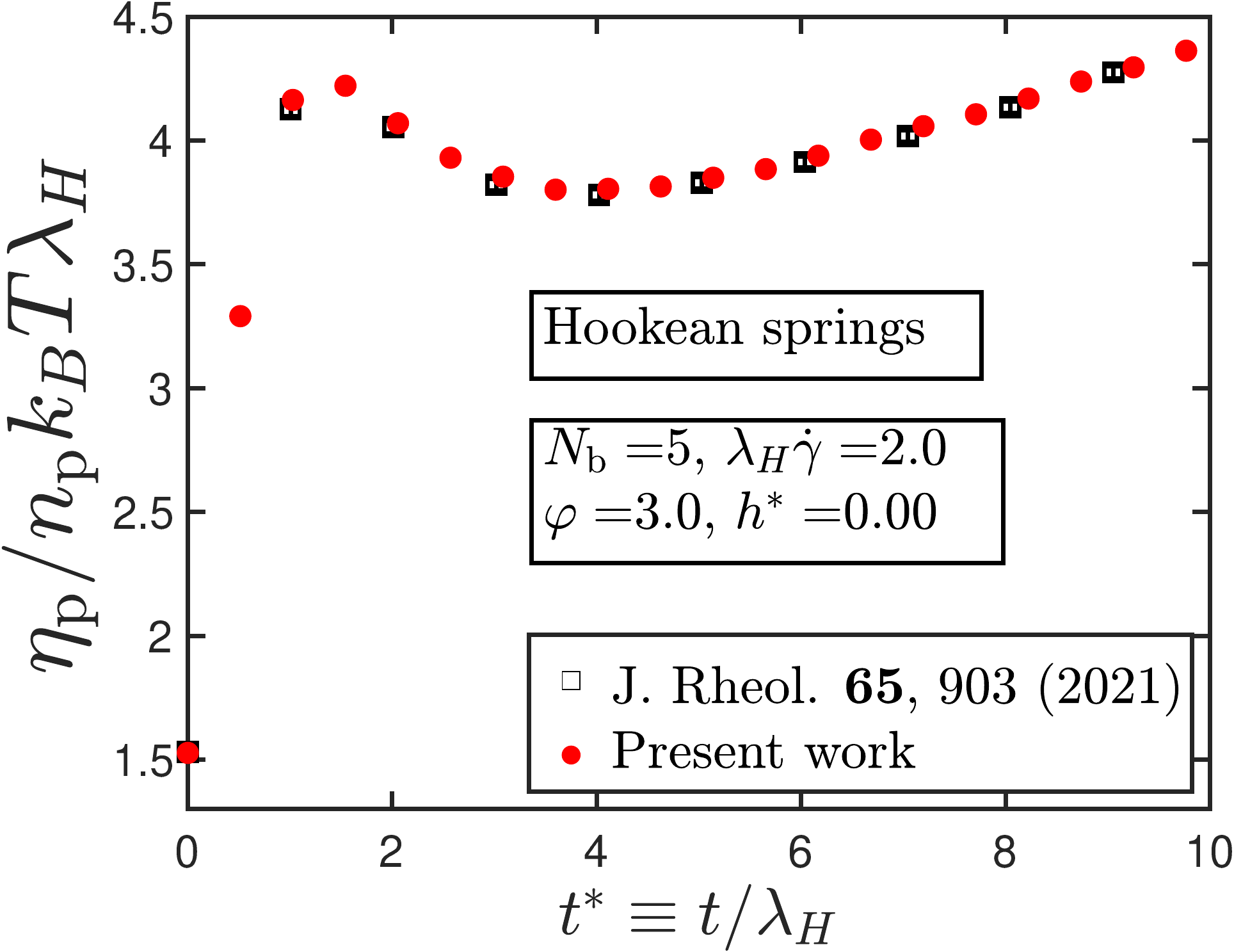}\\
(b)\\
\end{tabular}
\end{center}
\caption{ (Color online) Comparison of shear viscosity predicted by present work against those calculated for (a) FENE dumbbells with internal viscosity and hydrodynamic interactions in Ref.~\citenum{Kailasham2018}, and (b) freely-draining Rouse chains with internal viscosity in Ref.~\citenum{Kailasham2021}. Error bars, which represent standard error of the mean, are roughly of the same size or smaller than the symbols used.}
\label{fig:fene_db_ivhi}
\end{figure}

The viscometric functions for FENE dumbbells with internal viscosity and hydrodynamic interactions subjected to simple shear flow have been evaluated in Ref.~\citenum{Kailasham2018} by numerically integrating the governing stochastic differential equation obtained from an It\^{o} interpretation of the Fokker-Planck equation using a semi-implicit predictor-corrector algorithm. For the special case of a dumbbell, the divergence of the diffusion tensor is known analytically. Material function predictions for free-draining Rouse chains with internal viscosity have been presented in Ref.~\citenum{Kailasham2021}, which also uses the It\^{o} interpretation, a simple explicit Euler numerical integrator with the divergence of the diffusion tensor evaluated numerically, and the Giesekus expression for the stress tensor. We note that the Giesekus expression is not applicable for models with hydrodynamic interactions, and therefore cannot be used for the model considered in the present work.

In Fig.~\ref{fig:fene_db_ivhi}, the shear viscosity obtained using the methodology outlined in the present work are compared against data from Refs.~\citenum{Kailasham2018} and ~\citenum{Kailasham2021}. The excellent agreement between the results establishes the fidelity of the algorithm presented in this article.

\subsection{\label{sec:sjump} Stress jumps at the inception of steady shear flow}

\begin{figure}[t]
\begin{center}
\begin{tabular}{c}
\includegraphics[width=3.4in,height=!]{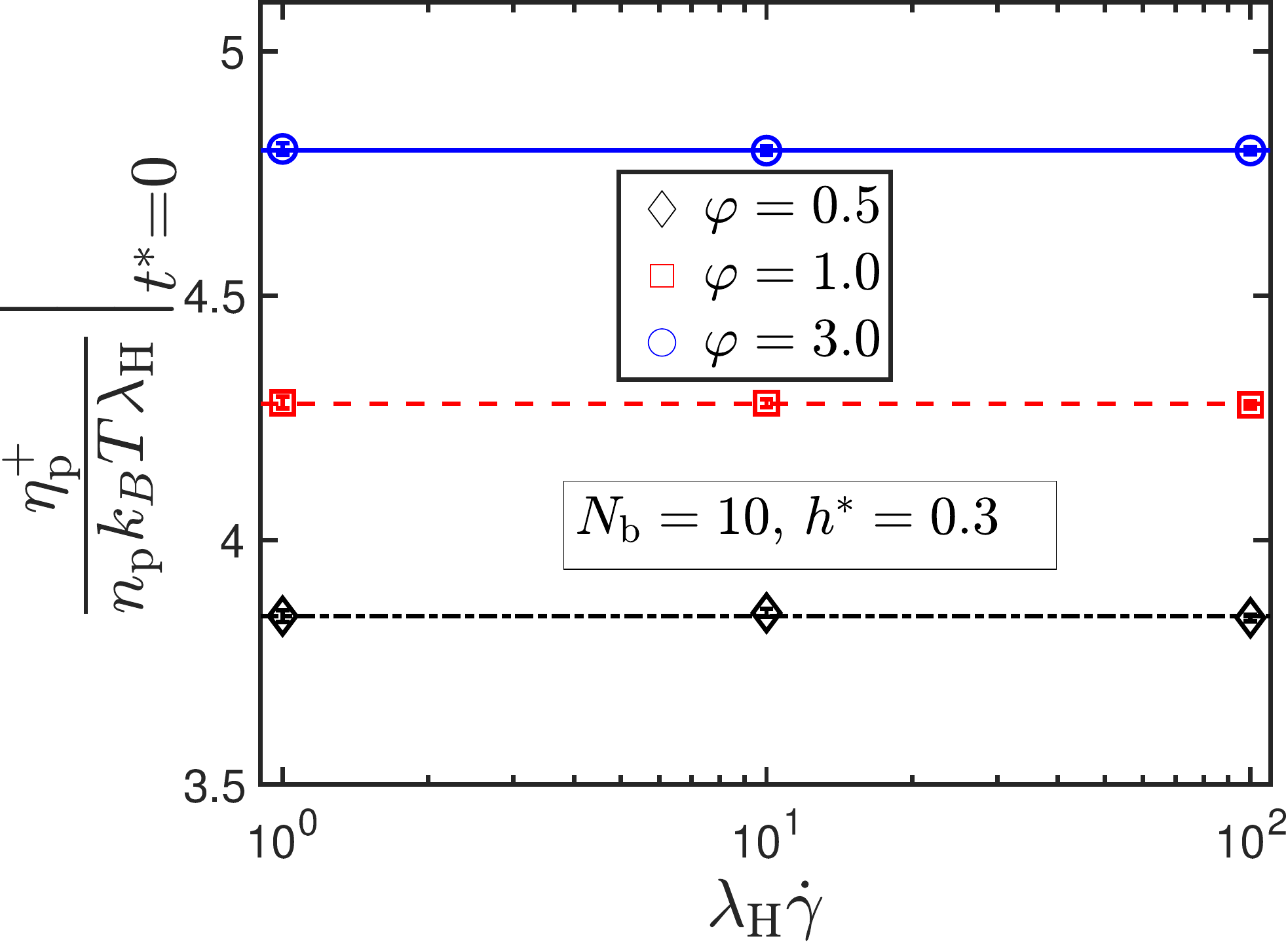}\\
(a)\\
\includegraphics[width=3.3in,height=!]{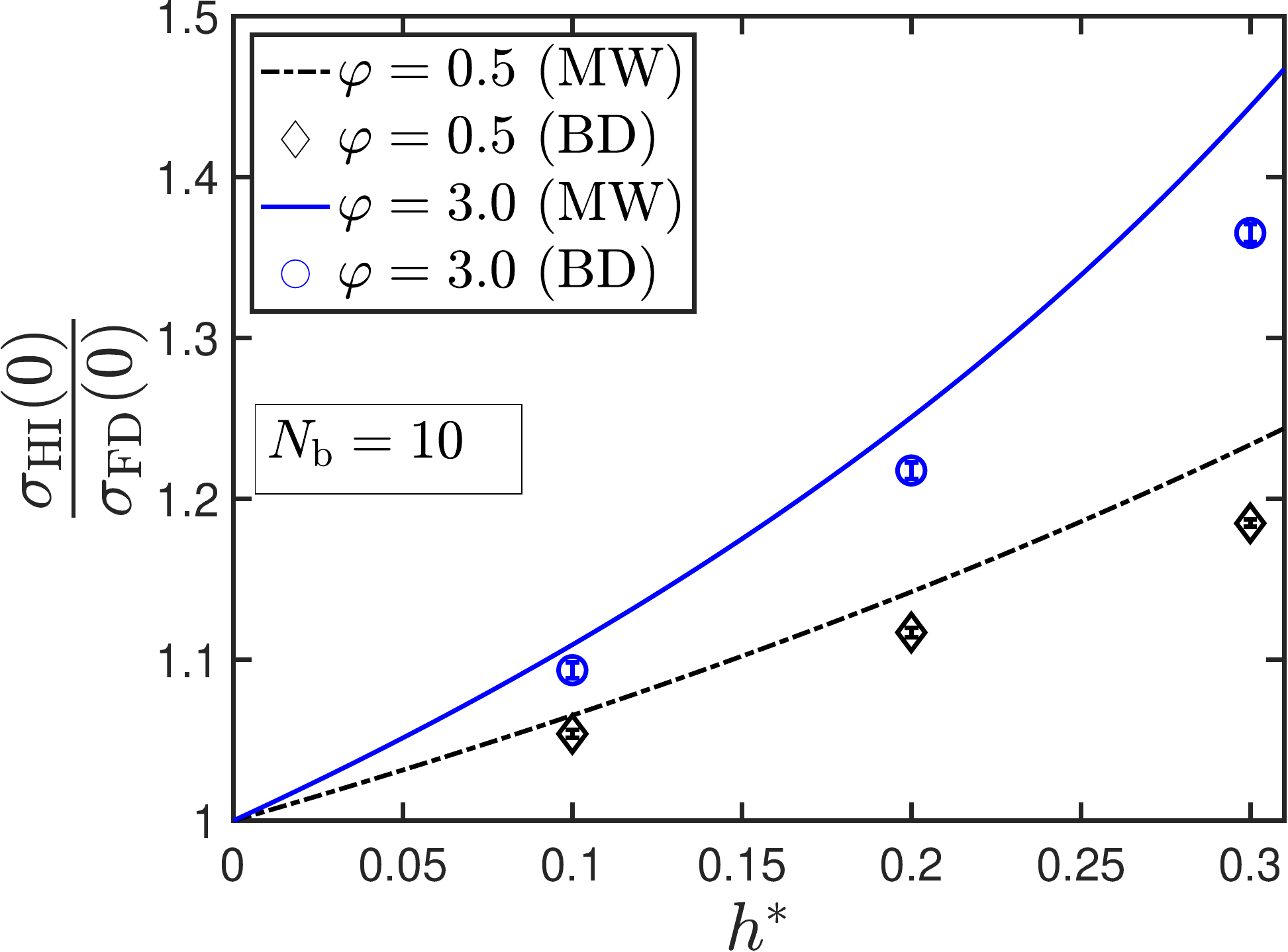}\\
(b)
\end{tabular}
\end{center}
\caption{ (Color online) (a) Shear-rate independence of stress jumps observed in BD simulations of a ten-bead chain with Hookean springs and a hydrodynamic interaction parameter of $h^{*}=0.3$, at various values of the internal friction parameter. The horizontal lines are error-weighted averages of the data points that they traverse. (b) Comparison of the stress jump ratio obtained from BD simulations (denoted by symbols) with the semi-analytical approximation of Manke and Williams~\cite{Manke1992} (indicated by lines). Error bars are roughly of the same size or smaller than the symbols used.}
\label{fig:sjump_figure}
\end{figure}

Having thus validated the numerical algorithm, this subsection presents results for the stress jump of bead-spring-dashpot chains with Hookean springs.
Polymer models with internal friction are known to exhibit a discontinuous jump in the viscosity at the inception of shear flow, and this is referred to as stress jump~\cite{Kailasham2018,Kailasham2021,Manke1988}. In Fig.~\ref{fig:sjump_figure}~(a), the magnitudes of stress jump for a ten-bead chain with Hookean springs and three different values of the internal friction parameter, computed over a range of shear rates are plotted. The stress jump increases with the internal friction parameter, and is found to be independent of the shear rate, consistent with prior observations in the literature~\cite{Gerhardt1994,Manke1992}. In the discussion and figures that follow in this section, the stress jump is computed at a dimensionless shear rate of $\lambda_{H}\dot{\gamma}=10.0$.

\begin{figure*}
\begin{center}
\begin{tabular}{cc}
\includegraphics[width=3.in,height=!]{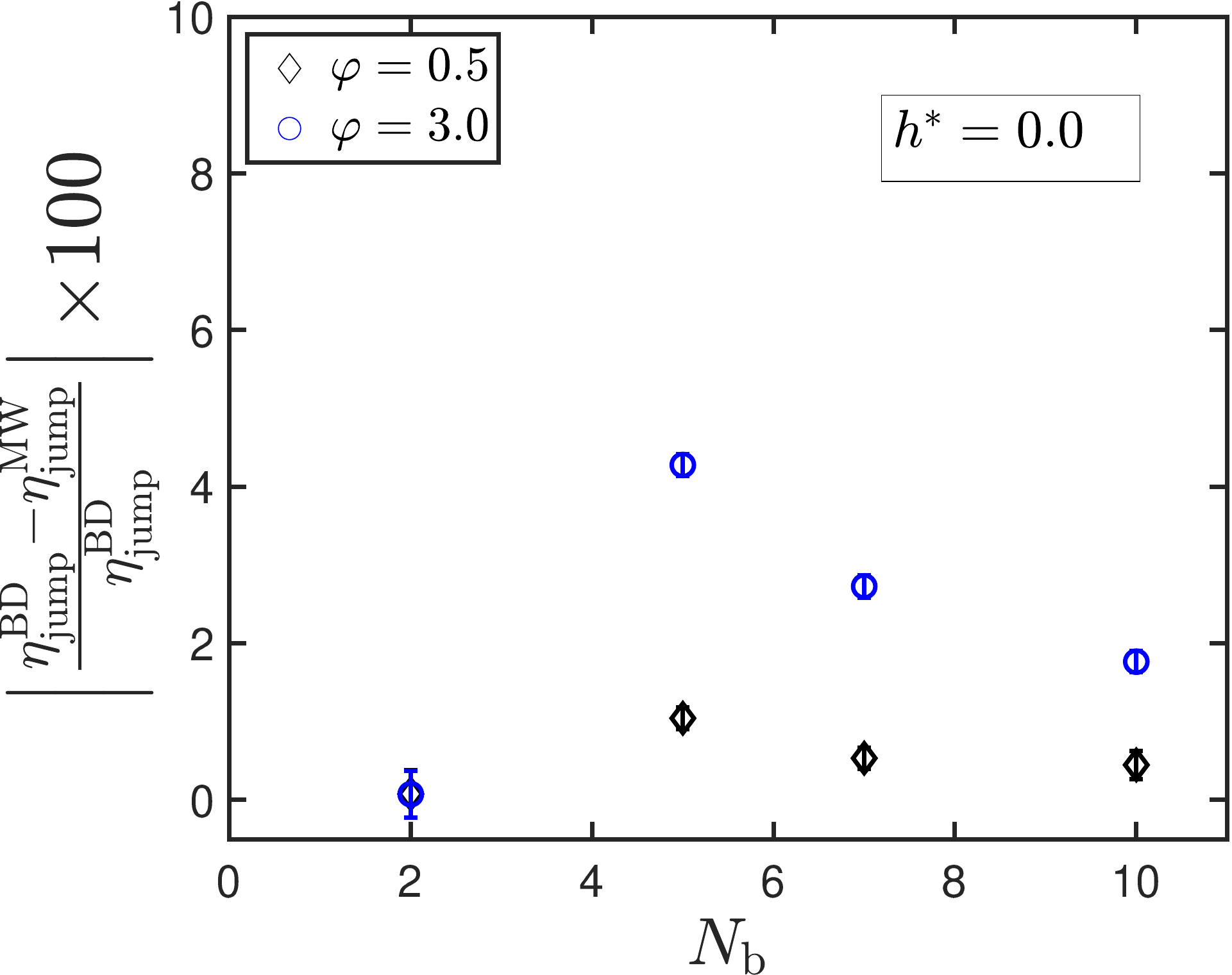}&
\includegraphics[width=3.in,height=!]{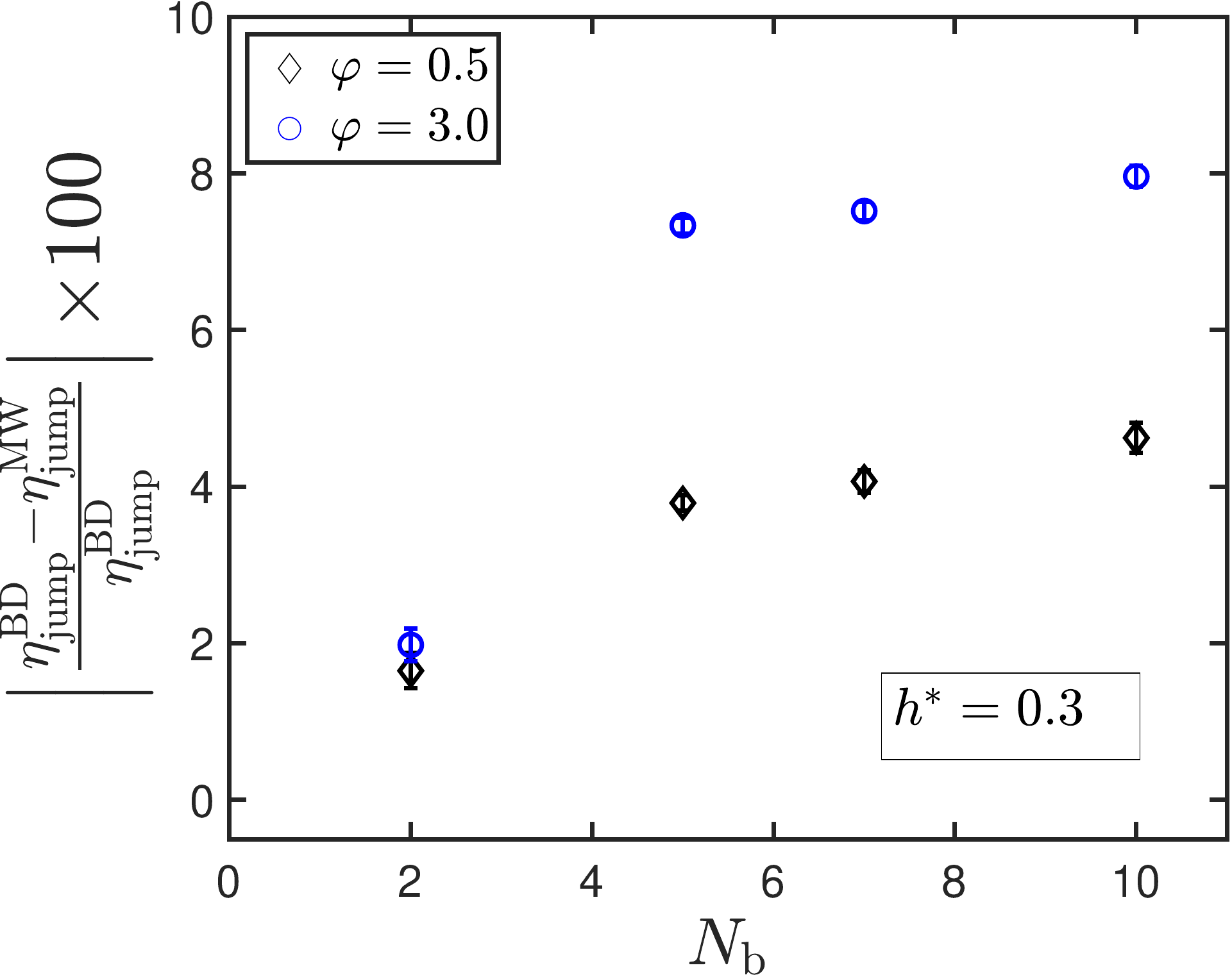}\\[5pt]
(a)&(b)\\
\includegraphics[width=3.in,height=!]{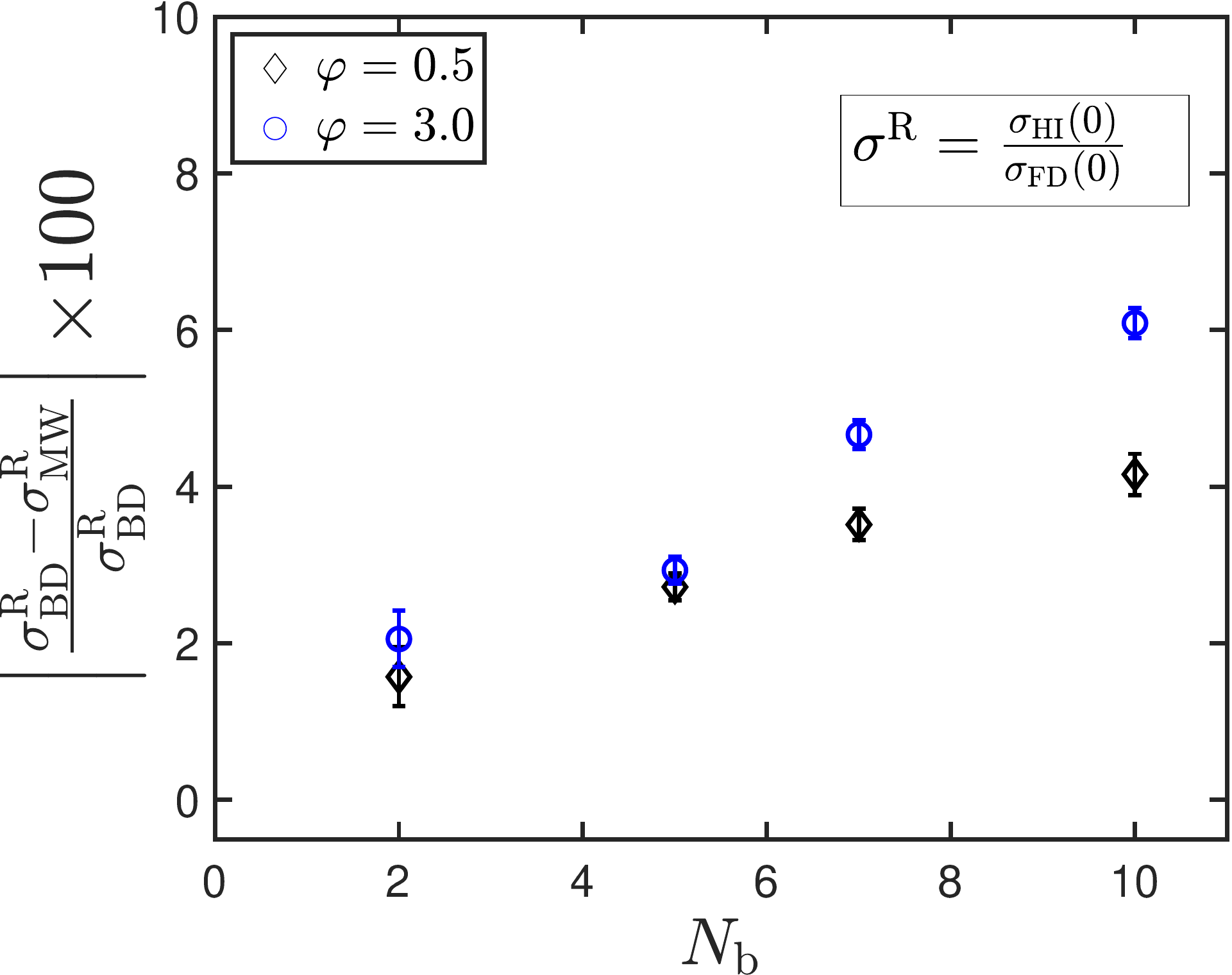}&
\includegraphics[width=3.in,height=!]{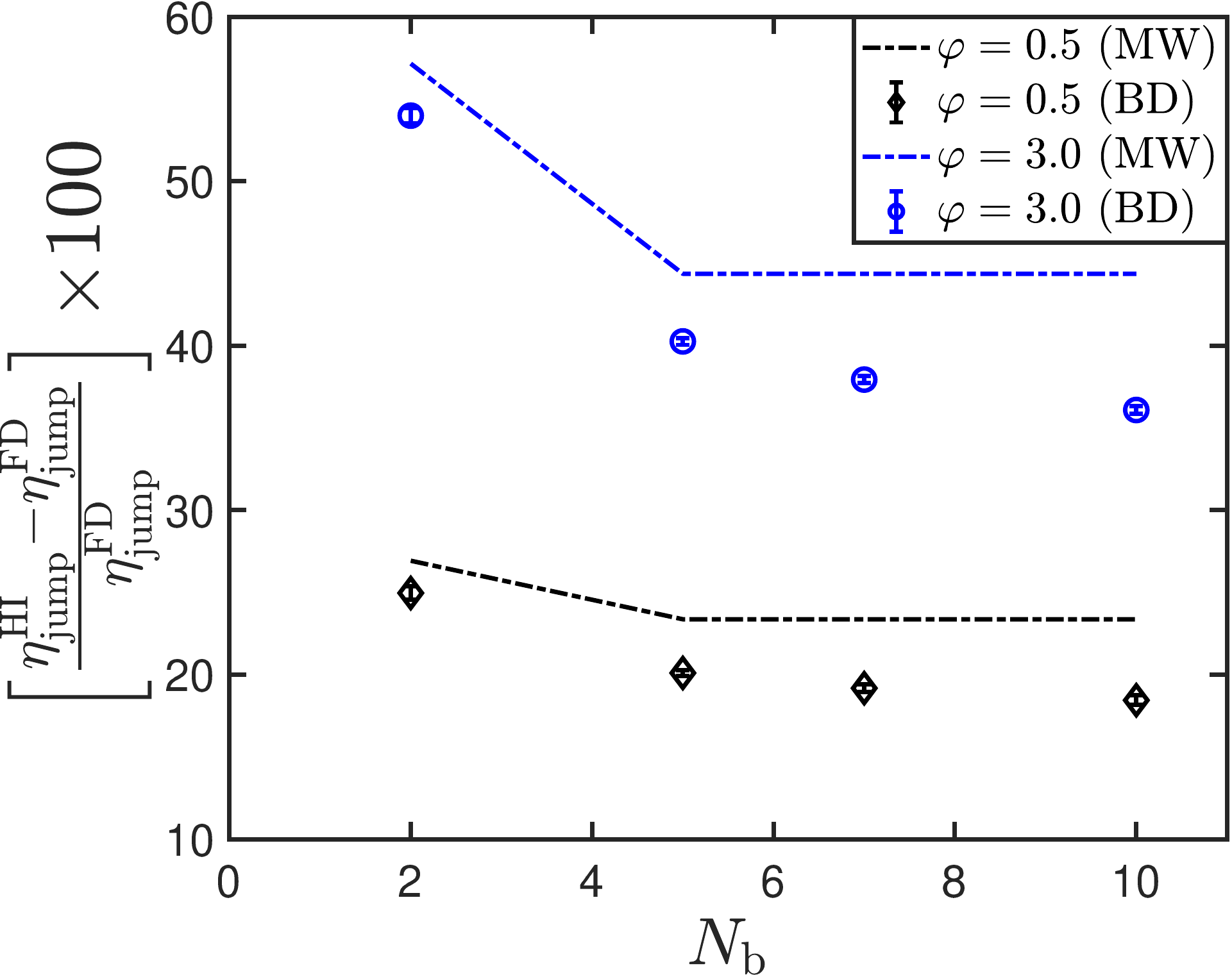}\\[5pt]
(c)&(d)
\end{tabular}
\end{center}
\caption{ (Color online) Percentage difference between exact BD simulations and semi-analytical predictions of Manke and Williams~\cite{Manke1992} as a function of the number of beads for chains with internal friction. The stress jumps for chains without and with hydrodynamic interactions are compared in (a) and (b), respectively. A comparison of the stress jump ratio predictions, $\sigma^{\text{R}}=\sigma_{\text{HI}}(0)/\sigma_{\text{FD}}(0)$, is provided in (c). The effect of hydrodynamic interactions on the magnification of the stress jump is illustrated in (d), where symbols denote BD simulations and the lines indicate the semi-analytical approximation~\cite{Manke1992}. Error bars are roughly of the same size or smaller than the symbols used.}
\label{fig:compare_mw_bd}
\end{figure*}

Using a preaveraged form of hydrodynamic interaction tensor,~\citet{Manke1992} derived a semi-analytical expression for the ratio of stress jumps for chains with and without hydrodynamic interactions, $\sigma^{\text{R}}=\sigma_{\text{HI}}(0)/\sigma_{\text{FD}}(0)$. In figure~\ref{fig:sjump_figure}~(b), the stress jump ratio evaluated over a range of $h^{*}$ values are compared against the \citet{Manke1992} prediction, for two different values of the IV parameter. It is found that the semi-analytical predictions agree qualitatively with the exact BD simulation results. A quantitative analysis of the percentage difference between the two predictions is presented in figures~\ref{fig:compare_mw_bd}~(a)-(c). In all the three cases, the simulation results differ from the theory by less than 10\%. For the free-draining case (fig.~\ref{fig:compare_mw_bd}~(a)), the percentage deviation increases with $N_{\text{b}}$ for $N_{\text{b}}<5$, before decreasing. This is consistent with the trend reported in~\citet{Kailasham2021}, and stems from the assumption used by~\citet{Manke1992} which considers that the terminal and interior connector vectors contribute equally to the stress jump. This assumption becomes increasingly accurate as the number of beads in the model is increased, since the terminal connector vectors represent a smaller fraction of the overall length of the chain. With the inclusion of HI, however, it is observed from \ref{fig:compare_mw_bd}~(b) that the deviation between theory and simulations does not decrease at larger chain lengths. This trend is also propagated in the estimation of the stress jump ratio, as seen from fig.\ref{fig:compare_mw_bd}~(c). The deviation is more easily perceived at higher values of the internal friction parameter, and could potentially be due to the `pre-averaging' assumption employed in~\citet{Manke1992}, which does not account for fluctuations in HI. In figure~\ref{fig:compare_mw_bd}~(d), the magnification of the stress jump due to hydrodynamic interactions, in comparison to the value for freely draining chains, is plotted as a function of $N_{\text{b}}$. Both the semi-analytical theory and simulations appear to predict that the magnification effect due to HI increases with the IV parameter and decreases slowly with $N_{\text{b}}$ before approaching a constant value. The role of fluctuations in HI in determining the magnitude of the stress jump is evident from this figure.

\subsection{\label{sec:steady_sr} Steady-shear viscometric functions}

\begin{figure}[h]
\begin{center}
\begin{tabular}{c}
\includegraphics[width=3.4in,height=!]{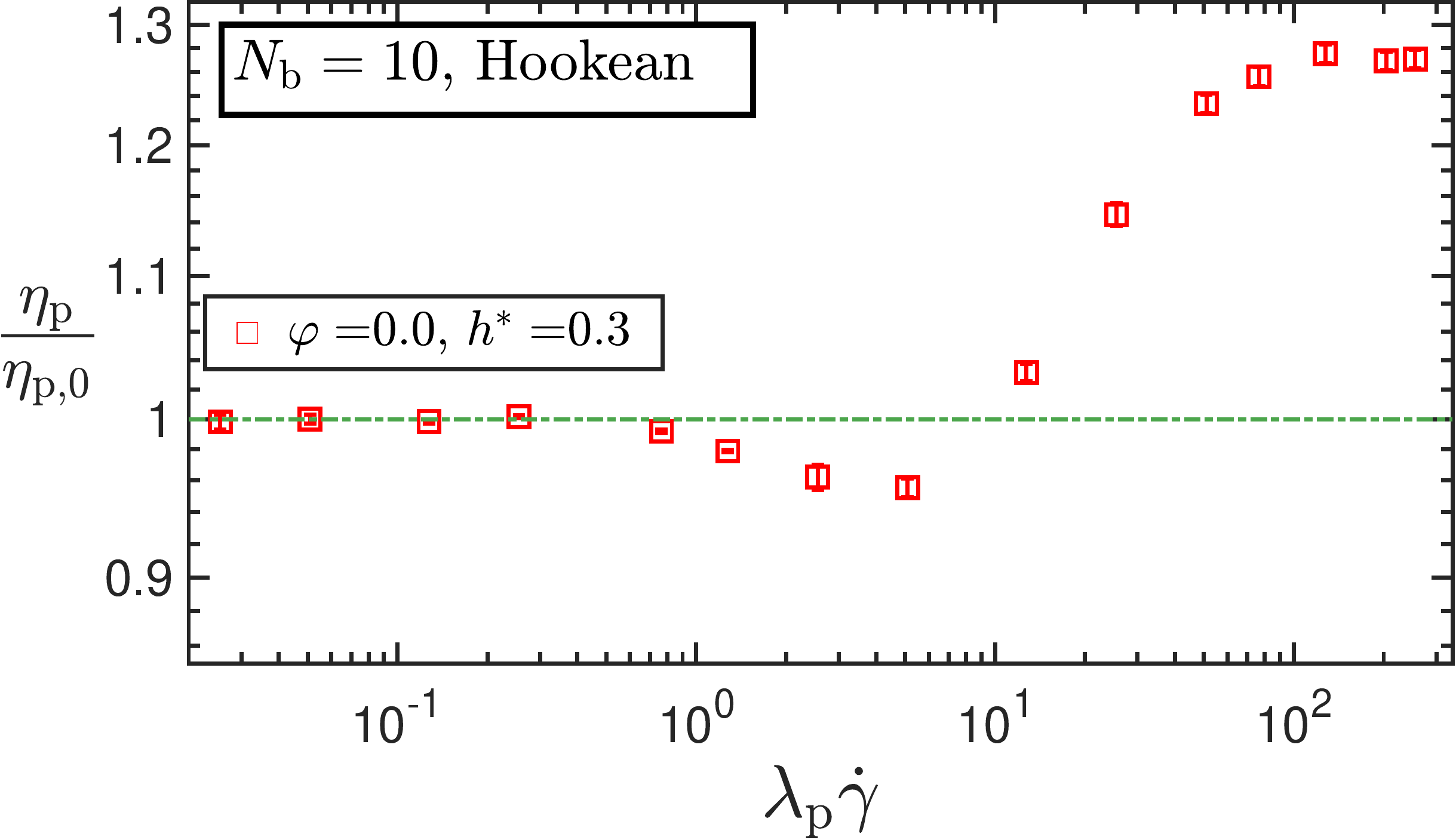}\\
(a)\\
\includegraphics[width=3.4in,height=!]{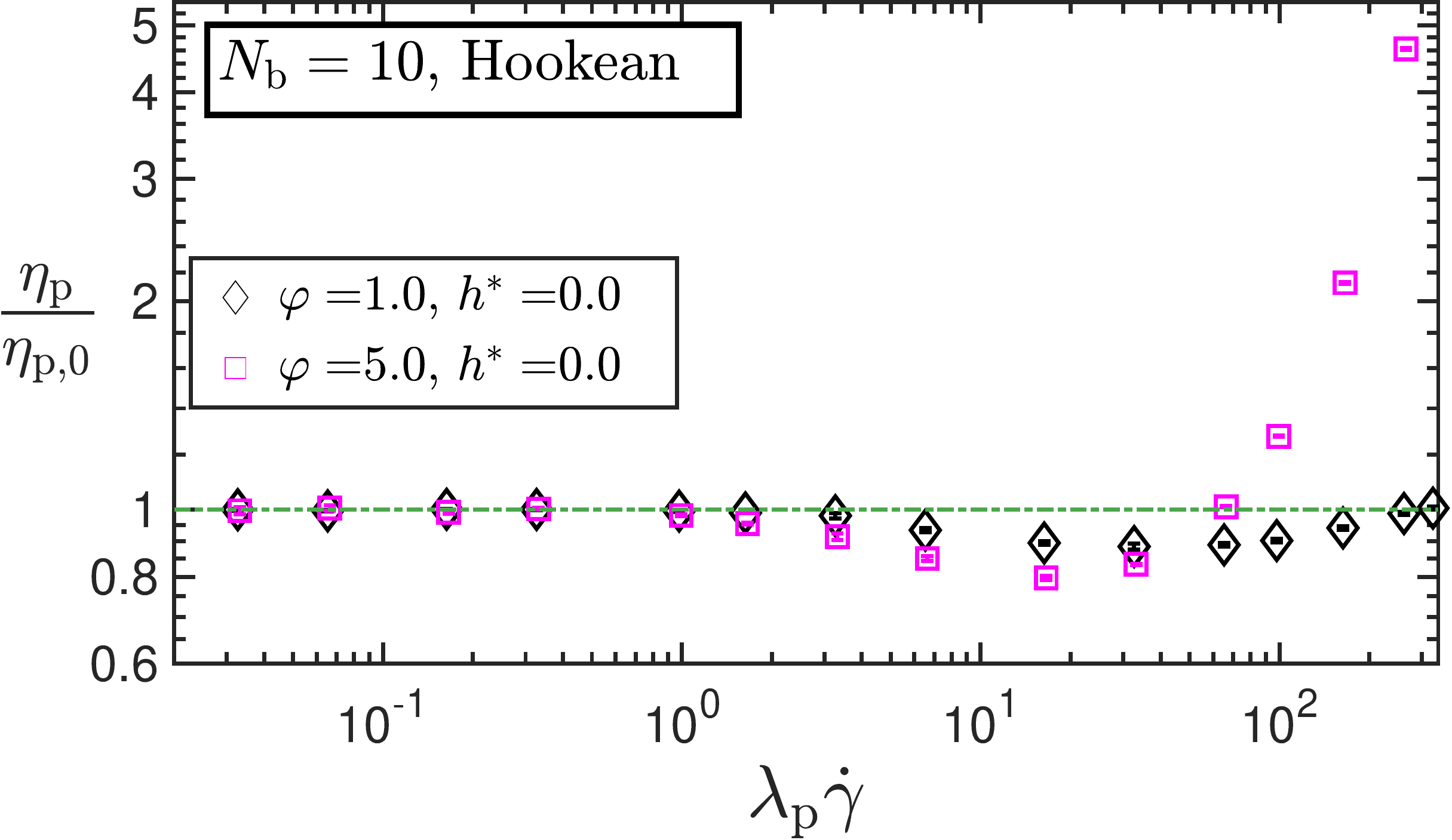}\\
(b)
\end{tabular}
\end{center}
\caption{ (Color online) Scaled shear viscosity as a function of the dimensionless shear rate, for a ten-bead Rouse chain with (a) hydrodynamic interactions ($h^{*}=0.3$) and (b) internal friction. Error bars are roughly of the same size or smaller than the symbols used.}
\label{fig:ref_case_rouse}
\end{figure}

In this section, we analyze the steady shear viscosity of polymer models with various combinations of three non-linear effects: finite extensibility, internal friction, and hydrodynamic interactions in a ten-bead chain. The effect of IV and HI, considered separately, on the steady-shear viscosity of a chain with Hookean springs is highlighted in Figs.~\ref{fig:ref_case_rouse}~(a) and (b). This figure will serve as the reference case for the rest of this section, as more combinations of the various nonlinear effects are considered. The simple Rouse model, with Hookean connecting springs between neighboring beads, predicts a constant value of the viscosity at all shear rates~\cite{Bird1987b}. With the inclusion of hydrodynamic interactions, the viscosity undergoes mild shear-thinning, followed by thickening and culminating in a high shear-rate plateau. This behaviour, reported previously in Refs.~\citenum{Kishbaugh1990,Zylka1991,Prabhakar2006}, is also illustrated in Fig.~\ref{fig:ref_case_rouse}~(a). Additionally, the zero-shear-rate viscosity of ten-bead Rouse chains with HI is lower than its free-draining counterpart.

\begin{figure}[t]
\begin{center}
\includegraphics[width=3.3in,height=!]{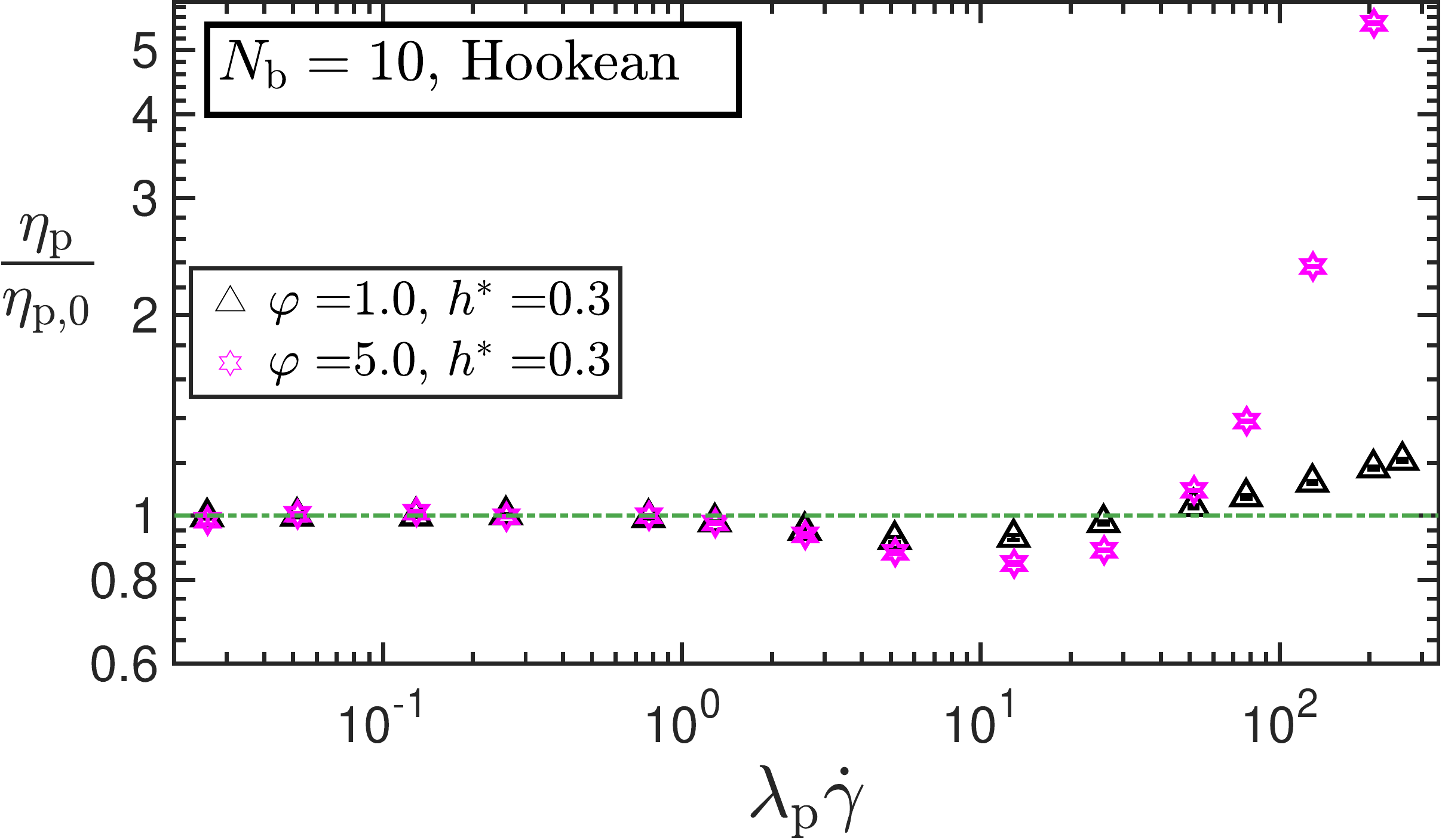}
\end{center}
\caption{ (Color online) Shear viscosity of Rouse chains with fluctuating hydrodynamic interactions ($h^{*}=0.3$), at two values of the internal friction parameter, $\varphi=1.0$ and $\varphi=5.0$. Error bars are roughly of the same size or smaller than the symbols used.}
\label{fig:rouse_iv_hi}
\end{figure}

The effect of internal friction on the material functions of free draining bead-spring-dashpot chains with Hookean springs has been examined in a previous work~\cite{Kailasham2021}, and we restate the key features with respect to the shear viscosity variation here. As seen from Fig.~\ref{fig:ref_case_rouse}~(b), the addition of IV effects to the Rouse model results in both a shear-thinning and thickening of the viscosity. An increase in the value of the IV parameter from $\varphi=1.0$ to $\varphi=5.0$ causes both the shear-thinning and thickening to increase. The critical shear rate at which the onset of shear-thickening is observed, however, remains largely unaffected by $\varphi$. Furthermore, the zero-shear rate viscosity for models with IV is also unaffected by $\varphi$, in agreement with previous theoretical and numerical predictions~\cite{Schieber1993,Kailasham2021}. A comparison between Fig.~\ref{fig:ref_case_rouse}~(a) and ~\ref{fig:ref_case_rouse}~(b) reveals that the magnitude of both shear-thinning and thickening induced by internal viscosity is larger than that due to hydrodynamic interactions, most noticeable for the $\varphi=5.0$ case. Additionally, while the effect of hydrodynamic interactions weakens at high shear rates due to the large inter-bead separations, no such weakening is expected for the internal friction force.

\begin{figure}[t]
\begin{center}
\includegraphics[width=3.3in,height=!]{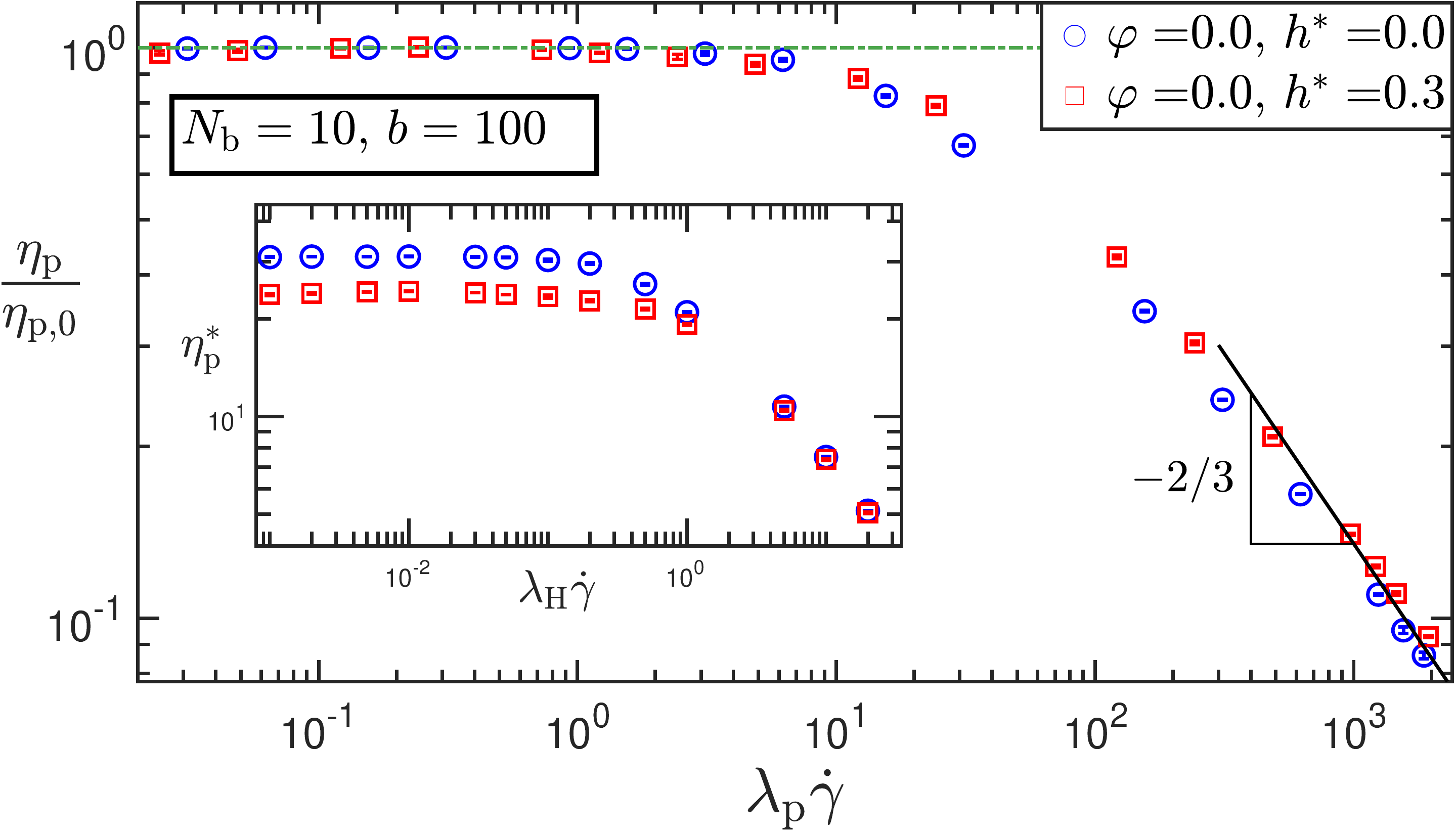}
\end{center}
\caption{ (Color online) Shear viscosity of a dilute solution (scaled by the zero-shear rate value) of ten-bead chains composed of finitely extensible springs without and with hydrodynamic interactions.  Inset shows the variation of the raw dimensionless viscosity as a function of the dimensionless shear rate for the two models. Error bars are roughly of the same size or smaller than the symbols used.}
\label{fig:fene_and_hi}
\end{figure}

In Fig.~\ref{fig:rouse_iv_hi}, the shear viscosity of Rouse chains with fluctuating IV and hydrodynamic interactions is plotted as a function of shear rate, for two different values of the internal friction parameter. A comparison with Fig.~\ref{fig:ref_case_rouse}~(b) reveals that the effect of inclusion of HI is least perceptible at low and moderate shear rates ($\lambda_{\text{p}}\dot{\gamma}\leq15$), with the shear viscosity undergoing a mild-thinning following the Newtonian plateau. The effect of HI, however, becomes more perceptible at higher shear rates. For freely-draining chains with $\varphi=1$, the scaled viscosity does not rise above unity over the range of shear rates considered. The inclusion of hydrodynamic interactions, however, aids the shear-thickening effect, causing the viscosity for the $\varphi=1$ to rise above unity at a smaller value of the shear rate. This cooperative interplay between internal viscosity and hydrodynamic interactions is also perceptible at higher values of the IV parameter, with the scaled viscosity in the presence of hydrodynamic interactions being higher than that for freely-draining chains at the same non-dimensional shear rate. 

The inclusion of finite extensibility causes the viscosity to decrease as a function of the shear rate, with a power-law exponent of $-2/3$, as illustrated in Fig.~\ref{fig:fene_and_hi}. This exponent is unaltered by the inclusion of HI. In the inset to Fig.~\ref{fig:fene_and_hi}, the viscosity and shear rates have not been scaled by $\eta_{\text{p,0}}$ and $\lambda_{\text{p}}$, respectively, but rather plotted in their dimensionless forms as directly obtained from simulations. The effect of hydrodynamic interactions is stronger at low shear rates, as seen from the reduction in the zero-shear-rate viscosity. At larger shear rates however, the calculated values of the viscosity for the two models coincide within error bars due to the weakening of hydrodynamic interactions. The interplay between finite extensibility and hydrodynamic interactions has been examined extensively in Refs.~\citenum{Kishbaugh1990} \& \citenum{Prabhakar2006}. At large enough values of the $b$ parameter, implying a high degree of extensibility, coarse grained models exhibit a thinning-thickening-thinning pattern in the shear viscosity, with the onset of the high shear-rate thinning attributed to the finite extensibility of the connecting springs. The singular limit $b\to\infty$ represents the Hookean spring, and there is no second shear-thinning regime for this special case, as the viscosity reaches a high-shear rate plateau, as already seen in Fig.~\ref{fig:ref_case_rouse}~(a). Below a threshold value of the $b$ parameter ($\approx 1000$), however, for this chain length, the thickening at intermediate shear rates vanishes, and the shear viscosity thins continuously. 

Aside from bead-spring-chain representations, a prominent choice for the modeling of dilute polymer solutions is bead-rod-chains, where the centres of friction are joined by rigid, inextensible rods. 
The equivalence between a rigid rod and a spring-dashpot in the limit of high internal viscosity ($\varphi\to\infty$) is an important topic, and has been discussed in detail in our previous work~\cite{Kailasham2018,Kailasham2021sm}. At the single-mode level, the stochastic differential equation for the connector vector joining the two beads in a spring-dashpot model is identical to that in a rigid rod in external potential, in the limit of $\varphi\to\infty$~\cite{Kailasham2021sm}. In this limit, the length of the dumbbell is constrained due to the high internal friction, but its orientation is free to execute diffusion on the surface of a sphere. In ref.~\citenum{Kailasham2021sm}, we have shown that a correct treatment of the internal friction force that accounts for fluctuations results in the diffusion tensor remaining finitely-valued even as $\varphi\to\infty$. A preaveraged treatment that ignores fluctuations in the IV force, however, predicts erroneously that $\bm{\mathcal{D}}\to\bm{0}$ as $\varphi\to\infty$. A detailed comparison between the stress relaxation modulus, $G(t)$, the stress jump, and the steady-shear viscosity predictions of the two models is presented in Sec.~III~D of ref.~\citenum{Kailasham2018}. For an ensemble of rigid dumbbells whose lengths are drawn from a FENE distribution function with $b=100$, $G(t)$ is sufficiently approximated by an ensemble of FENE-spring-dashpots with $\varphi=5$. The stress jump for the IV model is within $10\%$ of the rigid dumbbell ensemble. Furthermore, for an ensemble of FENE dumbbells with $b=1$ and $b=10$ subjected to steady shear flow, it is observed that using a value of either $\varphi=5$ or $\varphi=10$ results in a shear-thinning exponent of $-(1/3)$, which is identical to the rigid-rod prediction~\cite{Stewart1972}. The actual values of shear viscosity predicted by the two models are different, with the distinction most pronounced at lower shear rates. Therefore, while the equivalence between the bead-spring-dashpot and the rigid rod model is clear is at the dumbbell level, a comparison between the between the two models for a chain is not quite straigthforward.  Additionally, as seen from eq.~(\ref{eq:rev_1_4}),  there is a chain-wide propagation of momentum in the bead-spring-dashpot model, through the nearest neighbor interactions of the connector vectors due to internal friction, even in the absence of hydrodynamic interactions, that is absent in the bead-rod-chain model. This could be the major qualitative change while going from a dumbbell to chain level description that distinguishes the IV model from rigid rod one. A crucial difference in the shear viscosity profile of these two models is that while bead-spring-chains (with finite extensibility) predict an indefinite shear-thinning at high shear rates~\cite{Hsieh2004,Prabhakar2004}, the bead-rod-chain predicts a second Newtonian plateau at high shear rates, following a shear-thinning regime~\cite{Liu1989,Doyle1997,Petera1999,Hsieh2006}. This distinguishing feature remains true both in the presence and absence of fluctuating hydrodynamic interactions. Any discussion about the shear-thinning exponent in bead-rod-chain models, therefore, is applicable to the regime preceding the second Newtonian plateau. A detailed exposition on the shear-thinning phenomenon in dilute and semi-dilute polymer solutions, along with an extensive review of pertinent literature, is available in ref.~\citenum{Pan2018}. We would like to highlight refs.~\citenum{Pincus2020,Pincus2022} in this regard, which show that the viscometric, rheo-optical, and configurational properties of bead-rod-chains may be accurately mimicked by the use of suitably stiff FENE-Fraenkel springs in the conventional bead-spring-chain framework.

While modeling a polymer chain of fixed length, increasing the number of rods corresponds to an increase in the flexibility of the model. In a simulation of freely-draining bead-rod chains~\cite{Liu1989}, an increase in the number of rods from two to twenty results in a change in the magnitude of the shear-thinning exponent, from $(1/3)$ to $(1/2)$. An exponent of $-(6/11)$ has been observed in BD simulations of bead-rod chains by Doyle \textit{et al.}~\cite{Doyle1997} and in bead-spring chains by Jendrejack \textit{et al.}~\cite{Jendrejack2002}. A common method to increase the stiffness of bead-spring chain models is the introduction of a bending potential that penalizes torsional angle rotation. Multiparticle collision dynamics simulations of semiflexible polymer chains by~\citet{Ryder2006} suggest that a shear-thinning exponent of $-(1/3)$ is reached as the strength of the bending potential is increased. These observations, therefore, serve to hint that the shear-thinning exponent is a measure of the static flexibility of the chain. 

\begin{figure}[t]
\begin{center}
\begin{tabular}{c}
\includegraphics[width=3.4in,height=!]{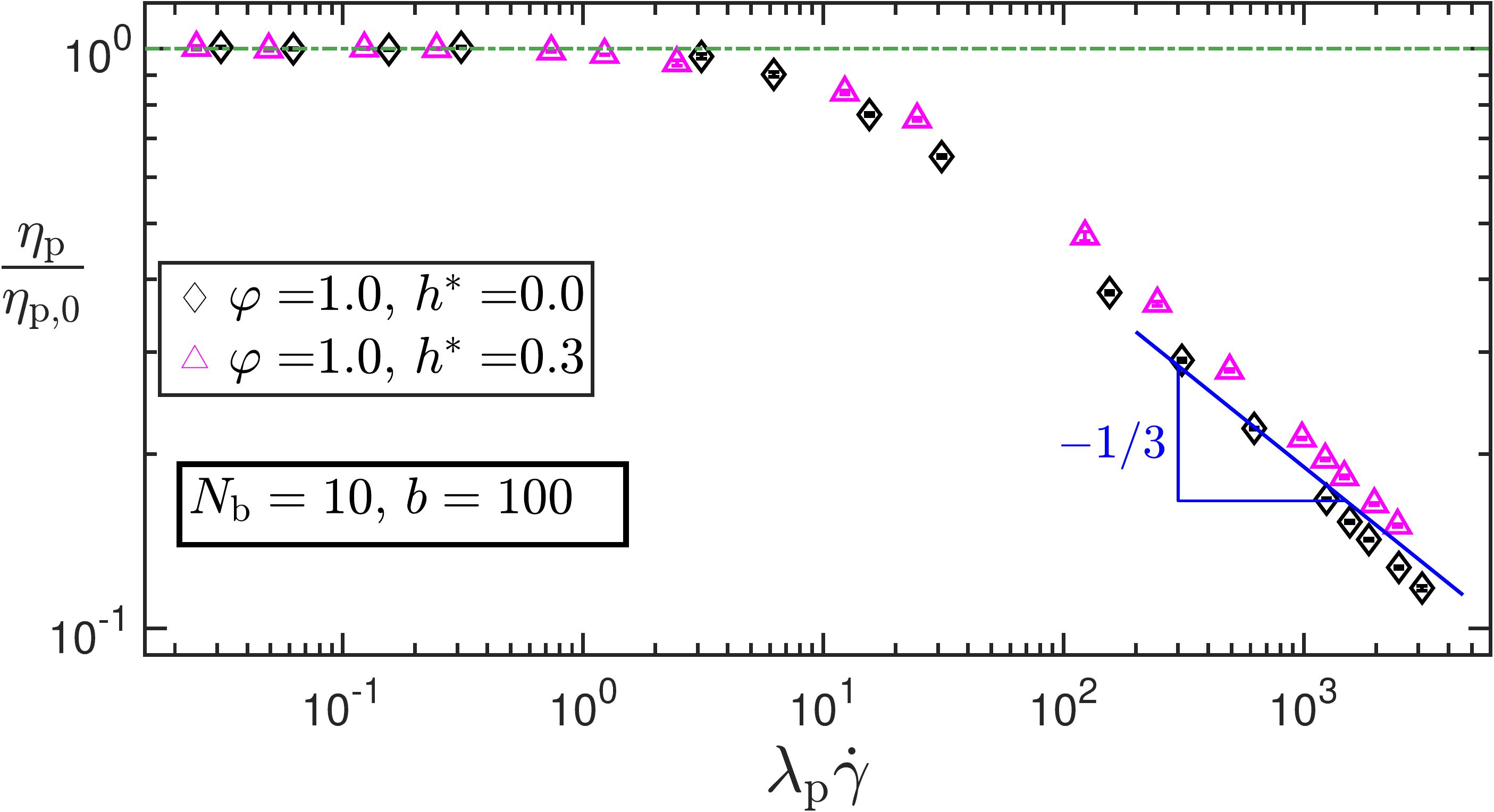}\\
(a)\\
\includegraphics[width=3.4in,height=!]{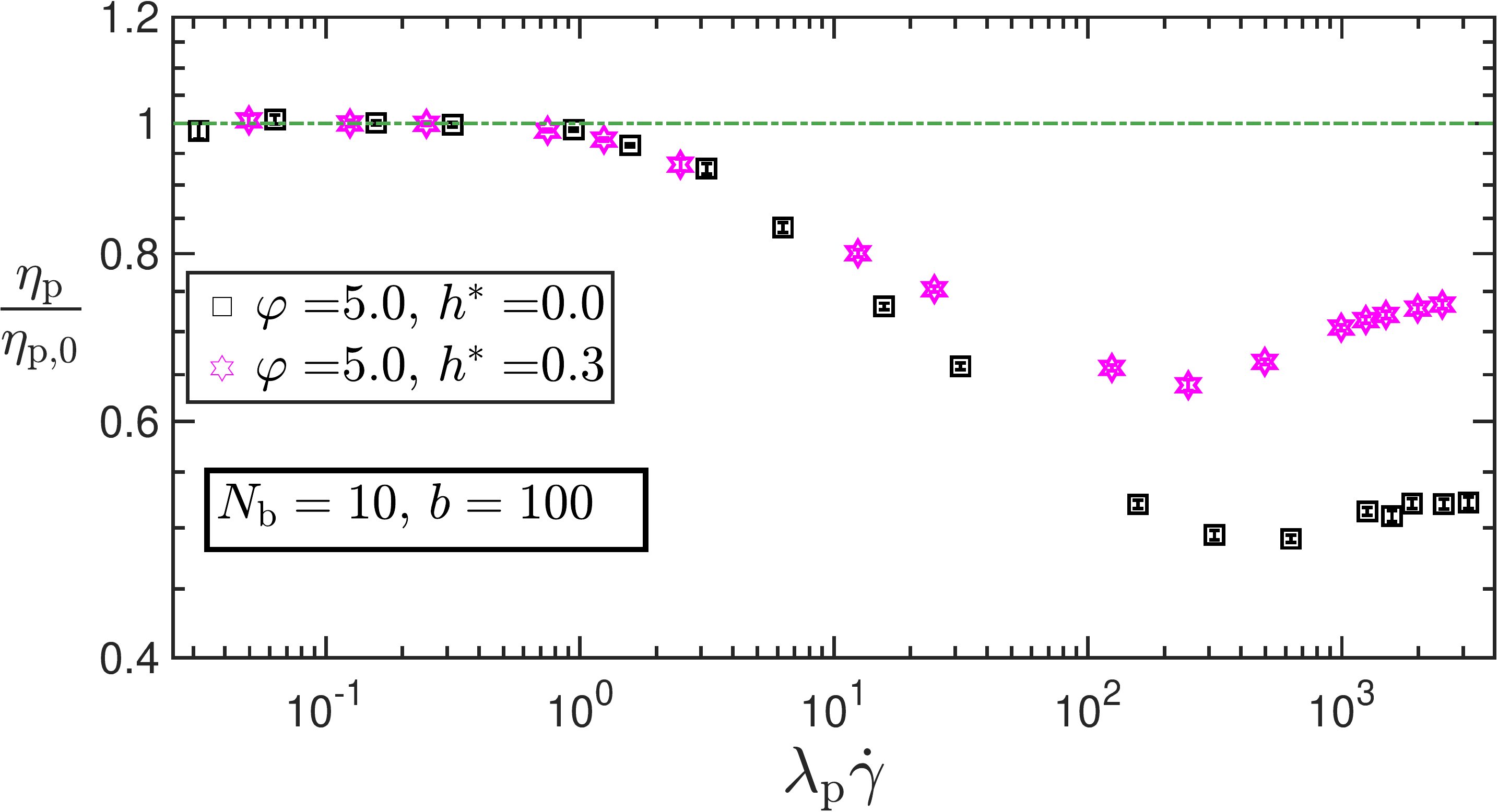}\\
(b)
\end{tabular}
\end{center}
\caption{ (Color online) Shear viscosity of a ten-bead chain with FENE springs and (a) $\varphi=1.0$ and (b) $\varphi=5.0$, with and without hydrodynamic interactions.  Error bars are roughly of the same size or smaller than the symbols used.}
\label{fig:fene_iv_hi_combo}
\end{figure}

A comparison between Figs.~\ref{fig:fene_and_hi} and \ref{fig:fene_iv_hi_combo}~(a) indicates that for the same level of discretization, the shear-thinning exponent changes from $-(2/3)$ to $-(1/3)$ upon the inclusion of a small value of the internal friction parameter ($\varphi=1.0$). The qualitative pattern remains the same, with the shear-thinning appearing to continue indefinitely. At $\varphi=1$, there is a slight spread in the viscosity data at higher shear rates when comparing free-draining chains and chains with HI. The zero-shear-rate viscosity is also reduced by the inclusion of HI (just as in fig.~\ref{fig:fene_and_hi}), but this effect is not visible in fig.~\ref{fig:fene_iv_hi_combo}~(a), where the viscosity has been normalized by $\eta_{\text{p,0}}$. For the $\varphi=5.0$ case [fig.~\ref{fig:fene_iv_hi_combo}~(b)], shear-thinning characterized by the $-(1/3)$ exponent is not observed over the range of shear rates examined, but we anticipate that at higher shear rates, the finite extensibility of the springs would result in a second shear-thinning regime. The bead-spring-dashpot chain therefore exhibits the shear-thinning signature of a single rigid dumbbell upon the inclusion of IV. A similar observation was made by McLeish and coworkers~\cite{Khatri2007rif} based on dynamic compliance calculations for a Rouse model with internal friction: they posit that ``... a polymer with many beads and degrees of freedom, but dominated by internal friction, acts like a dumbbell with a single degree of freedom, where the relaxation of faster or small-wavelength modes are effectively frozen out due to their slowness in changing conformation." While their comments were made in the context of the frequency response of a chain at equilibrium, we note from fig.~\ref{fig:fene_iv_hi_combo} that the freezing of slower modes could lead to dumbbell-like dynamics even away from equilibrium, in the presence of a flow-field. The chain-wide momentum propagation through nearest neighbor interactions of the connector vector due to internal friction (see eq.~(\ref{eq:rev_1_4})) could be a plausible reason for the entire chain to behave as a single entity at large shear rates, or short timescales.

For freely-draining bead-spring dashpot chains, an increase in the IV parameter to a value of $\varphi=5.0$ results in the appearance of a high-shear-rate plateau, following a shear-thinning-thickening pattern, as seen in figure~\ref{fig:fene_iv_hi_combo}~(b), a feature that is not observed in bead-spring-chain models without internal friction. A balance between the competing tendencies of the finite extensibility of the spring to decrease the viscosity at high shear rates, and that of IV to increase it could be a plausible explanation for this plateau. The simultaneous inclusion of fluctuating internal viscosity and hydrodynamic interactions in a model with finitely extensible springs results in the onset of a thickening in the viscosity at high shear rates, as seen from figure~\ref{fig:fene_iv_hi_combo}~(b). This could be attributed to the cooperative shear-thickening effects of IV and HI dominating over the finite-extensibility-induced shear-thinning.

\subsection{\label{sec:exp_compare} Comparison with experimental data}

\begin{figure}[t]
\begin{center}
\includegraphics[width=3.3in,height=!]{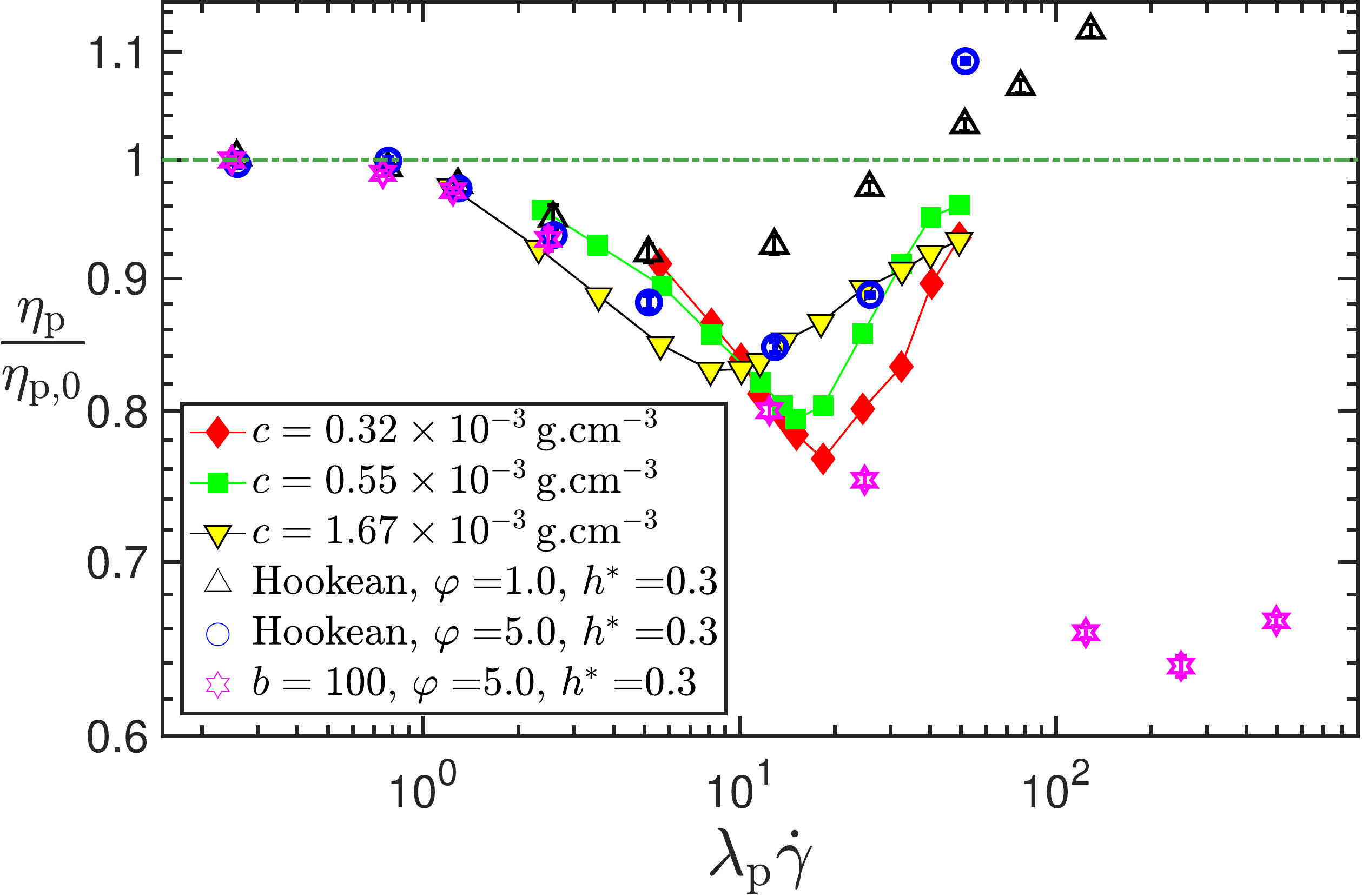}
\end{center}
\caption{ (Color online) Shear viscosity of dilute polymer solutions (scaled by their zero-shear rate value), as a function of the dimensionless shear rate. The solid symbols represent experimental data on polystyrene (MW=$8.4\times 10^{6}$ g/mol) in decaline, at various solution concentrations $c$, taken from Fig. 1a of ~\citet{Layec-Raphalen1976}. The hollow symbols are BD simulation results, with error bars roughly of the same size or smaller than the symbols used.}
\label{fig:exp_compare}
\end{figure}

Experimental evidence for shear-thickening is very scarce in the rheology literature, however, observations of shear-thickening have been reported in dilute solutions of polystyrene in decaline~\cite{Layec-Raphalen1976}, polyethylene in xylene and polypropylene in tetralin~\cite{Vrahopoulou1987a}, and polyisobutylene in polybutene~\cite{Bianchi1968}. The two most common explanations to rationalize this phenomenon invoke inter- and intramolecular interactions.

~\citet{Vrahopoulou1987a} observe that although entanglement effects are absent in their experiments on dilute polymer solutions, there exists the possibility of transient assembly and breakage of macrostructures in the presence of shear flow. The formation of such ``quasi-aggregates" under flow conditions has been attributed to shear-thickening, in the theoretical analyses by Simha et al.~\cite{Simha1949,Weissberg1951}, as well as Wolff and coworkers~\cite{Layec-Raphalen1976,Wolff1979,Dupuis1993}. ~\citet{Hatzikiriakos1996} perform Brownian dynamics simulations of Hookean dumbbells with anisotropic hydrodynamic drag in shear flow, wherein the dumbbells undergo a reversible association process to form $n-$mers. A qualitative agreement with the experimental results of ~\citet{Vrahopoulou1987a} is observed, and they posit the formation of tetramers to be the major reason for shear-thickening. While dilute and semidilute associative polymeric solutions~\cite{Tripathi2006,Jaishankar2015} have been known to result in shear-thickening due to the formation of intramolecular bonds, there does not appear to be evidence for aggregate formation in the literature on shear-thickening in dilute solutions of homopolymers cited above.

When intermolecular associations are discounted, the predominant framework used to explain theoretical predictions of shear-thickening is the ``upturn effect"~\cite{Peterlin1960}. This may be understood as follows: for coarse-grained models with greater than six beads, the Rouse viscosity is higher than the Zimm viscosity. At large shear rates, the effect of hydrodynamic interactions weakens due to larger inter-bead separation. The solution viscosity therefore tends to the Rouse value, resulting in shear-thickening, followed by the attainment of a high-shear rate plateau~\cite{Zylka1991}. With the inclusion of finitely extensible springs, the viscosity undergoes a thinning-thickening-thinning pattern, with the second shear-thinning regime appearing due to finite-extensibility effects~\cite{Kishbaugh1990,Prabhakar2006}. To the best of our knowledge, viscometric predictions of internal friction models have so far not been compared against experimentally observed shear-thickening. ~\citet{Layec-Raphalen1976} argue qualitatively that at high shear rates, the timescale for the rotation of the polymer chain becomes comparable to the deformation timescale of the molecule which is governed by internal friction. As a result, the molecule resists changes in its conformation, leading to an increase in the shear viscosity at higher shear rates for polymers with internal friction. This heuristic explanation aligns with the trends observed in Fig.~\ref{fig:rouse_iv_hi}, where the viscosity of free-draining Rouse chains with IV is seen to increase unboundedly at large shear rates. For chains with finitely extensible springs and IV, we expect that the shear-thickening would give way to thinning at large shear rates due to finite extensibility effects. We do not see the second shear-thinning regime in 
Fig.~\ref{fig:fene_iv_hi_combo}~(b) for the range of shear rates simulated in our work, but anticipate that going to higher shear rates would precipitate this effect.

In fig.~\ref{fig:exp_compare}, the scaled shear viscosity of dilute polystyrene solutions (at various concentrations) in decaline, reported by~\citet{Layec-Raphalen1976} is plotted as a function of the scaled shear rate, alongside BD simulation results for models with fluctuating IV and HI obtained in the present work. The experimental system is slightly above its $\theta$-point. The overall trend in both the simulation and experimental results is one of shear thinning followed by thickening. We introduce two quantities, $\eta_{\text{m}}$ and $\beta_{\text{m}}=\lambda_{\text{p}}\dot{\gamma}_{\text{m}}$, which represent, respectively, the minimum value of the scaled shear viscosity that is observed prior to its upturn, and the dimensionless shear rate at which the onset of shear-thickening occurs. It is clear from fig.~\ref{fig:rouse_iv_hi}, that an increase in the internal friction parameter leads to a depression in $\eta_{\text{m}}$, followed by a sharp increase in the viscosity, but has a minimal effect on $\beta_{\text{m}}$. On the other hand, the finite extensibility parameter ($b$) perceptibly affects both $\eta_{\text{m}}$ and $\beta_{\text{m}}$. It is therefore possible to get a better agreement with the experimental results by choosing appropriate values of $b$ and $\varphi$. Additionally, it is also important to consider excluded volume (EV) interactions when comparing against experimental data, because EV has been known to cause shear-thinning~\cite{Petera1999,Lyulin1999,Prakash2002,Liu2004,Prabhakar2004}.

We do note, however, that the experimental results are reported at various finite concentrations, while our model is applicable only for dilute polymer solutions, and cannot incorporate the effects of finite concentrations. A careful comparison against experiments would necessitate the construction of a numerical framework that considers multiple bead-spring-dashpot chains within a simulation box, accounts for the solvent quality through excluded volume interactions, and incorporates chain length effects through an appropriate choice of the finite extensibility parameter. The comparison presented in fig.~\ref{fig:exp_compare} is therefore purely qualitative in nature, mainly to highlight that the extent of shear thinning and thickening observed in dilute polymer solutions is comparable to that predicted by models with internal viscosity. A thorough investigation would necessitate systematic experiments to extricate and identify the various physical origins of shear-dependent viscosity in dilute homopolymer solutions. We hope that the present work would spur such experimental studies.

\section{\label{sec:conclusions} Conclusions}

A methodology for decoupling the connector vector velocities in coarse-grained polymer models with fluctuating internal friction and hydrodynamic interaction effects has been developed for the general case of a chain with $N_{\text{b}}$ beads. This expands the scope for the treatment of flexible polymer models with IV and HI effects, for which solutions were previously available only for the $N_{\text{b}}=2$ case~\cite{Hua19961473,Kailasham2018}. The relevant stochastic differential equations are obtained by attaching a kinetic interpretation~\cite{Hutter1998} to the governing Fokker-Planck equation, and integrated numerically using Brownian dynamics simulations. This method is validated by comparison against prior simulation results, where available, and is shown to be an order-of-magnitude faster than a previous, recursion-based technique~\cite{Kailasham2021} that is applicable only to free-draining models with internal friction. A thermodynamically consistent stress tensor expression for the model has been derived, and the divergence of the diffusion tensor appearing in this expression is evaluated using a random finite difference approach~\cite{Sprinkle2017,Sprinkle2019}. 

While the semi-analytical approximation for the stress jump at the inception of shear flow in free-draining bead-spring-dashpot chains~\cite{Manke1988} compares excellently against exact numerical simulations~\cite{Kailasham2021}, with the accuracy improving with the number of beads in the chain, the error in the corresponding approximation for such chains with pre-averaged hydrodynamic interactions~\cite{Manke1992} does not diminish with the $N_{\text{b}}$ when compared against exact simulation results computed in this paper that account for fluctuations in both internal viscosity and hydrodynamic interactions.

The steady-shear viscosity of ten-bead chains with various combinations of finite extensibility, IV, and HI effects are presented over a range of shear rates. We find that while both IV and HI induce a shear-thinning followed by a thickening in the viscosity, the thickening effect due to the former is more pronounced than the latter. The inclusion of a small value of the internal friction parameter (in free-draining, finitely extensible chains) results in a shear-thinning exponent of $-(1/3)$, mimicking a rigid dumbbell, and this behavior is unaltered by the inclusion of hydrodynamic interactions. At higher values of the internal friction parameter, however, one observes a shear-thickening or a plateau in the viscosity at high shear rates, depending on whether HI is accounted for or not. The interplay between finite extensibility, IV, and HI effects, therefore, result in a variety of shear-viscosity profiles.

We recognize that the present work does not consider the effects of excluded volume interactions. The inclusion of EV atop the various intramolecular interactions already considered in this work would certainly add more variety to what is already a rich tapestry of shear-thinning profiles. This additional dimension of the parameter space will be explored in a future publication. The key contributions of this work are: (a) the development of a solution algorithm that removes the one-to-all coupling between connector vector velocities in models with IV and HI, and (b) its efficient numerical integration using a kinetic interpretation. Given the continued interest from the biophysics community in understanding the roles of solvent-based and internal friction on the dynamics of proteins~\cite{Xia2021,Das2022,Mukherjee2022}, the present work also provides a mesoscopic simulation tool for answering such questions.

\section*{\label{sec:supp_mat} Supplementary Material}

Additional details pertaining to the study are presented in the Supplementary Material. Section~{SII} of the Supplementary Material establishes that the Fokker-Planck equation for a dumbbell, with fluctuating IV and HI derived using the methodology developed in the present work, corresponds to that obtained previously by an alternative route in ref.~\citenum{Kailasham2018}. Sec.~{SIII} establishes the equivalence between the present methodology and the decoupling algorithm developed in ref.~\citenum{Kailasham2021} for free-draining bead-spring-dashpot chains, and also compares the computational cost for the two approaches. Sec.~{SIV} illustrates that the governing equations for the present model satisfy the fluctuation dissipation theorem, and the time-step convergence of the BD simulation results presented in this work is established in Sec.~{SV}. Lastly, Sec.~{SVI} of the Supplementary Material contains the detailed steps for the derivation of the stress tensor expression used in the evaluation of shear viscosity from BD simulations.

\begin{acknowledgments}
R.K. thanks Prashant Patil, Aleksandar Donev, and Isaac Pincus for enlightening discussions. We also thank the anonymous referees for their insightful suggestions to improve the manuscript. This work was supported by the MonARCH and MASSIVE computer clusters of Monash University, and the SpaceTime-2 computational facility of IIT Bombay. R. C. acknowledges SERB for funding (Project No. MTR/2020/000230 under MATRICS scheme). We also acknowledge the funding and general support received from the IITB-Monash Research Academy.
\end{acknowledgments}

%

%

\end{document}


\beginsupplement
\title{Supplementary Material for: Shear viscosity for finitely extensible chains with fluctuating internal friction and hydrodynamic interactions}
\date{\today}
\author{R. Kailasham}
\email{rkailash@andrew.cmu.edu}
\affiliation{Department of Chemical Engineering, Carnegie Mellon University, Pittsburgh, Pennsylvania -  15213, USA}
\author{Rajarshi Chakrabarti}
\email{rajarshi@chem.iitb.ac.in}
\affiliation{Department of Chemistry, Indian Institute of Technology Bombay, Mumbai, Maharashtra -  400076, India}
\author{J. Ravi Prakash}
\email{ravi.jagadeeshan@monash.edu}
\affiliation{Department of Chemical Engineering, Monash University,
Melbourne, VIC 3800, Australia}

\maketitle

\section{\label{sec:intro} Introduction}

The main paper derives the governing stochastic differential equations for coarse-grained polymer models with finite extensibility (FE), fluctuating internal friction (IV) and hydrodynamic interaction (HI) effects and outlines a methodology for their numerical integration, using Brownian dynamics (BD) simulations, based on a kinetic interpretation of the governing Fokker-Planck equation. A thermodynamically consistent stress expression is used to predict the steady shear viscosity as a function of the shear rate for models with various combinations of the three effects, namely, FE, IV, and HI. Additional details corresponding to various aspects of the study are presented here.

This document is organized as follows. Section~\ref{sec:db_comparison} establishes that the Fokker-Planck equation for a dumbbell with fluctuating IV and HI derived using the methodology developed in the present work, is identical to that obtained in Ref.~\citenum{Kailasham2018}. Sec.~\ref{sec:fd_compare_approach} establishes the equivalence between the present methodology and the decoupling algorithm developed in Ref.~\citenum{kailasham2021rouse} for free-draining bead-spring-dashpot chains, and presents details on the computational cost and scaling for the two approaches. Sec.~\ref{sec:fdt_sat} illustrates that the governing equations for the present model satisfy the fluctuation dissipation theorem, and Sec.~\ref{sec:tstep_convergence} establishes that the BD simulation results presented in this work are time-step convergent. Lastly, Sec.~\ref{sec:stress_tens_derv} contains the derivation for the complete stress tensor expression used for the evaluation of shear viscosity from BD simulations. Summations are indicated explicitly, and the Einstein summation convention is not followed.
\vspace{-15pt}
\section{\label{sec:db_comparison} Comparison at dumbbell level}
The dimensionless version of the Fokker-Planck equation for a bead-spring-dashpot chain with fluctuating hydrodynamic interactions has been presented as Eq. (31) in the main text, and is reproduced here
 \begin{align}\label{eq:fp_ddot_dimless}
 \dfrac{\partial \Psi^{*}}{\partial t^{*}}&=-\sum_{j=1}^{N}\dfrac{\partial}{\partial \bm{Q}^{*}_{j}}\cdot\Biggl\{\Biggl[\sum_{k=1}^{N}\bm{M}_{jk}\cdot\left(\boldsymbol{\kappa}^{*}\cdot\bm{Q}^{*}_{k}\right)-\dfrac{1}{4}\sum_{k=1}^{N}\bm{L}_{jk}\cdot\left(\dfrac{\partial \phi^{*}}{\partial \bm{Q}^{*}_{k}}\right)+\dfrac{1}{4}\sum_{k=1}^{N}\dfrac{\partial}{\partial \bm{Q}^{*}_{k}}\cdot\bm{L}^{T}_{jk}\Biggr]\Psi^{*}\Biggr\} \\  \nonumber
 &+\dfrac{1}{4}\sum_{j,k=1}^{N}\dfrac{\partial}{\partial \bm{Q}^{*}_{j}}\dfrac{\partial}{\partial \bm{Q}^{*}_{k}}:\left[\bm{L}_{jk}^{T}\Psi^{*}\right]
 \end{align}
 where we have introduced the notation $\bm{L}_{jk}=2\bm{D}_{jk}$. For the case of a dumbbell, setting the number of springs, $N=1$ in Eq.~(\ref{eq:fp_ddot_dimless}) and denoting $\bm{Q}_1$ simply as $\bm{Q}$, we obtain
 \begin{align}\label{eq:fp_db1}
 \dfrac{\partial \Psi^{*}}{\partial t^{*}}&=-\dfrac{\partial}{\partial \bm{Q}^{*}}\cdot\Biggl\{\Biggl[\bm{M}_{11}\cdot\left(\boldsymbol{\kappa}^{*}\cdot\bm{Q}^{*}\right)-\dfrac{1}{4}\bm{L}_{11}\cdot\left(\dfrac{\partial \phi^{*}}{\partial \bm{Q}^{*}}\right)+\dfrac{1}{4}\dfrac{\partial}{\partial \bm{Q}^{*}}\cdot\bm{L}^{T}_{11}\Biggr]\Psi^{*}\Biggr\}\\ \nonumber
 &+\dfrac{1}{4}\dfrac{\partial}{\partial \bm{Q}^{*}}\dfrac{\partial}{\partial \bm{Q}^{*}}:\left[\bm{L}_{11}^{T}\Psi^{*}\right]
 \end{align}
 
 From the development presented in the main text
 \begin{align}\label{eq:db_simp1}
 \beta^{*}&=\left[1-\dfrac{\alpha}{Q^{*}}\left(\mathscr{A}+\mathscr{B}\right)\right]\\[5pt]
 \alpha&=\dfrac{3}{4}\sqrt{\pi}h^{*}
  \end{align}
  
\begin{align}
  \tilde{\bm{A}}_{11}&=2\left(\boldsymbol{\delta}-\zeta\Omega\right)\\[5pt]
 \tilde{\bm{Z}}_{11}&=2\beta^{*}\left(\dfrac{\bm{Q}^{*}\bm{Q}^{*}}{\bm{Q}^{*2}}\right)\\[5pt]
 \bm{Y}_{11}&=\left[\boldsymbol{\delta}-\dfrac{\epsilon\beta^{*}}{\epsilon\beta^{*}+1}\dfrac{\bm{Q}^{*}\bm{Q}^{*}}{\bm{Q}^{*2}}\right]\\[10pt]
 \bm{X}_{11}&=2\left[\boldsymbol{\delta}-\dfrac{\epsilon\beta^{*}}{\epsilon\beta^{*}+1}\dfrac{\bm{Q}^{*}\bm{Q}^{*}}{\bm{Q}^{*2}}\right]\cdot\left(\boldsymbol{\delta}-\zeta\Omega\right)\\[5pt]
 \bm{J}_{11}&=\boldsymbol{\delta}
 \end{align}
 and consequently,
 \begin{equation}\label{eq:db_simp2}
 \begin{split}
 \bm{J}^{-1}_{11}&=\boldsymbol{\delta}\\[5pt]
 \bm{M}_{11}&=\left[\boldsymbol{\delta}-\dfrac{\epsilon\beta^{*}}{\epsilon\beta^{*}+1}\dfrac{\bm{Q}^{*}\bm{Q}^{*}}{\bm{Q}^{*2}}\right]\\[5pt]
 \bm{L}_{11}&=2\left[\boldsymbol{\delta}-\dfrac{\epsilon\beta^{*}}{\epsilon\beta^{*}+1}\dfrac{\bm{Q}^{*}\bm{Q}^{*}}{\bm{Q}^{*2}}\right]\cdot\left(\boldsymbol{\delta}-\zeta\Omega\right)\\[5pt]
 &=2\Biggl\{\boldsymbol{\delta}\left[1-\dfrac{\mathscr{A}\alpha}{Q^{*}}\right]-\left(\dfrac{B\alpha}{Q^{*}}\right)\dfrac{\bm{Q}^{*}\bm{Q}^{*}}{\bm{Q}^{*2}}-\dfrac{\epsilon\beta^{*2}}{\epsilon\beta^{*}+1}\dfrac{\bm{Q}^{*}\bm{Q}^{*}}{\bm{Q}^{*2}}\Biggr\}\\[5pt]
 &=\bm{L}^{T}_{11}
 \end{split}
 \end{equation}
 Substituting Eq.~(\ref{eq:db_simp2}) in Eq.~(\ref{eq:fp_db1}) and simplifying,
   \begin{equation}\label{eq:fp_db2}
 \begin{split}
 \dfrac{\partial \Psi^{*}}{\partial t^{*}}&=-\dfrac{\partial}{\partial \bm{Q}^{*}}\cdot\Biggl\{\Biggl[\left[\boldsymbol{\delta}-\dfrac{\epsilon\beta^{*}}{\epsilon\beta^{*}+1}\dfrac{\bm{Q}^{*}\bm{Q}^{*}}{\bm{Q}^{*2}}\right]\cdot\left(\boldsymbol{\kappa}^{*}\cdot\bm{Q}^{*}-\dfrac{1}{2}\left(\boldsymbol{\delta}-\zeta\Omega\right)\cdot\left(\dfrac{\partial \phi^{*}}{\partial \bm{Q}^{*}}\right)\right)\\[5pt]
 &+\dfrac{1}{2}\dashuline{\dfrac{\partial}{\partial \bm{Q}^{*}}\cdot\left[\left(\boldsymbol{\delta}-\dfrac{\epsilon\beta^{*}}{\epsilon\beta^{*}+1}\dfrac{\bm{Q}^{*}\bm{Q}^{*}}{\bm{Q}^{*2}}\right)\cdot\left(\boldsymbol{\delta}-\zeta\Omega\right)\right]}\Biggr]\Psi^{*}\Biggr\}\\[5pt]
 &+\dfrac{1}{2}\dfrac{\partial}{\partial \bm{Q}^{*}}\dfrac{\partial}{\partial \bm{Q}^{*}}:\left[\left(\boldsymbol{\delta}-\dfrac{\epsilon\beta^{*}}{\epsilon\beta^{*}+1}\dfrac{\bm{Q}^{*}\bm{Q}^{*}}{\bm{Q}^{*2}}\right)\cdot\left(\boldsymbol{\delta}-\zeta\Omega\right)\Psi^{*}\right]
 \end{split}
 \end{equation}
 which is identical to the Fokker-Planck equation derived in Ref.~\citenum{Kailasham2018}, where the underlined divergence term is evaluated to be of the form $g_2\left(\bm{Q}^{*}/Q^{*}\right)$, where the scalar $g_2$ is a function of $\{\alpha,\epsilon,\mathscr{A},\mathscr{B},Q^{*}\}$ and its complete expression is given in Eqs. (A14) - (A18) of Appendix A in Ref.~\citenum{Kailasham2018}.
 We have therefore established that upon setting $N=1$, the general Fokker-Planck equation for bead-spring-chains with IV and HI derived in the present work reduces to the dumbbell expression available in the literature~\cite{Kailasham2018}. 

 \section{\label{sec:fd_compare_approach} Comparison against prior work on free-draining bead-spring dashpot chains}
 
The present work provides an exact solution to the bead-spring-dashpot chain model with fluctuating IV and HI, which has thus far remained unsolved for the general case of $N>1$. A comparison with prior art is therefore possible only in the free-draining limit, for which the solution was presented recently~\cite{kailasham2021rouse}. The dimensionless governing Fokker-Planck equation for free-draining bead-spring-dashpot chains derived in Ref.~\citenum{kailasham2021rouse} may be recast as
 \begin{align}\label{eq:fp_ddot_prior_work}
 \dfrac{\partial \Psi^{*}}{\partial t^{*}}&=-\sum_{j=1}^{N}\dfrac{\partial}{\partial \bm{Q}^{*}_{j}}\cdot\Biggl\{\Biggl[\sum_{k=1}^{N}\widehat{\bm{G}}_{jk}\cdot\left(\boldsymbol{\kappa}^{*}\cdot\bm{Q}^{*}_{k}\right)-\dfrac{1}{4}\sum_{k=1}^{N}\widehat{\bm{A}}_{jk}\cdot\left(\dfrac{\partial \phi^{*}}{\partial \bm{Q}^{*}_{k}}\right)+\dfrac{1}{4}\sum_{k=1}^{N}\dfrac{\partial}{\partial \bm{Q}^{*}_{k}}\cdot\widehat{\bm{A}}_{jk}^{\,T}\Biggr]\Psi^{*}\Biggr\}\\ \nonumber
 &+\dfrac{1}{4}\sum_{j,k=1}^{N}\dfrac{\partial}{\partial \bm{Q}^{*}_{j}}\dfrac{\partial}{\partial \bm{Q}^{*}_{k}}:\left[\widehat{\bm{A}}_{jk}^{\,T}\Psi^{*}\right]
 \end{align}
 where
 \begin{equation}
 \begin{split}
 \widehat{\bm{G}}_{jk}&=\delta_{jk}\boldsymbol{\delta}-\left(\dfrac{\varphi}{2\varphi+1}\right)\boldsymbol{U}_{jk}\\[5pt]
 \widehat{\boldsymbol{A}}_{jk}&=\boldsymbol{A}_{jk}-\left(\dfrac{\varphi}{2\varphi+1}\right)\boldsymbol{V}_{jk}
 \end{split}
 \end{equation}
 with the definitions of the quantities ($\bm{U}_{jk}$ and $\bm{V}_{jk}$) obtained using the decoupling methodology appearing in the Supplementary Material to Ref.~\citenum{kailasham2021rouse}. We define the block matrices $\bm{\mathcal{G}}$ and $\bm{\mathcal{A}}$ whose elements are $\widehat{\bm{G}}_{jk}$ and $\widehat{\boldsymbol{A}}_{jk}$, respectively.
 
\begin{figure}[t]
\begin{tabular}{c c}
\includegraphics[width=80mm]{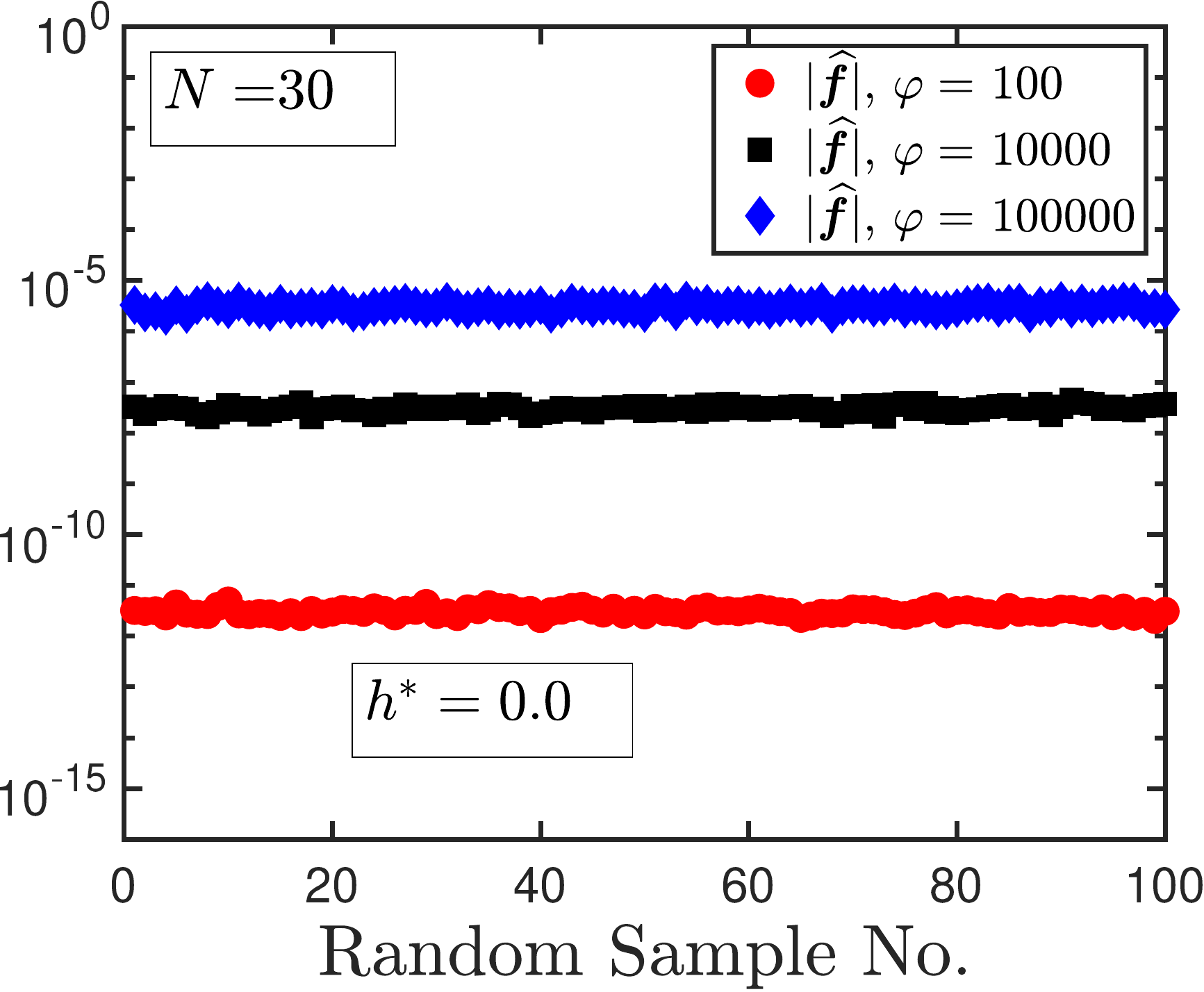}&
\includegraphics[width=80mm]{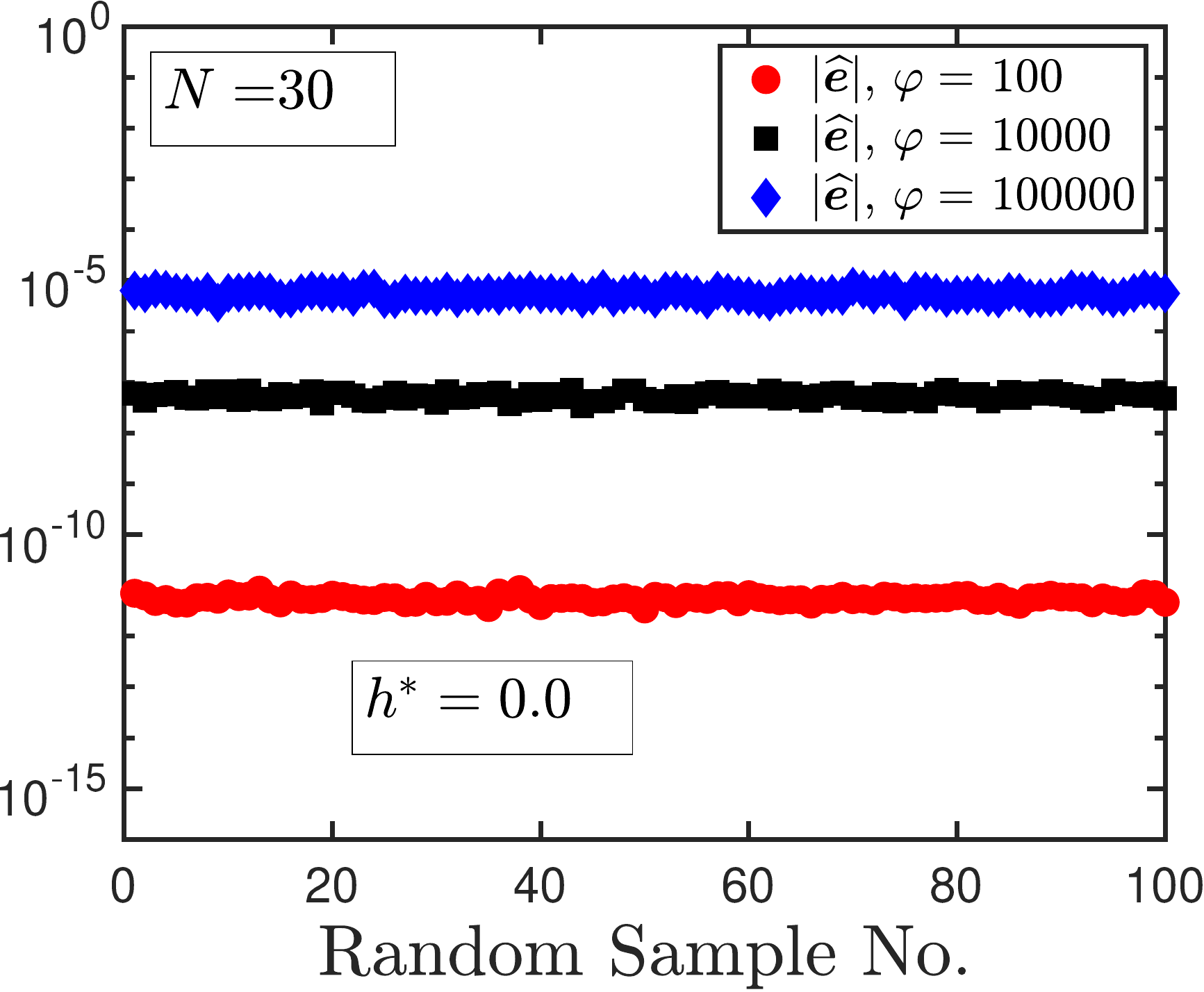} \\
 (a) &  (b)\\[5pt]
\end{tabular}
\caption{Difference between the coefficient matrices computed by the methodology described in the present work and that calculated using the algorithm given in Ref.~\citenum{kailasham2021rouse}. Part (a) represents the difference denoted by Eq.~(\ref{eq:f_hat_def}), and (b) corresponds to Eq.~(\ref{eq:e_hat_def})}
\label{fig:diff_approach}
\end{figure}

Equation~(\ref{eq:fp_ddot_prior_work}) is structurally identical to Eq.~(\ref{eq:fp_ddot_dimless}), and the differences between the coefficient matrices are quantified by evaluating
\begin{equation}\label{eq:f_hat_def}
||\bm{\mathcal{G}}-\bm{\mathcal{M}}||=|\bm{\widehat{f}}|\\[5pt]
\end{equation}
\begin{equation}\label{eq:e_hat_def}
||\bm{\mathcal{A}}-\bm{\mathcal{L}}||=|\bm{\widehat{e}}|
\end{equation}
where $|\cdots|$ denotes the 2-norm of an array of $9N^2$ elements, defined as follows
\begin{equation}\label{eq:norm_def}
\begin{split}
|\bm{\widehat{d}}|=\sqrt{\left(\widehat{d}_{1}\right)^2+\left(\widehat{d}_{2}\right)^2+\dots+\left(\widehat{d}_{9N^2}\right)^2}
\end{split}
\end{equation}
The differences computed in this manner are plotted in Fig.~\ref{fig:diff_approach}, for a thirty-spring chain at different values for the internal friction parameter. It is seen that both $|\bm{\widehat{f}}|$ and $|\bm{\widehat{e}}|$ increase with the internal friction parameter, and their magnitudes are nearly identical. The trend remains unaltered (not shown in plot) when a twenty-spring chain is considered. 

\begin{figure}[h]
\centering
\includegraphics[width=100mm]{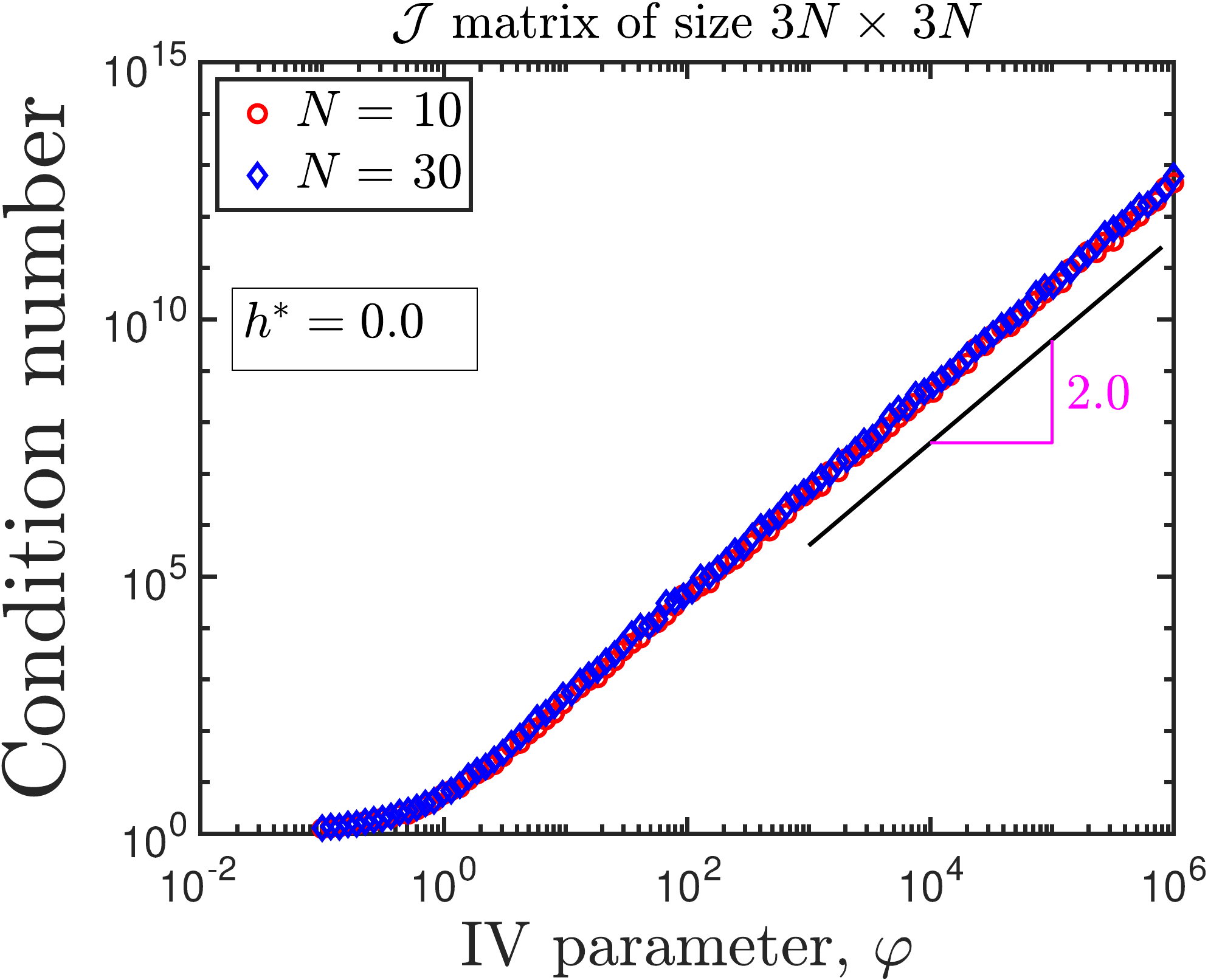}  
\caption{Condition number of the $N\times\,N$ block matrix $\bm{\mathcal{J}}$, whose elements $\bm{J}_{jk}$ are the $3\times3$ matrices given by Eq.~(23) of the main text, as a function of the internal friction parameter, $\varphi$, for two different values of the number of springs in the chain ($N$).}
\label{fig:cond_num_scaling}
\end{figure}

\begin{figure}
\begin{tabular}{c c}
\includegraphics[width=80mm]{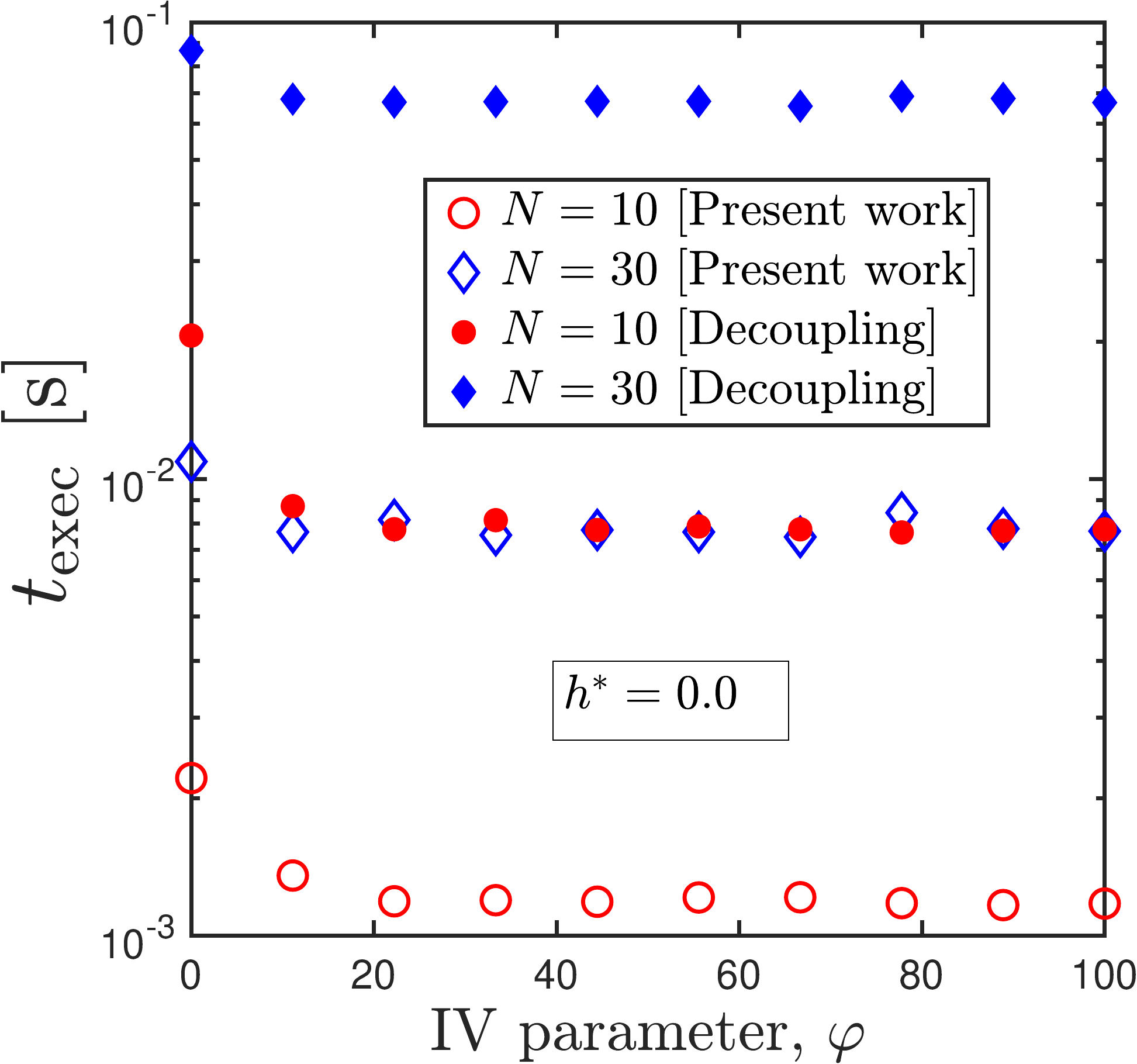}&
\includegraphics[width=78mm]{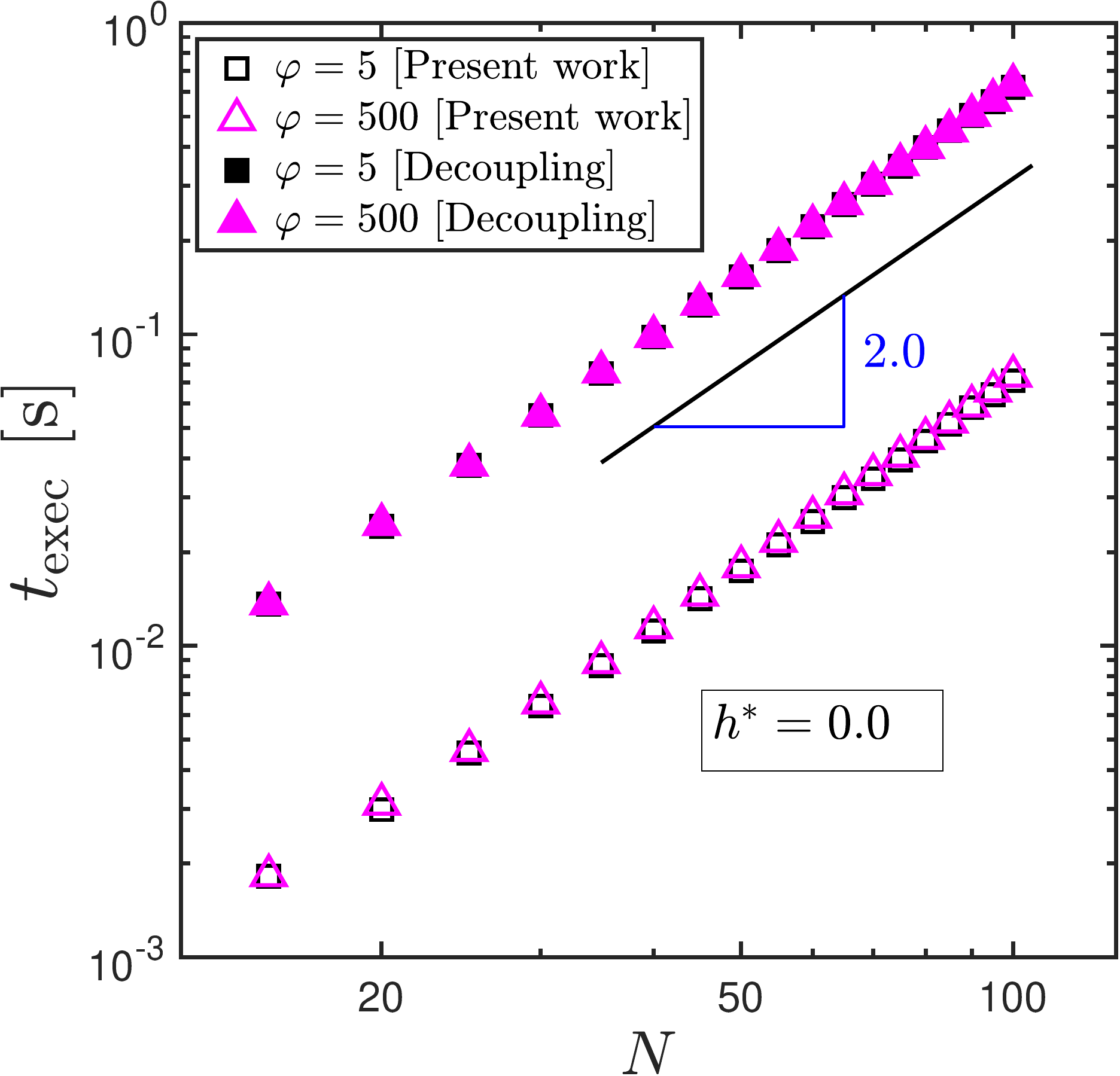} \\
 (a) &  (b)\\[5pt]
\end{tabular}
\caption{A comparison of the time required for the construction of the diffusion matrix, $\bm{\mathcal{L}}$ using two approaches, plotted against (a) the internal friction parameter, and (b) the number of springs in the chain. The execution time reported corresponds to the average time needed for the construction of $\bm{\mathcal{L}}$, for a hundred different randomly chosen initial configurations of the chain.}
\label{fig:timing_compare}
\end{figure}

\begin{figure}
\includegraphics[width=100mm]{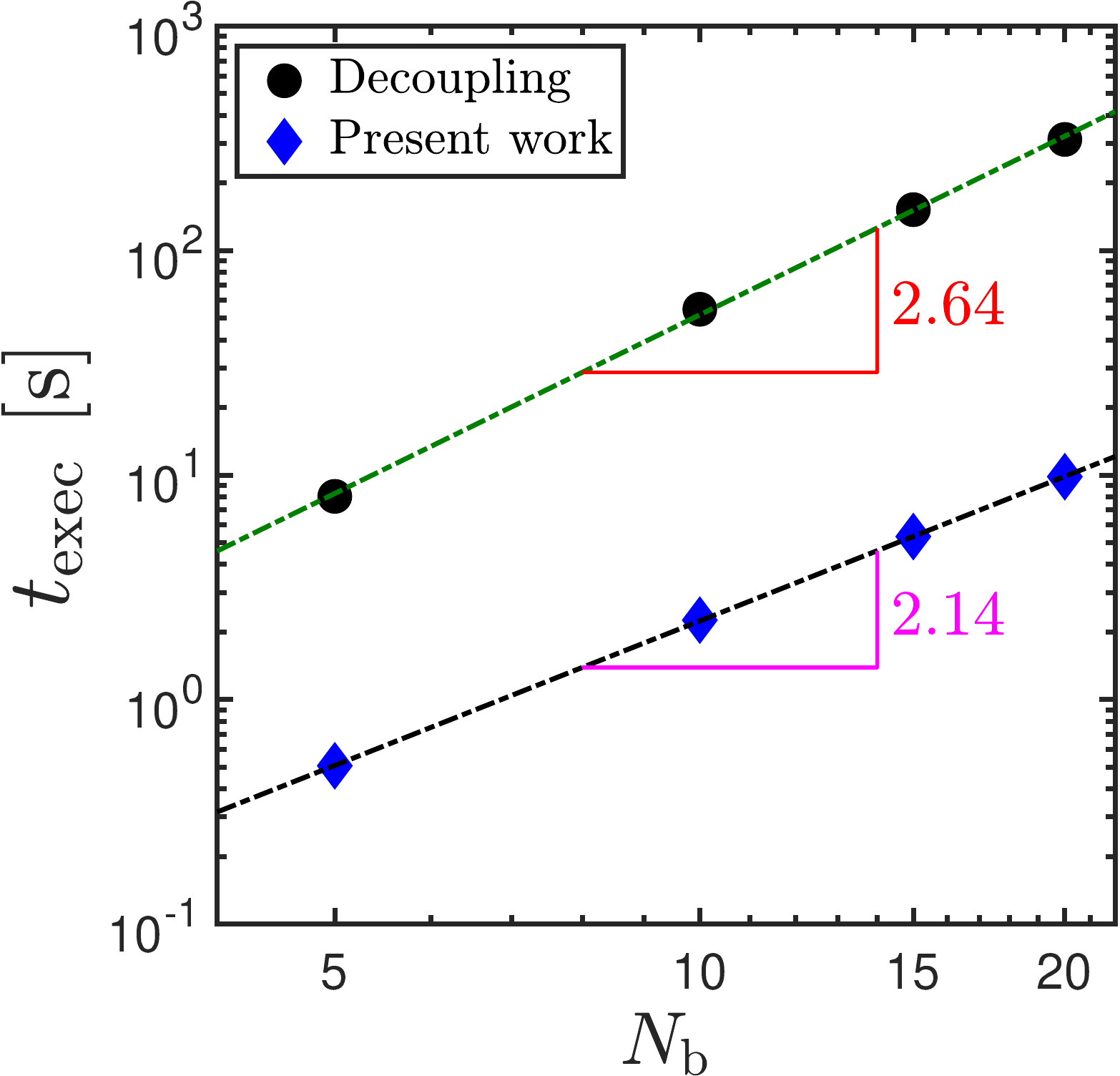}
\caption{A comparison of the execution time as a function of the number of beads, for equilibrium simulations of free-draining bead-spring-dashpot chains ($h^{*}=0.0$) with an internal friction parameter of $\varphi=5.0$.}
\label{fig:scale_compare}
\end{figure}

From Eqs.~(24) and (25) of the main text, it is clear that the construction of both $\bm{\mathcal{M}}$ and $\bm{\mathcal{L}}$ relies on the numerical computation of the inverse of $\bm{\mathcal{J}}$. In Fig.~\ref{fig:cond_num_scaling}, the condition number of $\bm{\mathcal{J}}$ is plotted as a function of the internal friction parameter, for two different values of the number of springs in the chain. It is observed that the condition number scales quadratically with the IV parameter for $\varphi>1$, and is practically insensitive to the chain length. We conclude, therefore, that the numerical error~\cite{press2007numerical} associated with the inverse calculation of a matrix whose condition number scales with the IV parameter, is responsible for the perceived increase in the difference between the matrix coefficients calculated in the present work and those evaluated in Ref.~\citenum{kailasham2021rouse} which does not rely on the numerical computation of matrix inverses. 

In Fig.~\ref{fig:timing_compare}, the CPU time cost for the construction of the diffusion matrix, $\bm{\mathcal{L}}$, using the two approaches mentioned above is examined as a function of the internal friction parameter and the chain length. From Fig.~\ref{fig:timing_compare}~(a), it is observed that the execution time is practically unaffected by the IV parameter for two different values of the chain length considered. The time required by the algorithm developed in the present work, though, is about an order of magnitude smaller than that needed by the methodology developed in Ref.~\citenum{kailasham2021rouse}. In Fig.~\ref{fig:timing_compare}~(b), the scaling of the execution time is plotted as a function of the chain length for two values of the internal friction parameter. While the computational cost for the two methods scale approximately as the square of the chain length, the method developed in the present work is about an order-of-magnitude faster than the decoupling-based approach used in Ref.~\citenum{kailasham2021rouse}.

A comparison of the computational cost of numerically integrating the governing stochastic differential equations for free-draining bead-spring-dashpots using the decoupling methodology~\cite{kailasham2021rouse}, and the algorithm described in the present work is presented in Fig.~\ref{fig:scale_compare}. There is a $\sqrt{N}$ difference in the scaling of the execution time between the two methodologies, with $t_{\text{exec}}\sim N^{2.64}_{\text{b}}$ for the decoupling approach, and $t_{\text{exec}}\sim N^{2.14}_{\text{b}}$ for the present work. Additionally, the present method is an order-of-magnitude faster.
 

 \section{\label{sec:fdt_sat} Satisfaction of fluctuation-dissipation theorem}

Internal friction and hydrodynamic interactions represent the dashpot- and solvent-mediated transfer of momentum between the beads in the chain, and do not feature in the Hamiltonian, $\mathcal{H}$, of the system, which contains contributions only from conservative intramolecular interactions. Consequently, it is expected that IV and HI do not affect the configurational distribution function of a chain at equilibrium, $\Psi_{\text{eq}}$, since $\Psi_{\text{eq}}\propto\exp[-\mathcal{H}/k_BT]$.  The fluctuation-dissipation theorem (FDT) prescribes a relation between the fluctuations in the chain configuration and the various sources of friction (dissipation) experienced by the chain. The FDT is imposed in our model formulation by requiring that $\bm{\mathcal{B}}\cdot\bm{\mathcal{B}}^{T}=\bm{\mathcal{D}}$ in the governing stochastic differential equation (see eq.~(36)) in the main text. 

 \begin{figure}[h]
\centering
\includegraphics[width=100mm]{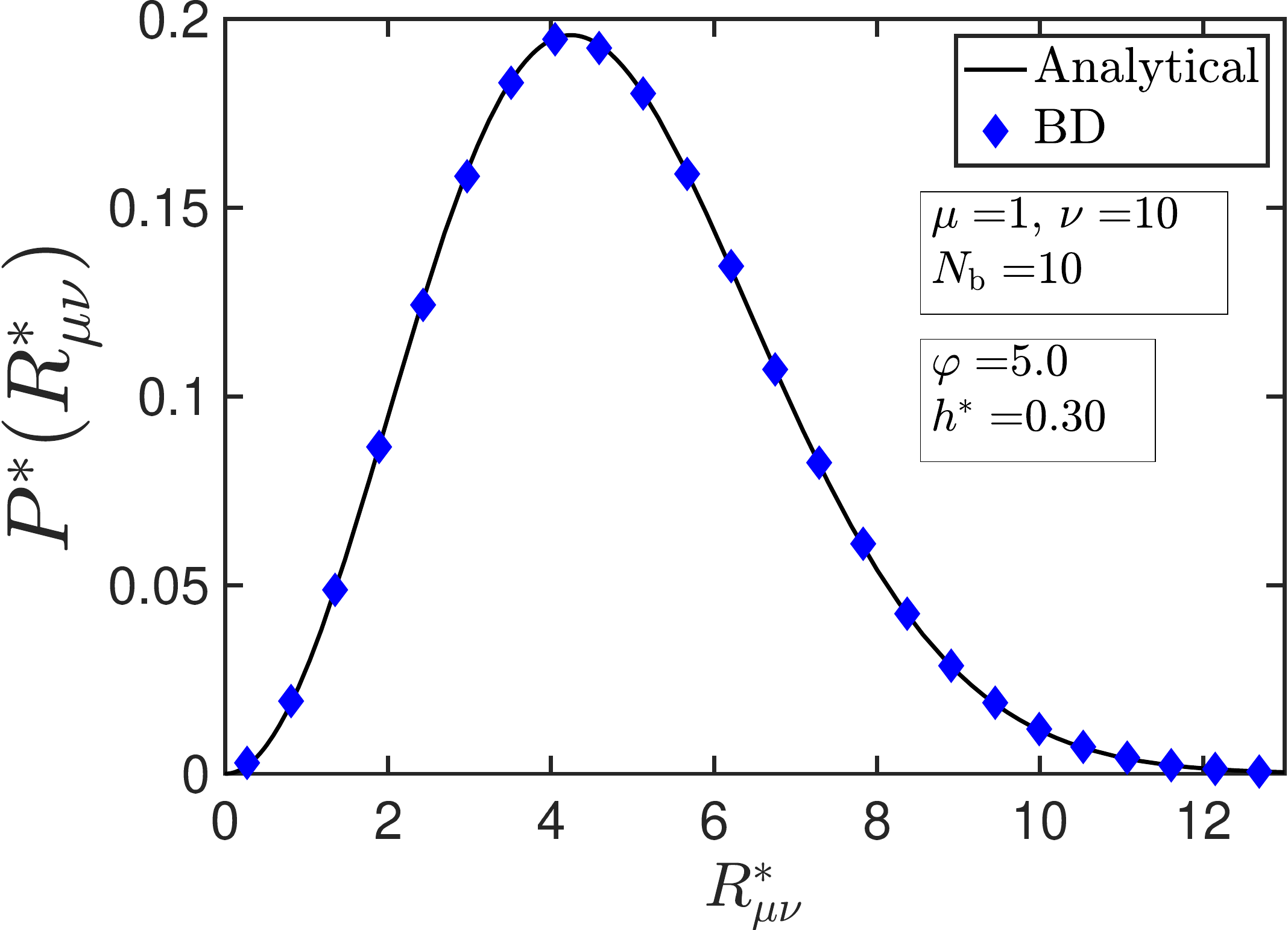}  
\caption{Probability distribution of the lengths of the end-to-end vector of a ten-bead chain with internal friction and hydrodynamic interactions. Symbols are BD simulation results, and the line represents the analytical result given by eq.~(\ref{eq:prob_dist_eqn}).}
\label{fig:fdt_fig}
\end{figure}

The probability distribution of the length of the vector running from bead $\mu$ to $\nu$ in a Rouse chain, $R^{*}_{\mu\nu}=|\bm{r}^{*}_{\nu}-\bm{r}^{*}_{\mu}|$, is known analytically~\cite{Bird1987b}, and may be written as follows
\begin{equation}\label{eq:prob_dist_eqn}
P^{*}\left(R^{*}_{\mu\nu}\right)=4\pi\,R^{*2}_{\mu\nu}\left[\dfrac{1}{2\pi|\nu-\mu|}\right]^{3/2}\exp\left[-\dfrac{R^{*2}_{\mu\nu}}{2|\nu-\mu|}\right]
\end{equation}

In Fig.~\ref{fig:fdt_fig}, the numerically computed probability distribution of the end-to-end vector of a ten-bead Rouse chain with internal friction and hydrodynamic interactions, evaluated by binning the output from $\mathcal{O}(5\times10^5)$ trajectories at the end of an equilibrium run of $t^{*}_{\text{max}}=10$ dimensionless times, is compared against the analytical result given by eq.~(\ref{eq:prob_dist_eqn}). The excellent agreement between the two results verifies that the fluctuation-dissipation theorem is indeed satisfied.

 \section{\label{sec:tstep_convergence} Establishing timestep convergence}
 
The calculation of shear viscosity at dimensionless shear rates $\lambda_{H}\dot{\gamma}<0.1$ is performed using the variance reduction algorithm~\cite{Wagner1997}, while that for $\lambda_{H}\dot{\gamma}\geq0.1$ does not use variance reduction. The choice of timestep ($\Delta t^{*}$) for integration is dictated by the value of the internal friction parameter, and the dimensionless shear rate. 

\begin{figure}[t]
\begin{tabular}{c c}
\includegraphics[width=80mm]{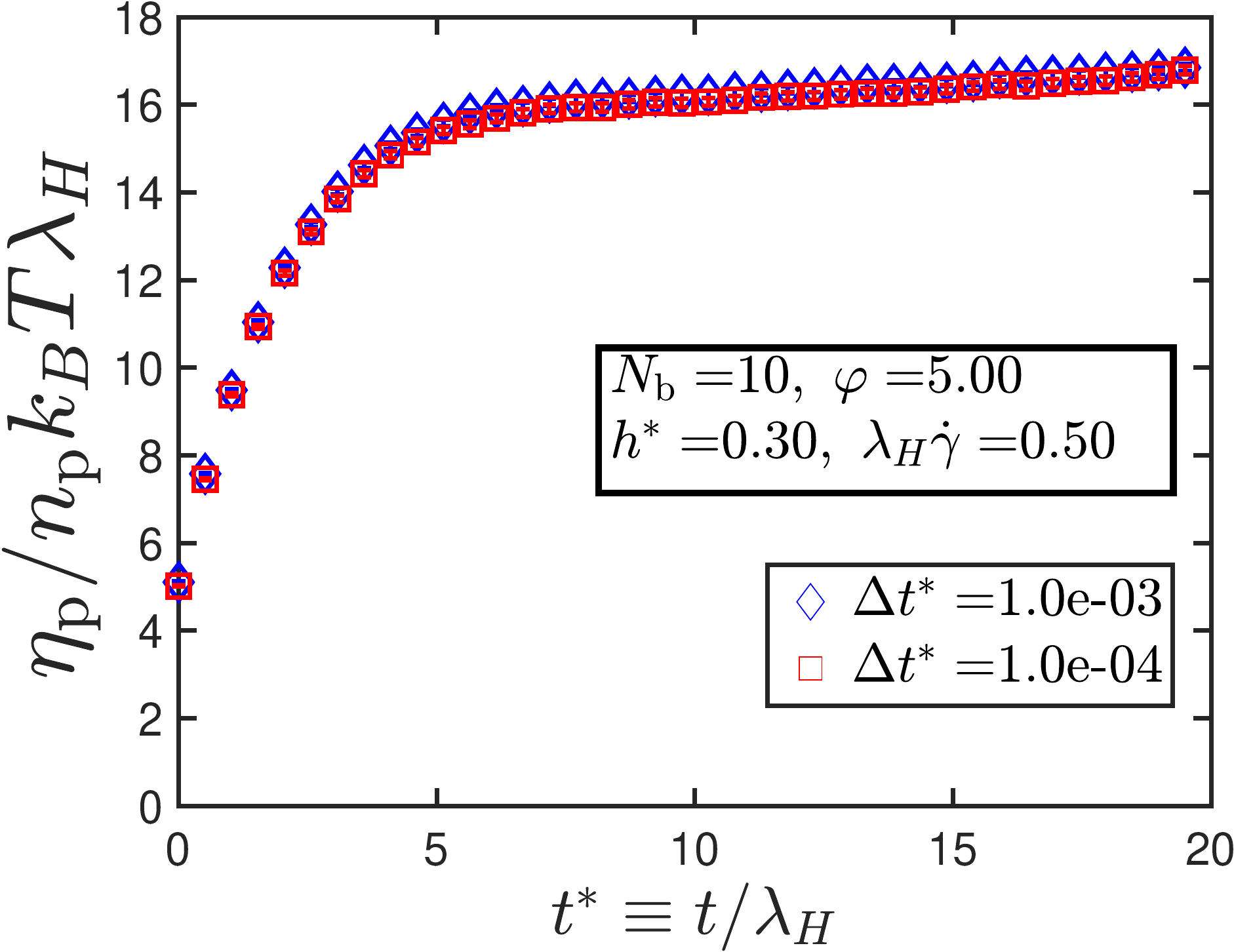}&
\includegraphics[width=80mm]{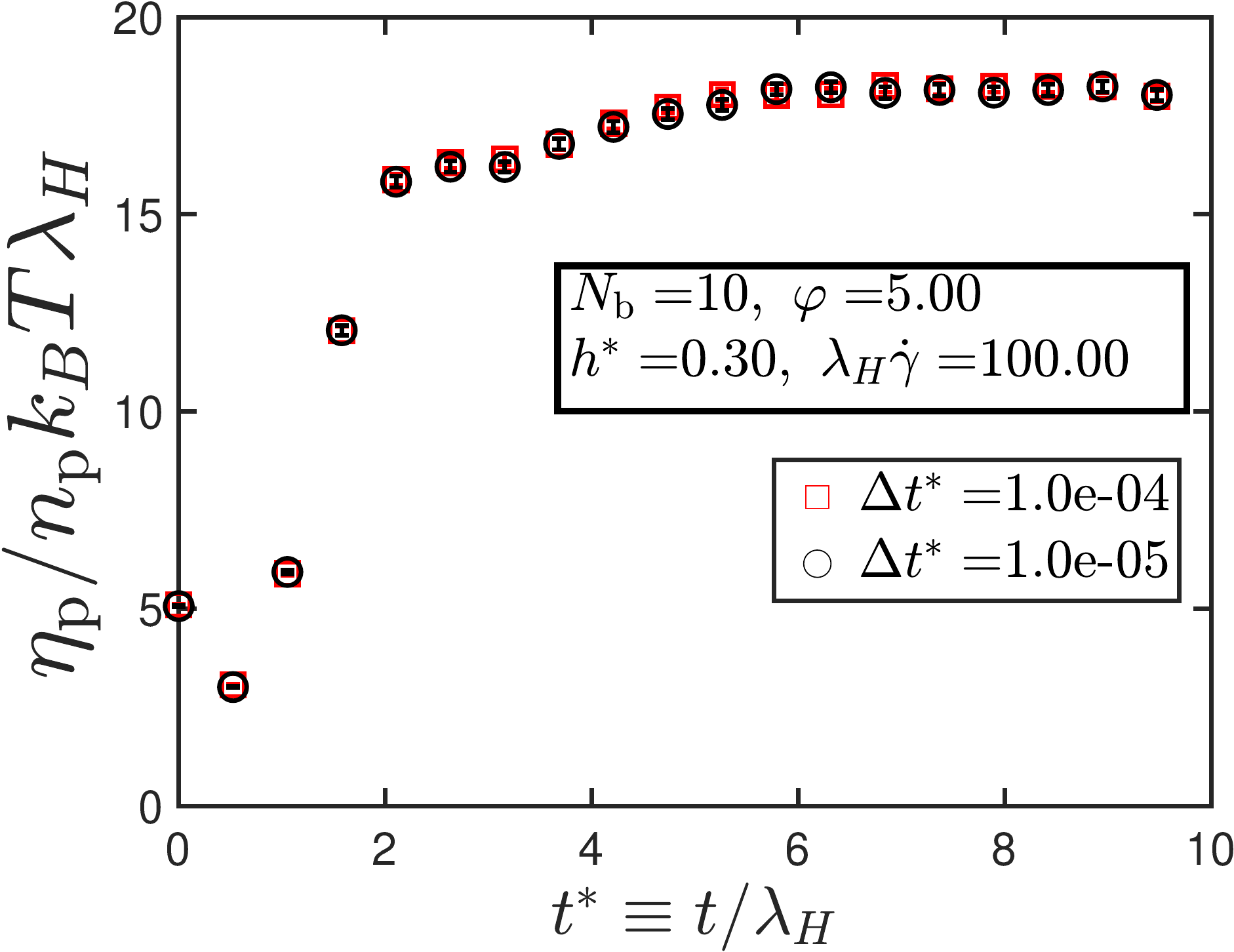} \\
 (a) &  (b)\\
\end{tabular}
\caption{Transient evolution of shear viscosity computed at two different timestep widths for a ten-bead chain with FENE springs of $b=100$ and fluctuating internal friction and hydrodynamic interactions, at (a) dimensionless shear rates of $\lambda_{H}\dot{\gamma}=0.5$, using variance reduction; and (b) $\lambda_{H}\dot{\gamma}=100.0$.}
\label{fig:tstep_conv}
\end{figure}

In Fig.~\ref{fig:tstep_conv}, the transient evolution of the shear viscosity at two different shear rates, computed using timestep widths that differ by a factor of ten, is plotted for the highest values of the internal friction and hydrodynamic interaction parameters considered in this work ($\varphi=5.0, h^{*}=0.3$). Given the agreement between the results at the different timesteps, we consider our numerical computations to be timestep convergent.
 
 \section{\label{sec:stress_tens_derv} Derivation of stress tensor expression}
 
The first step in the derivation of a usable stress tensor expression for the model under investigation is the conversion of the Kramers-Kirkwood expression, 
 \begin{equation}\label{eq:kram_kirk_textbook}
\boldsymbol{\tau}_{\text{p}}=-{n_{\text{p}}}\sum_{\nu=1}^{N_{\text{b}}}\left<\bm{R}_{\nu}\bm{F}_{\nu}^{(h)}\right>,
\end{equation}
given in terms of the bead position vectors into its equivalent form using the connector vector notation. It is noted that $\bm{R}_{\nu}=\bm{r}_{\nu}-\bm{r}_{\text{c}}$ is the position of the $\nu^{\text{th}}$ bead with respect to the centre of mass of the chain, and $\bm{F}_{\nu}^{(h)}$ is the hydrodynamic drag force on the $\nu^{\text{th}}$ bead.
From Ref.~\citenum{ravibook}, we have
\begin{equation}
\bm{F}_{\nu}^{(h)}=k_BT\dfrac{\partial \ln\psi}{\partial \bm{r}_{v}}-\bm{F}_{\nu}^{(\phi)}-\bm{F}^{(\text{IV})}_{\nu}
\end{equation}
The configurational distribution function written in terms of the bead positions, $\psi\left(\bm{r}_{1},\bm{r}_{2},..\bm{r}_{N_{\text{b}}}\right)$ is equivalent to that written in terms of the internal coordinates $\Psi\left(\bm{Q}_{1},\bm{Q}_{2},...,\bm{Q}_{N}\right)$ under conditions of homogeneous flow, where the distribution does not depend explicitly on the coordinates of the centre of mass. We may then write, using entry (F) of Table 15.4-1 in~\citet{Bird1987b}, 
\begin{equation}\label{eq:lnpsi_rel}
\dfrac{\partial \ln\psi}{\partial \bm{r}_{v}}=\sum_{k=1}^{N}\bar{B}_{k\nu}\dfrac{\partial \ln\Psi}{\partial \bm{Q}_{k}}
\end{equation}
with $\bar{B}_{k\nu}=\delta_{k+1,\nu}-\delta_{k\nu}$, $N=N_{\text{b}}-1$. We de not account for excluded volume (EV) interactions in our model, and therefore, the total conservative force on a bead consists only of contributions from the springs, i.e., $\bm{F}_{\nu}^{(\phi)}=\bm{F}_{\nu}^{(\text{S})}$. We may then write
\begin{equation}\label{eq:force_rel}
\bm{F}_{\nu}^{(\text{S})}=-\sum_{k=1}^{N}\bar{B}_{k\nu}\bm{F}_{k}^{(\text{c})}
\end{equation}
where
\begin{equation}
\bm{F}_{k}^{(\text{c})}=Hf\left(Q_{k}\right)\bm{Q}_{k}
\end{equation}
with
\begin{equation}\label{eq:f_fac_def}
f\left(Q_{k}\right)= \left\{
    \begin{array}{l}
      1\,\,\qquad\qquad\qquad\text{for Hookean springs,}\\[2pt]
      \dfrac{1}{1-\left(Q_k/Q_0\right)^2}\quad\text{for FENE springs}.
    \end{array} \right.
\end{equation}
Similarly, we have
\begin{equation}\label{eq:ivforce_rel}
\bm{F}^{(\text{IV})}_{\nu}=-\sum_{k=1}^{N}\bar{B}_{k\nu}\bm{F}_{k}^{(\text{IV})}
\end{equation}
with
\begin{equation}\label{eq:def_iv_force}
\bm{F}_{k}^{(\text{IV})}=K\dfrac{\bm{Q}_k\bm{Q}_k}{\bm{Q}^2_k}\cdot\llbracket\dot{\bm{Q}}_{k}\rrbracket
\end{equation}
Using eqs.~(\ref{eq:lnpsi_rel}),~(\ref{eq:force_rel}), and~(\ref{eq:ivforce_rel}), $\bm{F}_{\nu}^{(h)}$ may be rewritten as
\begin{equation}\label{eq:hyd_q}
\bm{F}_{\nu}^{(h)}=\sum_{k=1}^{N}\bar{B}_{k\nu}\left[k_BT\dfrac{\partial \ln\Psi}{\partial \bm{Q}_{k}}+\bm{F}_{k}^{(\text{c})}+\bm{F}_{k}^{(\text{IV})}\right]
\end{equation}
From Eq. (11.6-4) in~\citet{Bird1987b}, 
\begin{equation}
\bm{R}_{\nu}\equiv\bm{r}_{\nu}-\bm{r}_{\text{c}}=\sum_{j=1}^{N}B_{\nu j}\bm{Q}_{j}
\end{equation}
with $B_{\nu j}=(j/N) - \Theta(j-\nu)$ with $\Theta(j-\nu)$ denoting a Heaviside step function.
Thus we have
\begin{equation}
\bm{R}_{\nu}\bm{F}_{\nu}^{(h)}=\sum_{k=1}^{N}\sum_{j=1}^{N}B_{\nu j}\bm{Q}_{j}\bar{B}_{k\nu}\left[k_BT\dfrac{\partial \ln\Psi}{\partial \bm{Q}_{k}}+\bm{F}_{k}^{(\text{c})}+\bm{F}_{k}^{(\text{IV})}\right]
\end{equation}
and
\begin{equation}\label{eq:intermed_rf}
\sum_{\nu=1}^{N_{\text{b}}}\left<\bm{R}_{\nu}\bm{F}_{\nu}^{(h)}\right>=\sum_{\nu=1}^{N_{\text{b}}}\sum_{k=1}^{N}\sum_{j=1}^{N}\left(\bar{B}_{k\nu}B_{\nu j}\right)\left<\bm{Q}_{j}\left[k_BT\dfrac{\partial \ln\Psi}{\partial \bm{Q}_{k}}+\bm{F}_{k}^{(\text{c})}+\bm{F}_{k}^{(\text{IV})}\right]\right>
\end{equation}
Using 
\begin{equation}
\sum^{N_{\text{b}}}_{\nu=1}\bar{B}_{k\nu}B_{\nu j}=\delta_{kj}
\end{equation}
and combining eq.~(\ref{eq:intermed_rf}) with eq.~\ref{eq:kram_kirk_textbook}, the Kramers-Kirkwood expression is rewritten as
\begin{equation}\label{eq:con_vec_kram_kirk}
\boldsymbol{\tau}_{\text{p}}=-{n_{\text{p}}}\sum_{k=1}^{N_{\text{b}}-1}\left<\bm{Q}_{k}\left[k_BT\,\dfrac{\partial \ln \Psi}{\partial\bm{Q}_{k}}+\bm{F}^{(\text{c})}_{k}+\bm{F}_{k}^{(\text{IV})}\right]\right>
\end{equation}
which is equivalent to eq.~(42) in the main text. The decoupled expression for $\llbracket\dot{\bm{Q}}_{k}\rrbracket$ has been derived in eq.~(27) of the main text, and is reproduced below
\begin{equation}\label{eq:qdot_indiv_decoupled}
\llbracket\dot{\bm{Q}}_k\rrbracket=\sum_{j=1}^{N}\bm{M}_{kj}\cdot\left(\boldsymbol{\kappa}\cdot\bm{Q}_{j}\right)-\dfrac{2k_BT}{\zeta}\sum_{j=1}^{N}\bm{D}_{kj}\cdot\left(\dfrac{\partial \ln \Psi}{\partial \bm{Q}_j}\right)-\dfrac{2}{\zeta}\sum_{j=1}^{N}\bm{D}_{kj}\cdot\left(\dfrac{\partial \phi}{\partial \bm{Q}_{j}}\right)
\end{equation}
Combining eq.~(\ref{eq:qdot_indiv_decoupled}) with eq.~(\ref{eq:def_iv_force}) and simplifying, eq.~(\ref{eq:intermed_rf}) becomes
\begin{equation}\label{eq:kk_expanded}
\begin{split}
&\sum_{\nu=1}^{N_{\text{b}}}\left<\bm{R}_{\nu}\bm{F}_{\nu}^{(h)}\right>=k_BT\sum_{k=1}^{N}\left<\bm{Q}_{k}\dfrac{\partial \ln \Psi}{\partial\bm{Q}_{k}}\right>+\sum_{k=1}^{N}\left<\bm{Q}_{k}\bm{F}^{(\text{c})}_{k}\right>\\[5pt]
&+K\sum_{k,j=1}^{N}\left<\left(\dfrac{\bm{Q}_{k}\bm{Q}_{k}\bm{Q}_{k}}{Q^2_{k}}\right)\cdot\bm{M}_{kj}\cdot\left(\boldsymbol{\kappa}\cdot\bm{Q}_{j}\right)\right>-\dfrac{2k_BTK}{\zeta}\sum_{k,j=1}^{N}\left<\left(\dfrac{\bm{Q}_{k}\bm{Q}_{k}\bm{Q}_{k}}{Q^2_{k}}\right)\cdot\bm{D}_{kj}\cdot\left(\dfrac{\partial \ln\Psi}{\partial \bm{Q}_{j}}\right)\right>\\[5pt]
&-\dfrac{2K}{\zeta}\sum_{k,j=1}^{N}\left<\left(\dfrac{\bm{Q}_{k}\bm{Q}_{k}\bm{Q}_{k}}{Q^2_{k}}\right)\cdot\bm{D}_{kj}\cdot\bm{F}^{(\text{c})}_{j}\right>
\end{split}
\end{equation}
The terms on the RHS of eq.~(\ref{eq:kk_expanded}) are simplified as shown below. Firstly,
\begin{equation}\label{eq:brown2}
\begin{split}
\left<\bm{Q}_{k}\dfrac{\partial \ln \Psi}{\partial\bm{Q}_{k}}\right>&=\left[\sum_{\mu,\omega}\int\left(\dfrac{\partial \Psi}{\partial Q_k^{\mu}}\right)Q^{\omega}_{k}\,d\bm{\mathcal{Q}}\right]\bm{e}_{\mu}\bm{e}_{\omega}
\end{split}
\end{equation}
where $d\bm{\mathcal{Q}}\equiv\,d\boldsymbol{Q}_1d\boldsymbol{Q}_2\cdots\,d\boldsymbol{Q}_{N}$ and $\bm{e}_{\mu}$ and $\bm{e}_{\omega}$ denote the unit vectors in Cartesian coordinates. Then,
\begin{equation}\label{eq:brown3}
\begin{split}
\int\left(\dfrac{\partial \psi}{\partial Q_k^{\mu}}\right)Q^{\omega}_{k}\,d\bm{\mathcal{Q}}&=\int\left[\dfrac{\partial}{\partial Q^{\mu}_{k}}\left(Q^{\omega}_{k}\Psi\right)-\psi\dfrac{\partial Q^{\omega}_{k}}{\partial Q^{\mu}_k}\right]d\bm{\mathcal{Q}}=\dashuline{\int\dfrac{\partial}{\partial Q^{\mu}_{k}}\left(Q^{\omega}_{k}\Psi\right)d\bm{\mathcal{Q}}}-\delta^{\mu\omega}\int\Psi\,d\bm{\mathcal{Q}}
\end{split}
\end{equation}
The underlined term in Eq.~(\ref{eq:brown3}) vanishes due to the Gauss divergence theorem, and the integral in the second term is unity due to the normalization condition, and we thus have
\begin{equation}\label{eq:brown4}
\begin{split}
\left<\bm{Q}_{k}\dfrac{\partial \ln \Psi}{\partial\bm{Q}_{k}}\right>=-\boldsymbol{\delta}
\end{split}
\end{equation} 
We may similarly write
\begin{equation}\label{eq:diff_integr}
\left<\left(\dfrac{\bm{Q}_{k}\bm{Q}_{k}\bm{Q}_{k}}{Q^2_{k}}\right)\cdot\bm{D}_{kj}\cdot\left(\dfrac{\partial \ln\Psi}{\partial \bm{Q}_{j}}\right)\right>=\left[\sum_{\mu,\omega}\sum_{m,n}\int\dfrac{Q^{\mu}_kQ^{\omega}_kQ^{m}_k}{Q^2_k}D^{mn}_{kj}\dfrac{\partial \Psi}{\partial Q^{n}_{j}}d\bm{\mathcal{Q}}\right]\bm{e}_{\mu}\bm{e}_{\omega}
\end{equation}
and
\begin{equation}\label{eq:split_3term}
\begin{split}
\dfrac{Q^{\mu}_kQ^{\omega}_kQ^{m}_k}{Q^2_k}D^{mn}_{kj}\dfrac{\partial \Psi}{\partial Q^{n}_{j}}&=\dfrac{\partial}{\partial Q^{n}_{j}}\left[\left(\dfrac{Q^{\mu}_kQ^{\omega}_kQ^{m}_k}{Q^2_k}\right)D^{mn}_{kj}\Psi\right]-\Psi D^{mn}_{kj}\dfrac{\partial}{\partial Q^{n}_{j}}\left[\dfrac{Q^{\mu}_kQ^{\omega}_kQ^{m}_k}{Q^2_k}\right]\\[5pt]
&-\Psi\left(\dfrac{Q^{\mu}_kQ^{\omega}_kQ^{m}_k}{Q^2_k}\right)\dfrac{\partial}{\partial Q^{n}_{j}}D^{mn}_{kj}
\end{split}
\end{equation}
The first term on the RHS of eq.~(\ref{eq:split_3term}) would vanish upon integration due to the Gauss divergence theorem, and is not treated further. The second term, upon simplification, may be written as
\begin{equation}\label{eq:sec_term}
\begin{split}
\Psi D^{mn}_{kj}\dfrac{\partial}{\partial Q^{n}_{j}}\left[\dfrac{Q^{\mu}_kQ^{\omega}_kQ^{m}_k}{Q^2_k}\right]=\dfrac{\delta_{jk}\Psi}{Q^4_k}\Biggl[&Q^2_{k}\left(D^{mn}_{kj}Q^{\mu}_kQ^{\omega}_k\delta^{mn}+D^{mn}_{kj}Q^{\mu}_kQ^{m}_k\delta^{n\omega}+D^{mn}_{kj}Q^{\omega}_kQ^{m}_{k}\delta^{n\mu}\right)\\[5pt]
&-2D^{mn}_{kj}Q^{\mu}_kQ^{\omega}_kQ^{m}_kQ^{n}_k\Biggr]
\end{split}
\end{equation}
Noting that
\begin{equation}\label{eq:vec_transl}
\begin{split}
\sum_{\mu,\omega}\sum_{m,n}D^{mn}_{kj}Q^{\mu}_kQ^{\omega}_k\delta^{mn}\bm{e}_{\mu}\bm{e}_{\omega}&=\text{tr}\left(\bm{D}_{kj}\right)\bm{Q}_{k}\bm{Q}_{k}\\[5pt]
\sum_{\mu,\omega}\sum_{m,n}D^{mn}_{kj}Q^{\mu}_kQ^{m}_k\delta^{n\omega}\bm{e}_{\mu}\bm{e}_{\omega}&=\left(\bm{Q}_{k}\bm{Q}_{k}\right)\cdot\bm{D}_{kj}\\[5pt]
\sum_{\mu,\omega}D^{mn}_{kj}Q^{\omega}_kQ^{m}_{k}\delta^{n\mu}\bm{e}_{\mu}\bm{e}_{\omega}&=\bm{D}^{T}_{kj}\cdot\left(\bm{Q}_{k}\bm{Q}_{k}\right),
\end{split}
\end{equation}
we may write
\begin{equation}\label{eq:part_1_diff}
\begin{split}
&\left[\sum_{\mu,\omega}\sum_{m,n}\int \Psi D^{mn}_{kj}\dfrac{\partial}{\partial Q^{n}_{j}}\left[\dfrac{Q^{\mu}_kQ^{\omega}_kQ^{m}_k}{Q^2_k}\right] d\bm{\mathcal{Q}}\right]\bm{e}_{\mu}\bm{e}_{\omega}=\left<\text{tr}\left(\bm{D}_{kk}\right)\dfrac{\bm{Q}_k\bm{Q}_k}{Q^2_k}\right>+\left<\dfrac{\bm{Q}_k\bm{Q}_k}{Q^2_k}\cdot\bm{D}_{kk}\right>\\[5pt]
&+\left<\bm{D}_{kk}\cdot\dfrac{\bm{Q}_k\bm{Q}_k}{Q^2_k}\right>-2\left<\bm{D}_{kk}:\dfrac{\bm{Q}_{k}\bm{Q}_{k}\bm{Q}_{k}\bm{Q}_{k}}{Q^4_k}\right>
\end{split}
\end{equation}
where we have used the fact that $\bm{D}_{kk}$ is a symmetric matrix, based on the symmetricity of the block diffusion matrix $\bm{\mathcal{D}}$.

Furthermore,
\begin{equation}\label{eq:part_2_diff}
\begin{split}
&\left[\sum_{\mu,\omega}\sum_{m,n}\int \Psi\left(\dfrac{Q^{\mu}_kQ^{\omega}_kQ^{m}_k}{Q^2_k}\right)\dfrac{\partial}{\partial Q^{n}_{j}}D^{mn}_{kj} d\bm{\mathcal{Q}}\right]\bm{e}_{\mu}\bm{e}_{\omega}=\left<\left(\dfrac{\bm{Q}_k\bm{Q}_k}{Q^2_k}\right)\bm{Q}_{k}\cdot\left[\dfrac{\partial}{\partial \bm{Q}_{j}}\cdot\bm{D}^{T}_{kj}\right]\right>
\end{split}
\end{equation}

From eqs.~(\ref{eq:split_3term}),~(\ref{eq:part_1_diff}), and ~(\ref{eq:part_2_diff}), we have
\begin{equation}
\begin{split}
&\left<\left(\dfrac{\bm{Q}_{k}\bm{Q}_{k}\bm{Q}_{k}}{Q^2_{k}}\right)\cdot\bm{D}_{kj}\cdot\left(\dfrac{\partial \ln\Psi}{\partial \bm{Q}_{j}}\right)\right>=-\left<\left[\text{tr}\left(\bm{D}_{kk}\right)-2\bm{D}_{kk}:\dfrac{\bm{Q}_k\bm{Q}_k}{Q^2_k}\right]\dfrac{\bm{Q}_k\bm{Q}_k}{Q^2_k}\right>\\[5pt]
&+\left<\dfrac{\bm{Q}_k\bm{Q}_k}{Q^2_k}\cdot\bm{D}_{kk}\right>+\left<\bm{D}_{kk}\cdot\dfrac{\bm{Q}_k\bm{Q}_k}{Q^2_k}\right>-\left<\left(\dfrac{\bm{Q}_k\bm{Q}_k}{Q^2_k}\right)\bm{Q}_{k}\cdot\left[\dfrac{\partial}{\partial \bm{Q}_{j}}\cdot\bm{D}^{T}_{kj}\right]\right>
\end{split}
\end{equation}

The ensemble average in the third term on the RHS of eq.~(\ref{eq:kk_expanded}) is simplified as follows
\begin{equation}\label{eq:flow_term}
\begin{split}
\left<\left(\dfrac{\bm{Q}_{k}\bm{Q}_{k}\bm{Q}_{k}}{Q^2_{k}}\right)\cdot\bm{M}_{kj}\cdot\left(\boldsymbol{\kappa}\cdot\bm{Q}_{j}\right)\right>&=\left<\left(\bm{M}_{kj}\cdot\boldsymbol{\kappa}\right)^{T}:\left[\dfrac{\bm{Q}_{k}\bm{Q}_{j}\bm{Q}_{k}\bm{Q}_{k}}{Q^2_{k}}\right]\right>
\end{split}
\end{equation}

The ensemble average in the last term on the RHS of eq.~(\ref{eq:kk_expanded}) may be rewritten as follows
\begin{equation}\label{eq:force_term}
\begin{split}
\left<\left(\dfrac{\bm{Q}_{k}\bm{Q}_{k}\bm{Q}_{k}}{Q^2_{k}}\right)\cdot\bm{D}_{kj}\cdot\bm{F}^{(\text{c})}_{j}\right>&=\left<\left(\dfrac{\bm{Q}_k\bm{Q}_k}{Q^2_k}\right)\left[\bm{D}^{T}_{kj}\cdot\bm{Q}_{k}\right]\cdot\bm{F}^{(\text{c})}_{j}\right>=\left<\left[\bm{D}_{kj}:\bm{Q}_{k}\bm{F}^{(\text{c})}_{j}\right]\dfrac{\bm{Q}_k\bm{Q}_k}{Q^2_k}\right>
\end{split}
\end{equation}

Substituting eqs.~(\ref{eq:flow_term}) and (\ref{eq:force_term}) into eq.~(\ref{eq:kk_expanded}), and plugging the resulting expression into eq.~(\ref{eq:kram_kirk_textbook}), the dimensional stress tensor expression is obtained as
\begin{equation}\label{eq:stensor_dimensional}
\begin{split}
&\boldsymbol{\tau}_{\text{p}}={n_{\text{p}}}k_BT\boldsymbol{\delta}-{n_{\text{p}}}\sum_{k=1}^{N}\left<\bm{Q}_{k}\bm{F}^{(\text{c})}_{k}\right>-{n_{\text{p}}K}\sum_{k,j=1}^{N}\left<\left(\bm{M}_{kj}\cdot\boldsymbol{\kappa}\right)^{T}:\left[\dfrac{\bm{Q}_{k}\bm{Q}_{j}\bm{Q}_{k}\bm{Q}_{k}}{Q^2_{k}}\right]\right>\\[5pt]
&+\dfrac{2{n_{\text{p}}}k_BTK}{\zeta}\sum_{k,j=1}^{N}\Biggl[\left<\dfrac{\bm{Q}_k\bm{Q}_k}{Q^2_k}\cdot\bm{D}_{kk}\right>+\left<\bm{D}_{kk}\cdot\dfrac{\bm{Q}_k\bm{Q}_k}{Q^2_k}\right>-\left<\left(\dfrac{\bm{Q}_k\bm{Q}_k}{Q^2_k}\right)\bm{Q}_{k}\cdot\left[\dfrac{\partial}{\partial \bm{Q}_{j}}\cdot\bm{D}^{T}_{kj}\right]\right>\\[5pt]
&-\left<\left[\text{tr}\left(\bm{D}_{kk}\right)-2\bm{D}_{kk}:\dfrac{\bm{Q}_k\bm{Q}_k}{Q^2_k}\right]\dfrac{\bm{Q}_k\bm{Q}_k}{Q^2_k}\right>\Biggr]+\dfrac{2{n_{\text{p}}}K}{\zeta}\sum_{k,j=1}^{N}\left<\left[\bm{D}_{kj}:\bm{Q}_{k}\bm{F}^{(\text{c})}_{j}\right]\dfrac{\bm{Q}_k\bm{Q}_k}{Q^2_k}\right>
\end{split}
\end{equation}
The dimensionless form of eq.~(\ref{eq:stensor_dimensional}) is given by eq.~(43) in the main text.
%